\documentclass[11pt,twoside]{article} 

\usepackage{times}
\usepackage{fancyhdr}
\usepackage{cite}
\usepackage{graphicx}

\newcommand{\ii}{\mathrm{i}}
\newcommand{\dd}{\mathrm{d}}

\date{\today}
 
\title{Heavy Quark Diffusion as a Probe of the Quark-Gluon Plasma}

\author{Ralf Rapp and Hendrik van Hees \\
Cyclotron Institute and Physics Department, Texas A\&M
  University \\ College Station, Texas 77843-3366, U.S.A.} 

\begin{document}

\pagestyle{fancy}
\fancyhead{}
\fancyhead[EC]{Ralf Rapp and Hendrik van Hees}
\fancyhead[EL,OR]{\thepage}
\fancyhead[OC]{Heavy Quark Diffusion as a Probe of the Quark-Gluon Plasma}
\fancyfoot{}
\renewcommand\headrulewidth{0.5pt}
\addtolength{\headheight}{2pt}
 
\maketitle

\begin{abstract}
  In this article we report on recent research on the properties of
  elementary particle matter governed by the strong nuclear force, at
  extremes of high temperature and energy density. At about $10^{12}$
  Kelvin, the theory of the strong interaction, Quantum Chromodynamics
  (QCD), predicts the existence of a new state of matter in which the
  building blocks of atomic nuclei (protons and neutrons) dissolve into
  a plasma of quarks and gluons. The Quark-Gluon Plasma (QGP) is
  believed to have prevailed in the Early Universe during the first few
  microseconds after the Big Bang. Highly energetic collisions of heavy
  atomic nuclei provide the unique opportunity to recreate, for a short
  moment, the QGP in laboratory experiments and study its properties.
  After a brief introduction to the basic elements of QCD in the vacuum,
  most notably quark confinement and mass generation, we discuss how
  these phenomena relate to the occurrence of phase changes in strongly
  interacting matter at high temperature, as inferred from
  first-principle numerical simulations of QCD (lattice QCD). This will
  be followed by a short review of the main experimental findings at the
  Relativistic Heavy Ion Collider (RHIC) at Brookhaven National
  Laboratory. The data taken in collisions of gold nuclei thus far
  provide strong evidence that a QGP has indeed been produced, but with
  rather remarkable properties indicative for an almost perfect liquid
  with unprecedentedly small viscosity and high opacity. We then discuss
  how heavy quarks (charm and bottom) can be utilized to quantitatively
  probe the transport properties of the strongly-coupled QGP (sQGP). The
  large heavy-quark mass allows to set up a Brownian motion approach,
  which can serve to evaluate different approaches for heavy-quark
  interactions in the sQGP. In particular, we discuss an implementation
  of lattice QCD computations of the heavy-quark potential in the
  QGP. This approach generates ``pre-hadronic'' resonance structures in
  heavy-quark scattering off light quarks from the medium, leading to
  large scattering rates and small diffusion coefficients. The resonance
  correlations are strongest close to the critical temperature ($T_c$),
  suggesting an intimate connection to the hadronization of the QGP. The
  implementation of heavy-quark transport into Langevin simulations of
  an expanding QGP fireball at RHIC enables quantitative comparisons
  with experimental data. The extracted heavy-quark diffusion
  coefficients are employed for a schematic estimate of the shear
  viscosity, corroborating the notion of a strongly-coupled QGP in the
  vicinity of $T_c$.
\end{abstract}

\section{Introduction}
\subsection{Elementary Particles and Forces}
The quest for the elementary constituents from which the matter around
us is built has always fascinated mankind. In the fifth century B.C.,
Greek philosophers introduced the notion of an {\em indivisible} entity
of matter, the $\alpha$$\tau$$o$$\mu$$o$$\sigma$ (atom). More than 2000
years passed before this concept was systematized in the nineteenth
century in terms of the chemical elements as the building blocks of the
known substances. The large variety of the chemical elements, however,
called for a deeper substructure within these atoms, which were soon
revealed as bound states of negatively charged electrons ($e^-$) and
positively charged atomic nuclei, held together by their mutual
attraction provided by the electromagnetic force. The nuclei, while very
small in size (but carrying about 99\% of the atom's mass), were found
to further decompose in positively charged protons ($p$) and uncharged
neutrons ($n$), both of approximately equal mass, $M_{p,n}\simeq
0.94$~GeV/$c^2$. This was a great achievement, since at this point all
matter was reduced to 3 particles: $p$, $n$, $e^-$.  There was still the
problem of the stability of the atomic nucleus, since packing together
many positive charges (protons) in a small region of space obviously
implies a large electric repulsion. The solution to this problem
triggered the discovery of the Strong Nuclear Force acting between
nucleons (protons and neutrons); it turned out to be a factor of
$\sim$100 stronger than the electromagnetic one, but with a very short
range of only a few femtometer (1~fm=10$^{-15}$~m). In the 1950's and
1960's, rapid progress in particle accelerator technology opened new
energy regimes in collision experiments of subatomic particles (e.g.,
$p$-$p$ collisions). As a result, many more particles interacting via
the Strong Force (so-called hadrons) were produced and discovered,
including ``strange'' hadrons characterized by a for strongly
interacting particles untypically long lifetime. Again, this
proliferation of states (the ``hadron zoo'') called for yet another
simplification in terms of hadronic substructure. Gell-Mann introduced
three types of ``quarks''~\cite{GellMann:1964nj} as the elementary
constituents from which all known hadrons could be built; they were
dubbed up ($u$), down ($d$) and strange ($s$) quarks, with fractional
electric charges +2/3, -1/3 and -1/3, respectively. In this scheme,
hadrons are either built from 3 quarks (forming baryons, e.g.,
$p=(uud)$, $n=(udd)$), or a quark and an antiquark (forming mesons,
e.g., $\pi^+=(u\bar d)$ or $K^0=(d\bar s$)). Three more heavy quark
``flavors'', carrying significantly larger masses than the light quarks
($u$, $d$, $s$), were discovered in the 1970's -- charm ($c$) and bottom
($b$) -- as well as in 1995 -- the top ($t$) quark.

The discovery of an increasingly deeper structure of the fundamental
matter particles is intimately related to the question of their mutual
forces which, after all, determine how the variety of observed composite
particles is built up. The understanding of fundamental {\em
  interactions} is thus of no less importance than the identification of
the matter constituents. The modern theoretical framework to provide a
unified description are Quantum Field Theories (QFTs), which combine the
principles of Quantum Mechanics with those of Special Relativity. In
QFTs, charged matter particles (fermions of half-integer spin) interact
via the exchange of field quanta (bosons with integer spin). The QFT of
the Electromagnetic Force is Quantum Electro-Dynamics (QED), where the
associated field quantum is the photon ($\gamma$) coupling to electric
charges (positive) and anticharges (negative). The coupling constant (or
charge) of QED is rather small, $\alpha_{\rm em}=e^2/4\pi$=1/137, which
allows one to organize theoretical calculations in a series of terms
characterized by increasing powers of $\alpha_{\rm em}$, so-called
perturbation theory. The smallness of $\alpha_{\rm em}$ then allows for
precise perturbative calculations of electromagnetic observables with an
accuracy exceeding ten significant digits for select quantities,
rendering QED one of the most successful theories in physics.

\begin{figure}[!t]
\begin{center}
\includegraphics[width=0.5\linewidth]{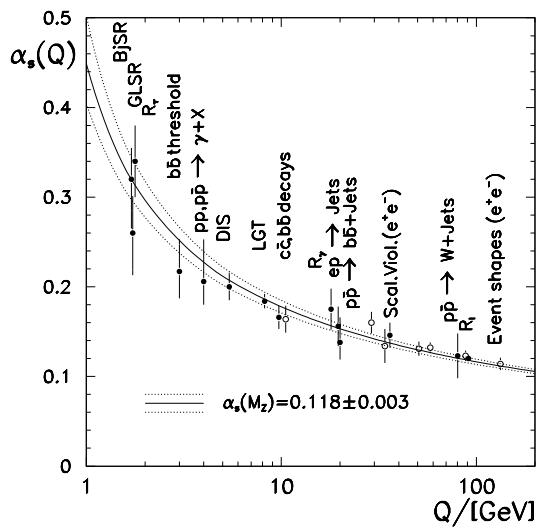}
\end{center}
\caption{Dependence of the QCD coupling constant, $\alpha_s=g^2/4\pi$,
  on the momentum transfer, $Q$ (or inverse distance $1/r\sim Q$), of
  the interaction. Figure taken from Ref.~\cite{Schmelling:1996wm}.}
\label{fig_alphas}
\end{figure}

The QFT of the strong nuclear force, Quantum Chromo-Dynamics (QCD), has
been developed in the early 1970's~\cite{Gross:1973id,Politzer:1973fx}.
The chromo (=color) charge of quarks comes in three variants: red, green
and blue (plus their anticharges), rendering QCD a mathematically more
involved theory. In particular, the force quanta (``gluons'') themselves
carry a nonzero (color-) charge~\cite{Fritzsch:1973pi}, giving rise to
gluon self-interactions. The latter are closely related to another
remarkable property of QCD, namely the ``anti-screening'' of its charges
in the vacuum: quantum fluctuations, i.e., the virtual quark-gluon cloud
around a color charge, induces an increase of the effective charge with
increasing distance, a phenomenon known as asymptotic freedom which
manifests itself in the running coupling constant (or charge) of QCD,
$\alpha_s(Q)$, cf.~Fig.~\ref{fig_alphas}.
On the one hand, the interactions at small distances, $r$ (which, by
means of Heisenberg's uncertainty principle, corresponds to large
momentum transfers, $Q\sim1/r$, in a scattering process), are
comparatively weak and perturbation theory is applicable (much like in
QED). In this regime, QCD is well tested, being in excellent agreement
with experiment (albeit not at the same level of precision as QED; even
at very large $Q$, $\alpha_s$ is still a factor of $\sim$10 larger than
$\alpha_{\rm em}$).  On the other hand, the coupling constant grows
toward small $Q$ entering the realm of ``strong QCD'' where new {\em
  nonperturbative} phenomena occur.  Most notably these are the
Confinement of color charges and the ``Spontaneous Breaking of Chiral
Symmetry'' (SBCS). The former refers to the fact that quarks and gluons
have never been observed as individual particles, but only come in
``colorless'' baryons (where the 3 quarks carry an equal amount of the 3
different color charges) or mesons (where quark and antiquark carry
color charge and anticharge).  Spontaneous Chiral Symmetry Breaking is
closely related to the complex structure of the QCD vacuum; the latter
is filled with various condensates of quark-antiquark and gluon
fields. In particular, the scalar quark condensate of up and down quarks
can be quantified by a vacuum expectation value, $\langle 0|\bar q
q|0\rangle\simeq (-250~{\rm MeV})^3$, translating into a total pair
density of about 4 per fm$^3$.\footnote{In nuclear and particle physics
  it is common practice to use units of $\hbar$ (Planck's constant), $c$
  (speed of light) and $k$ (Boltzmann constant); in these units,
  energies, e.g., can be converted into inverse distance by division
  with $\hbar c \simeq197.33$~MeV\,fm; energies are also equivalent to
  temperature.} Thus, the QCD vacuum is a rather dense state, and the
quarks inside the hadrons propagating through it acquire an effective
mass, $m_{u,d}^*\simeq 350$~MeV, which is much larger than their bare
mass, $m_{u,d}^0\simeq 5$-$10 \; {\rm MeV}$.  The QCD condensates are
thus the main source of the visible (baryonic) mass in the Universe. The
theoretical understanding of the mechanisms underlying Confinement and
SBCS, and their possible interrelation, constitutes a major challenge in
contemporary particle and nuclear physics research.  Currently, the only
way to obtain first-principle information on this nonperturbative realm
of QCD is through numerical lattice-discretized computer simulations
(lattice QCD). However, even with modern day computing power, the
numerical results of lattice QCD computations for observable quantities
are often hampered by statistical and systematic errors (e.g., due to
finite volume and discretization effects); the use of effective models
is thus an indispensable tool for a proper interpretation and
understanding of lattice QCD results, and to provide connections to
experiment.

\subsection{Elementary Particle Matter and the Quark-Gluon Plasma} 
A particularly fascinating aspect of the Strong Force is the question of
what kind of matter (or phases of matter) it gives rise to. The
conventional phases of matter (such as solid, liquid and gas phases of
the chemical elements and their compounds, or even electron-ion plasmas)
are, in principle, entirely governed by the electromagnetic
force. Matter governed by the Strong Force is ``readily'' available only
in form of atomic nuclei: the understanding of these (liquid-like)
droplets of nuclear matter (characterized by a mass density of $\sim
1.67 \cdot 10^{15} {\rm g}/{\rm cm}^3$) is a classical research
objective of nuclear physics. But what happens to nuclear matter under
extreme compression and/or heating? What happens to the (composite)
nucleons?  Is it possible to produce truly elementary-particle matter
where nucleons have dissolved into their quark (and gluon) constituents
(as may be expected from the asymptotic freedom, i.e., small coupling
constant, of the quark and gluon interactions at short distance)?  Does
the condensate structure of the QCD vacuum melt, similar to the
condensate of Cooper pairs in a superconductor at sufficiently high
temperature?  Are there phase transitions associated with these
phenomena?  The investigation of these questions not only advances our
knowledge of strong QCD (including the fundamental problems of
confinement and mass generation), but also directly relates to the
evolution of the early universe, as well as to the properties of
extremely compact stellar objects (so-called neutron stars). Lattice-QCD
computations at finite temperature indeed predict that hadronic matter
undergoes a transition to a state of matter where quarks and gluons are
no longer confined into hadrons.  The temperature required to induce
this transition into the ``Quark-Gluon Plasma'' (QGP) is approximately
$kT\sim 0.2~{\rm GeV} = 2\cdot 10^8$~eV, or $T\sim 10^{12} K$. The
Universe is believed to have passed through this transition at about
10$\mu s$ (=0.00001$s$) after its birth. This was, however, $\sim$ 15
billion years ago, and the question arises how one can possibly study
the QGP, or more generally the phase diagram of QCD matter,
today~\cite{BraunMunzinger:2008tz}. Clearly, without input from
experiment, this would be a hopeless enterprise.

\begin{figure}[!t]
\begin{center}
\includegraphics[width=0.7\linewidth]{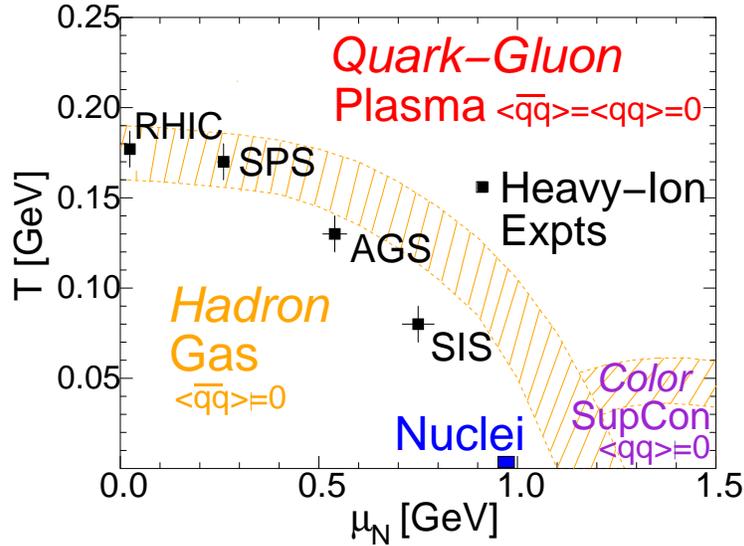}\hspace*{3mm}
\end{center}
\caption{Schematic view of the phase diagram of strongly interacting
  matter, in terms of the nucleon (or baryon) chemical potential,
  $\mu_N$, and temperature, $T$. The former determines the net baryon
  density in the system. The shaded bands are schematic dividers of the
  different phases as expected from theoretical model calculations. At
  vanishing $\mu_N$, current lattice QCD calculations indicate a
  crossover transition from hot hadronic matter to the QGP at a
  (pseudo-) critical temperature of
  $T_c$=160-190~MeV~\cite{Cheng:2006qk,Aoki:2006br}.  Normal nuclear
  matter (as present in atomic nuclei) is located on the $T$=0 axis at
  $\mu_N\simeq970$~MeV (corresponding to a nucleon density of
  $\varrho_N\simeq0.16$~fm$^{-3}$, the nuclear saturation density). At
  larger $\mu_N$ (and small $T$$\le$~50-100~MeV), one expects the
  formation of a Color-Superconductor~\cite{Rapp:1997zu,Alford:1997zt},
  i.e., cold quark matter with a BCS-type condensate of quark Cooper
  pairs, $\langle 0|qq|0\rangle\ne 0$. The ``data'' points are empirical
  extractions of $(\mu_N,T)$-values from the observed production ratios
  of various hadron species ($\pi$, $p$, $K$, $\Lambda$, etc.) in
  heavy-ion experiments at different beam
  energies~\cite{BraunMunzinger:2003zd}.  }
\label{fig_phasedia}
\end{figure}

It turns out that by colliding heavy atomic nuclei at high energies, one
can create highly excited strongly interacting matter in the laboratory.
The incoming kinetic energy of the colliding nuclei is largely
converted into compression and thermal energy, and by varying the
collision energy one is able to produce a wide range of different matter
types as characterized by their baryon density, $\varrho_B$ and
temperature, $T$.  This is illustrated in a schematic phase diagram of
strongly interacting matter in Fig.~\ref{fig_phasedia}.
In the present article we will mainly focus on heavy-ion collisions at
the highest currently available energies. These experiments are being
conducted at the Relativistic Heavy-Ion Collider (RHIC) at Brookhaven
National Laboratory (BNL, Upton, New York): gold (Au) nuclei, fully
stripped of their electrons, are accelerated in two separate beam pipes
to an energy of $E=100$~GeV per nucleon, before being smashed together
head on at four collision points where two large (PHENIX and STAR) and
two smaller (BRAHMS and PHOBOS) detector systems have been positioned.
With each gold nucleus consisting of $A$=197 nucleons (as given by the
atomic mass number of gold), a total (center-of-mass) energy of $E_{\rm
  cm}\simeq 200~A~{\rm GeV} \simeq 40~{\rm TeV} = 4\cdot 10^{13}~{\rm
  eV}$ is brought into the collision zone. Note that the energy of the
accelerated nuclei exceeds their rest mass by more than a factor of 100
(recall that the rest mass of the nucleon is
$M_N\simeq0.94$~GeV/$c^2$). In a central Au-Au collision at RHIC
approximately 5000 particles are produced (as observed in the
detectors), emanating from the collision point with velocities not far
from the speed of light. Most of these particles are pions, but
essentially all known (and sufficiently long-lived) hadrons made of $u$,
$d$ and $s$ quarks are observed.  The key challenge is then to infer
from the debris of produced particles the formation and properties of
the matter - the ``fireball'' - that was created in the immediate
aftermath of the collision.  While the typical lifetime of the fireball
is only about $\sim10^{-22}$s, it is most likely long enough to form
locally equilibrated strongly interacting matter which allows for a
meaningful analysis of its properties in terms of thermodynamic
concepts, and thus to study the QCD phase diagram as sketched in
Fig.~\ref{fig_phasedia}. This has been largely deduced from the
multiplicities and momentum spectra of produced hadrons, which allow to
determine typical temperatures and collective expansion velocities of
the exploding fireball, at least in the later (hadronic) phases of its
evolution. A possibly formed QGP will, however, occur in the earlier
(hotter and denser) phases of a heavy-ion reaction. The identification
and assessment of suitable QGP signatures is at the very forefront of
contemporary research. Hadrons containing heavy quarks (charm and
bottom, $Q=c,b$) have been identified as particularly promising probes
of the QGP. The basic idea is as follows: since charm- and bottom-quark
masses, $m_{c}\simeq1.5$~GeV/$c^2$ and $m_{b}\simeq4.5$~GeV/$c^2$, are
much larger than the typical temperatures, $T\simeq T_c\simeq0.2{\rm
  GeV}$, of the medium formed in a heavy-ion collision, they are (i)
only produced very early in the collision (upon first impact of the
colliding nuclei) and, (ii) not expected to thermalize during the
lifetime of the fireball. Furthermore, the largest changes of their
momentum spectra occur when the collision rate and momentum transfer are
the highest. This is facilitated by a large density and temperature
(i.e., in the early phases of a heavy-ion reaction), but is crucially
dependent also on the interaction strength. Both aspects are embodied
into the notion of transport coefficients.  The main objective of this
article is to provide a theoretical description of heavy-quark transport
in the QGP, and to test the results in applications to RHIC data.

The remainder of this article is organized as follows: In
Sec.~\ref{sec_qgp} we present a general overview of the physics of the
Strong Force and the facets of its different matter phases.  We start in
Sec.~\ref{ssec_qcd} by introducing basic features of Quantum
Chromodynamics (QCD), the quantum field theory describing the strong
interactions between quarks and gluons, the elementary building blocks
of hadrons.  In Sec.~\ref{ssec_phasedia} we briefly review our current
understanding of strongly interacting matter and its phase diagram as
theoretically expected from both numerical lattice QCD computations and
model analysis. In Sec.~\ref{ssec_urhic} we elucidate the main ideas and
achievements of the experimental high-energy heavy-ion programs as
conducted at the Relativistic Heavy Ion Collider (RHIC) at the
Brookhaven National Laboratory (New York), as well as at the Super
Proton Synchrotron (SPS) and the future Large Hadron Collider (LHC) at
the European Organization for Nuclear Research (CERN, Geneva,
Switzerland). We summarize what has been learned about hot and dense
strongly interacting matter thus far, and which questions have emerged
and/or remained open. This leads us to the central part of this article,
Sec.~\ref{sec_hq}, where we discuss in some detail the theoretical
developments and phenomenological applications in using heavy quarks
(charm and bottom) as a probe of the Quark-Gluon Plasma. In
Sec.~\ref{ssec_hq-int} we concentrate on the theoretical understanding
of heavy-quark (HQ) interactions in the QGP. We mostly address elastic
scattering, within both perturbative and nonperturbative approaches,
where the latter are divided into a resonance model and $T$-matrix
calculations utilizing potentials extracted from lattice QCD. The HQ
interactions in the medium are used to compute pertinent self-energies
and transport coefficients for drag and diffusion. In
Sec.~\ref{ssec_transport} the latter are implemented into a Brownian
motion framework of a Fokker-Planck equation which is particularly
suitable for describing the diffusion of a heavy particle in a heat
bath.  In Sec.~\ref{ssec_obs} these concepts are applied to heavy-ion
collisions, by implementing a Langevin simulation of HQ transport into
realistic QGP fireball expansions for Au-Au collisions at RHIC. To make
contact with experiment, the quarks have to be hadronized which involves
a quark coalescence approach at the phase transition, in close
connection to successful phenomenology in light hadron spectra.  This is
followed by an analysis of transverse momentum spectra of heavy mesons
and their electron decay spectra for which experimental data are
available. In Sec.~\ref{ssec_sqgp} we recapitulate on the ramifications
of the theoretical approach in a broader context of Quark-Gluon Plasma
research and heavy-ion phenomenology, and outline future lines of
investigation.  Sec.~\ref{sec_concl} contains a brief overall summary
and conclusions.

\section{The Quark-Gluon Plasma and Heavy-Ion Collisions}
\label{sec_qgp}

\subsection{The Strong Force and Quantum Chromodynamics}
\label{ssec_qcd}
\begin{figure}[!t]
\begin{center}
\begin{minipage}{0.25 \linewidth}
\includegraphics[width=\textwidth]{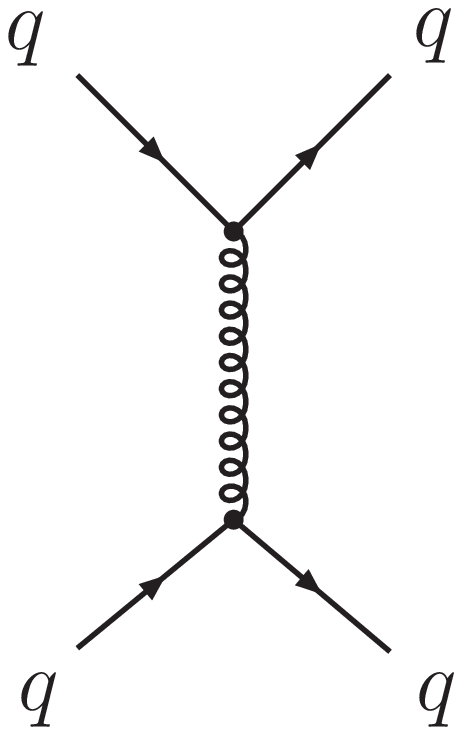}
\end{minipage} \hspace*{1.5cm}
\begin{minipage}{0.4 \linewidth}
\includegraphics[width=\textwidth]{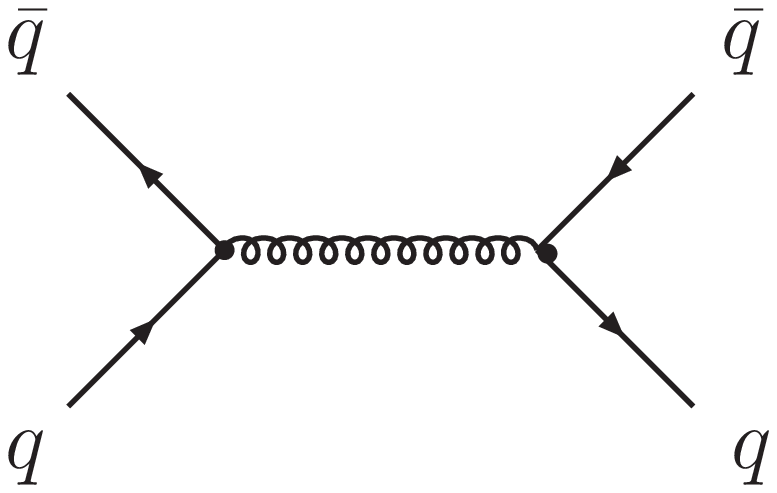}
\end{minipage}\newline
\begin{minipage}{0.4\linewidth}
\includegraphics[width=\textwidth]{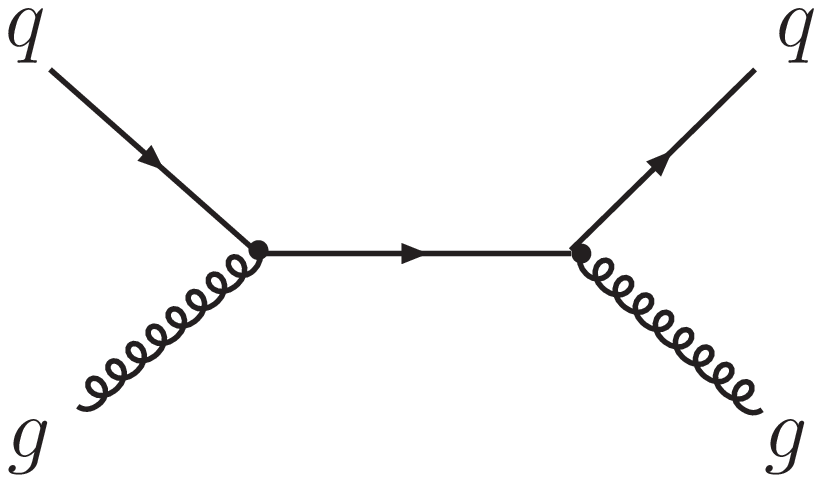}
\end{minipage}\hspace*{0.7cm}
\begin{minipage}{0.25\linewidth}
\includegraphics[width=\textwidth]{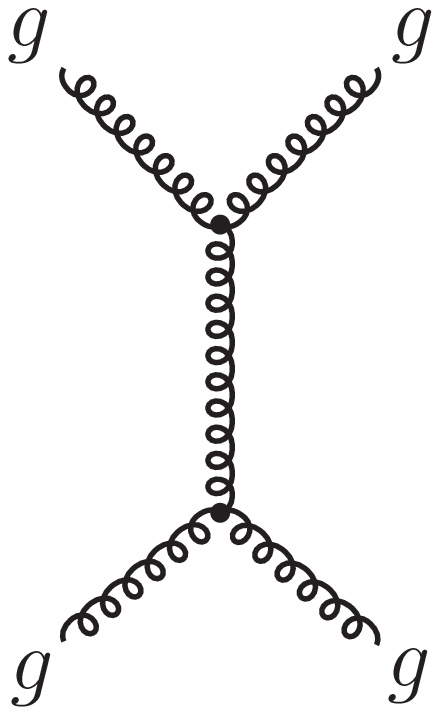}
\end{minipage}
\begin{minipage}{0.25\linewidth}
\includegraphics[width=\textwidth]{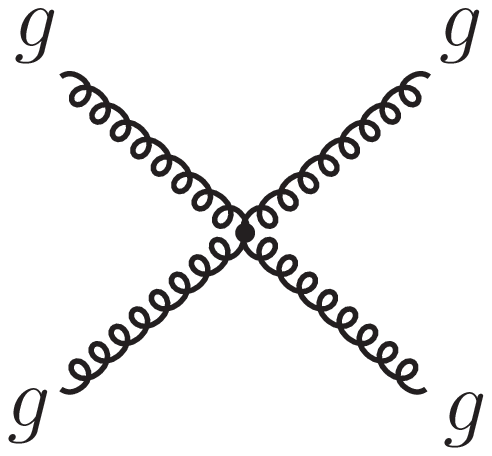}
\end{minipage}
\end{center}
\caption{Feynman diagrams for quark-quark scattering via the exchange of
  a gluon (top left) quark-antiquark scattering via annihilation into a
  gluon (top right), quark-gluon scattering and gluon-gluon scattering
  (bottom panels).}
\label{fig_gluo-ex}
\end{figure}

\begin{figure}[!t]
\begin{center}
\includegraphics[width=0.6\linewidth]{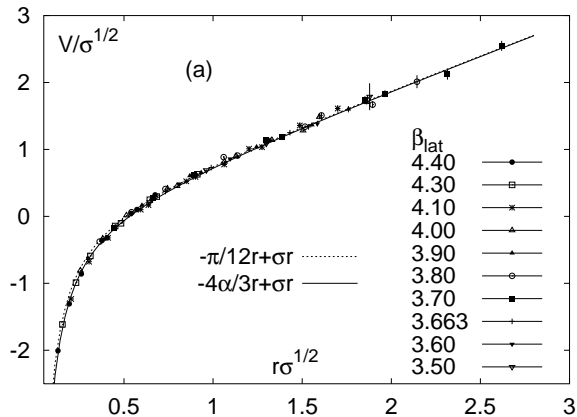}
\end{center}
\caption{The static potential between a heavy quark and antiquark in
  vacuum, as a function of their distance, as computed in lattice
  QCD~\cite{Kaczmarek:2005ui}. Potential and distance are given in units
  of the string tension, $\sqrt{\sigma}\simeq 
  0.42~{\rm GeV}\simeq$~2.12~fm$^{-1}$ (where the conversion has been done
  using $\hbar c \simeq 0.197$~GeV~fm).}
\label{fig_hq-pot}
\end{figure}

The basic quantity which, in principle, completely determines the theory
of the strong interaction, is the Lagrangian of QCD,
\begin{equation}
{\cal L_{\rm QCD}} = 
\bar q \ (\ii \not\!\!{D}- \hat{m}_q) \ q -\frac{1}{4} G_{\mu\nu} G^{\mu\nu} \ ,
\label{lqcd}
\end{equation}
where $q$ and $\bar q$ denote the elementary matter fields, quarks and
antiquarks. The quark fields are specified by several quantum numbers:
(i) color charge (red, green or blue), (ii) flavor (up, down, strange,
charm, bottom and top) and (iii) spin ($\pm\frac{1}{2}\hbar$).  The mass
matrix $\hat{m}_q={\rm diag}(m_u, m_d, m_s, m_c, m_b, m_t)$ is a simple
diagonal matrix in flavor space. It roughly separates QCD into a
light-flavor ($u$, $d$ with bare masses
$m_{u,d}$$\simeq$0.005~GeV/$c^2$) and a heavy-flavor sector ($c$, $b$,
$t$ with $m_c$$\simeq$1.3~GeV/$c^2$, $m_b$$\simeq$4.5~GeV/$c^2$,
$m_t$$\simeq$175~GeV/$c^2$), while the strange-quark mass is somewhat in
between ($m_s$$\simeq$0.12~GeV/$c^2$).  The interactions of the quarks
are encoded in the covariant derivative, $\not\!\!{D}=\not\!\!\partial -
\ii g \not\!\!\!A$, where $A$ denotes the gluon field, the carrier of
the Strong Force, and the gauge coupling $g$ quantifies the interaction
strength as referred to in Fig.~\ref{fig_alphas}. A pictorial
representation of these interactions can be given in terms of Feynman
diagrams: quarks interact via the exchange of gluons,
cf.~Fig.~\ref{fig_gluo-ex}.
The gluons are massless particles with spin 1; most notably, they also
carry color charge (which comes in 8 different charge-anticharge
combinations, e.g., red-antigreen, red-antiblue, etc.), giving rise to
gluon self-interactions. These are encoded in the last term of
Eq.~(\ref{lqcd}) where $G_{\mu\nu}=\partial_\mu A_\nu -\partial_\nu
A_\mu + g[A_\mu,A_\nu]$; the commutator term, $[A_\mu,A_\nu]= A_\mu\cdot
A_\nu - A_\nu\cdot A_\mu$ is a consequence of the 3$\times$3 matrix
structure in color-charge space and generates 3- and 4-gluon interaction
vertices (as depicted in the bottom right panel of
Fig.~\ref{fig_gluo-ex}).  Each interaction vertex brings in a factor of
$g$ into the calculations of a given diagram (except for the 4-gluon
vertex which is proportional to $g^2$). All diagrams (more precisely,
the pertinent scattering amplitudes) shown in Fig.~\ref{fig_gluo-ex} are
thus proportional to $\alpha_s=g^2/4\pi$, while more complicated
diagrams (involving additional quark/gluon vertices) are of higher order
in $\alpha_s$. This forms the basis for perturbation theory: for small
$\alpha_s$, higher order diagrams are suppressed, ensuring a rapid
convergence of the perturbation series.  Perturbative QCD (pQCD) indeed
works very well for reactions at large momentum transfer, $Q$, where
$\alpha_s(Q)$ is small, cf.~Fig.~\ref{fig_alphas}, while it breaks down
for small $Q$.  One possibility to obtain information on the
nonperturbative interactions is to investigate the potential between two
static quark charges, i.e., the potential between a heavy quark and
antiquark. In the color neutral channel, a phenomenological ansatz can
be written as a combination of a Coulombic attraction at short distances
which merges into a linearly rising potential at large distance
(signifying confinement),
\begin{equation}
V_{Q\bar Q}(r) = -\frac{4}{3} \frac{\alpha_s}{r} + \sigma r \ ,
\label{Vqqbar}
\end{equation}
where $\sigma\simeq 1$~GeV/fm denotes the ``string tension''.  Such a
potential provides a good description of the observed spectra of heavy
quarkonium states, i.e., charm-anticharm and bottom-antibottom quark
bound states. In recent years, the heavy-quark potential has been
computed with good precision in lattice QCD, which fully confirmed the
phenomenological ansatz, Eq.~(\ref{Vqqbar}), (see, e.g.,
Fig.~\ref{fig_hq-pot}). Subsequently, the potential approach, in
combination with expansions organized in powers of the inverse HQ mass,
$m_{c,b}$ (rather than the coupling constant), has been developed into
an effective theory of low-energy QCD, cf. Ref.~\cite{Brambilla:2004wf}
for a review.

Besides the (external) quark masses, QCD has only one (intrinsic)
dimensionful scale which is generated by quantum effects (loop
corrections). The latter give rise to the running coupling constant,
\begin{equation}
  \alpha_s(Q)= \frac{1}{(11N_c-2 N_f) \ln(Q^2/\Lambda_{\rm QCD}^2)} \ ,
\end{equation}
where $N_c$=3 is the number of color charges and $N_f$ the number of
active quark flavors at given $Q$ (i.e., the number of flavors with $m_q
\le Q$).  It is tempting to interpret the value of $\Lambda_{\rm
  QCD}=0.2$~GeV as the dividing line between perturbative and
nonperturbative regimes of QCD. In practice, however, the scale for the
onset of nonperturbative effects is significantly larger, typically
given by the hadronic mass scale of $\sim$1~GeV.  To understand the
emergence of this scale, it is important to realize that the QCD vacuum
structure is rather rich, characterized by quark and gluon
condensates. E.g., in the light scalar quark-antiquark channel ($\bar
uu$ and $\bar dd$) a strongly attractive force leads to the spontaneous
formation and condensation of $\bar qq$ pairs (reminiscent to a Bose
condensate)\footnote{The origin of this force is most likely not the
  perturbative exchange of gluons, but nonperturbative gluon
  configurations - so-called instantons - which correspond to tunneling
  events between topologically different vacua and are characterized by
  a 4-dimensional ``radius'' of $\rho\simeq(0.6~{\rm GeV})^{-1}$, cf.
  Ref.~\cite{Schafer:1996wv} for a review}.  An important consequence of
the condensate formation is that the light quarks acquire an effective
mass when propagating through the condensed vacuum, which is given by
the condensate as $m_q^*\simeq G \langle 0|\bar qq|0\rangle \simeq
0.4$~GeV, where $G$ is an (instanton-induced) effective quark coupling
constant.  Note that this mass exceeds the bare quark masses by about a
factor of $\sim$100, being the major source of the proton mass, $M_p\sim
3m_q^*$, and thus of the visible mass in the Universe.  Formally, the
presence of the constituent quark mass (and quark condensate) is closely
related to the phenomenon of ``Spontaneous Breaking of Chiral Symmetry''
(SBCS): in the limit of vanishing bare-quark masses (which is a good
approximation for the very light $u$ and $d$ quarks), the QCD Lagrangian
is invariant under rotations in isospin and chirality (handedness),
i.e., transformations that change $u$ into $d$ quarks and left-handed
into right-handed quarks. This invariance is equivalent to the
conservation of isospin and chiral quantum numbers of a quark. However,
the constituent quark mass breaks the chiral symmetry (i.e., massive
quarks can change their chirality so that it is no longer conserved).
SBCS not only manifests itself in the QCD ground-state, but also in its
excitation spectrum, i.e., hadrons. Therefore, hadronic states which
transform into each other under ``chiral rotations'' (so-called chiral
multiplets or partners) are split in mass due to SBCS. Prominent
examples in the meson spectrum are $\pi$(140)-$\sigma$(400-1200) and
$\rho$(770)-$a_1$(1260), or $N$(940)-$N^*$(1535) in the nucleon
spectrum.

In the following Section, we will discuss how the presence of strongly
interacting matter affects the nonperturbative structure of the QCD
vacuum and the interactions therein.

\subsection{Strongly Interacting Matter and the QCD Phase Diagram}
 \label{ssec_phasedia}

\begin{figure}[!t]
\hspace{-0.7cm}
\includegraphics[width=0.53\linewidth]{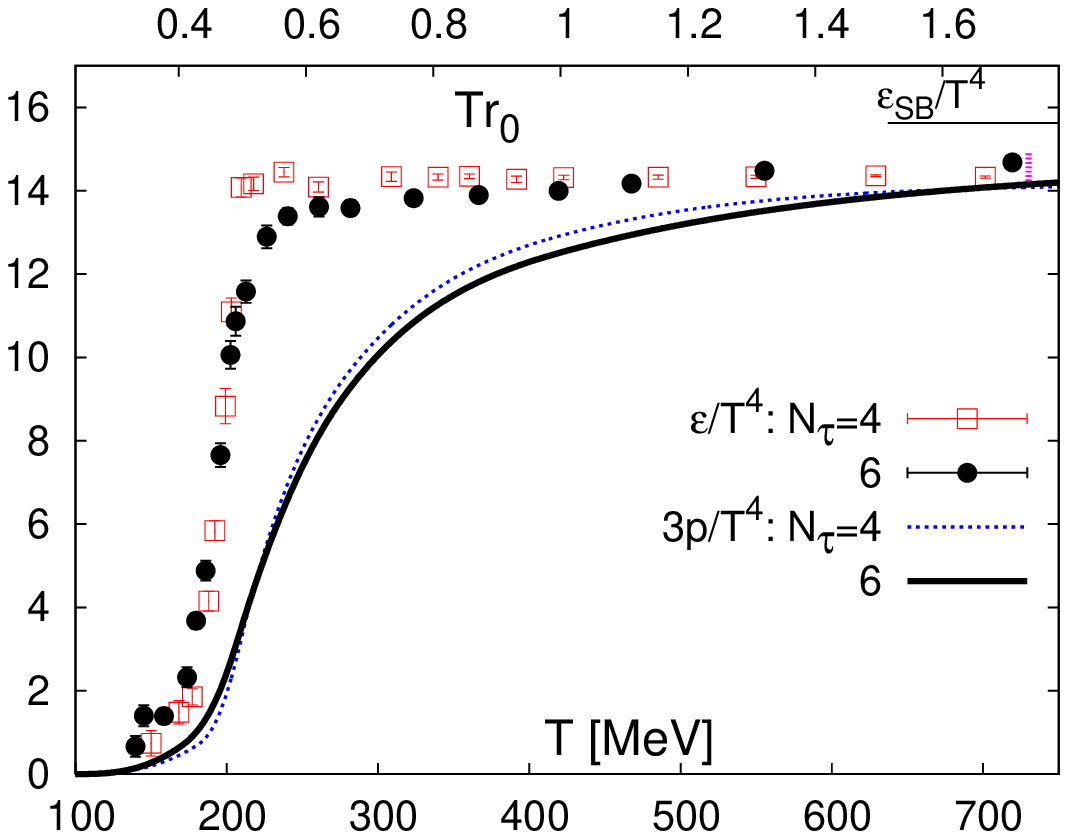}
\includegraphics[width=0.53\linewidth]{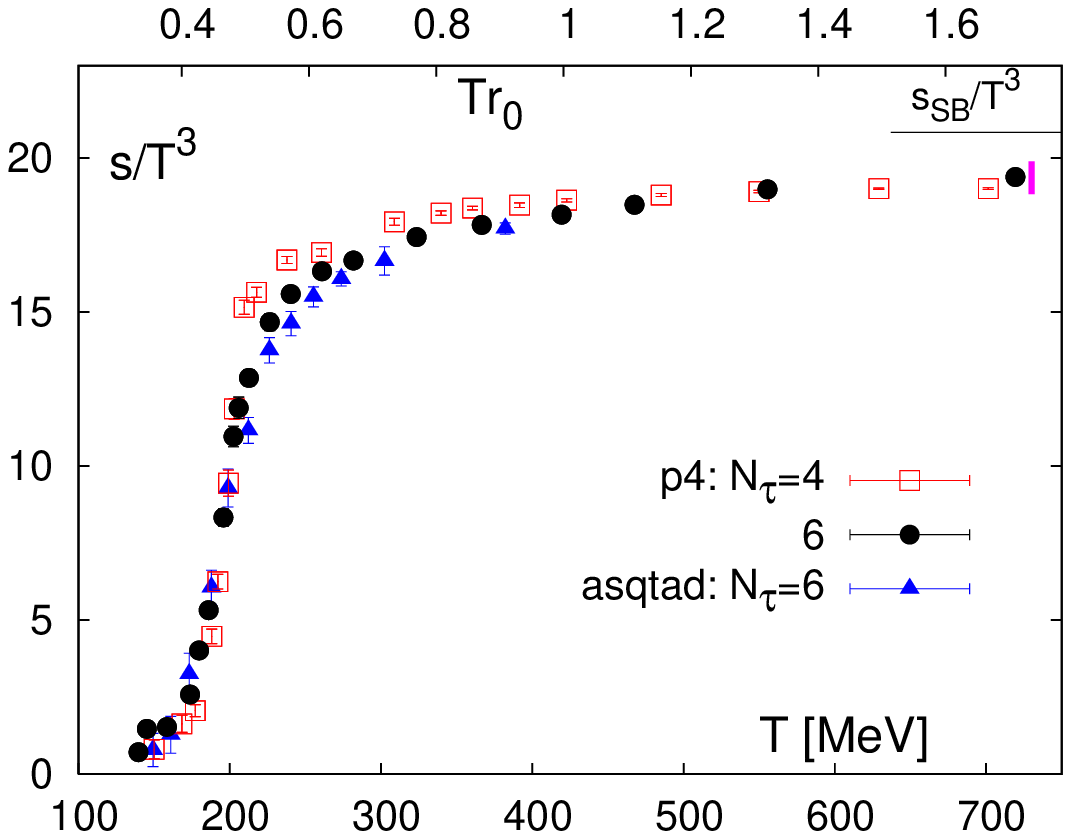}
\caption{The equation of state of strongly interacting matter as
  computed in lattice QCD~\cite{Cheng:2007jq} in terms of the energy
  density (left panel) and entropy density (right panel) as a function
  of temperature (at $\mu_B=0$). The computations use bare light and
  strange quark masses close to their physical values.}
\label{fig_eos-lqcd}
\end{figure}

 When heating a condensed state, the general expectation is that the
 condensate eventually ``melts'' (or ``evaporates''), and that the
 interactions are screened due to the presence of charged particles in
 the medium. Transferring this expectation to the QCD vacuum implies
 that, at sufficiently large temperature, the condensates should vanish
 and the quarks and gluons are released from their hadronic bound states
 (deconfinement), forming the Quark-Gluon Plasma (QGP).  Numerical
 lattice QCD (lQCD) computations of the thermodynamic partition function
 at finite temperature have substantially quantified this notion over
 the last two decades or so. The pertinent equation of state (EoS),
 i.e., the pressure, energy and entropy density, indeed exhibits a
 rather well defined transition, as shown in Fig.~\ref{fig_eos-lqcd} for
 a lQCD calculation with close to realistic input for the bare light and
 strange quark masses.
At high temperatures ($T>3T_c$) the equation of state is within
$\sim$15\% of the values expected for an ideal gas, known as the
Stefan-Boltzmann (SB) limit. The SB values are given by
$\varepsilon_{q\bar q}=\frac{7}{8} d_{q\bar q} \frac{\pi^2}{30} T^4$ and
$\varepsilon_{g} = d_{g} \frac{\pi^2}{30} T^4$ with degeneracies
$d_{q\bar q} = N_c N_s N_{\bar q} N_f =12 N_f$ for quarks plus
antiquarks ($N_c$=3 for red, green and blue colors, $N_s$=2 for spin up
and down and $N_{\bar q}$=2 for anti-/quarks; $N_f$ is the number of
massless flavors), and $d_{g}=N_s N_c=16$ for gluons ($N_c$=8
color-anticolor combinations, $N_s$=2 transverse spin polarizations); 
the relative factor of 7/8 is due to the difference of Fermi vs. Bose
distribution functions for quarks vs. gluons. For $N_f$=3 one finds
$\varepsilon_{\rm SB}=\varepsilon_{q\bar
  q}+\varepsilon_{g}\simeq15.6~T^4$, as indicated by the
``$\varepsilon_{\rm SB}$'' limit in the upper right corner of the left
panel in Fig.~\ref{fig_eos-lqcd}. Likewise, using $P=s T - \varepsilon$
(for quark chemical potential $\mu_q=0$),
one finds for the SB limit of the entropy density
$s=(10.5N_f+16)\frac{4\pi^2}{90} T^3 \simeq 20.8~T^3$, cf.~right panel
of Fig.~\ref{fig_eos-lqcd}.

\begin{figure}[!t]
\hspace{-0.4cm}
\includegraphics[width=0.51\linewidth]{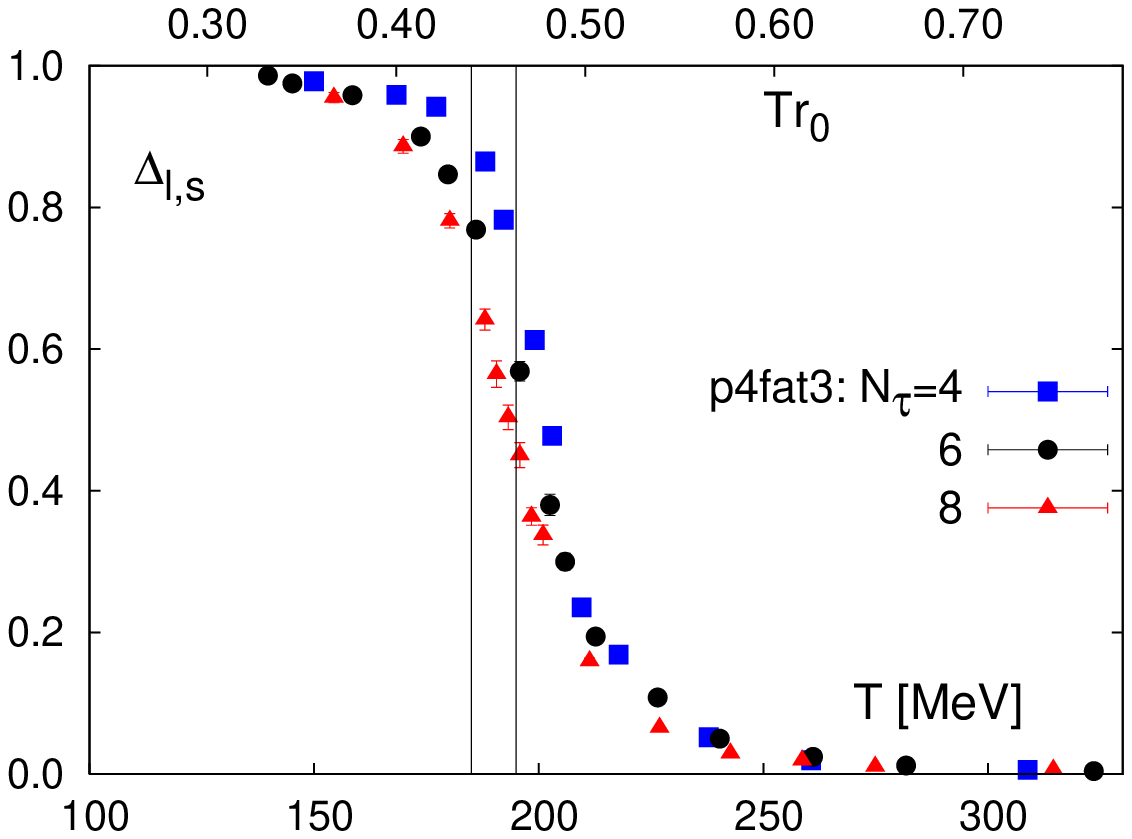}
\includegraphics[width=0.51\linewidth]{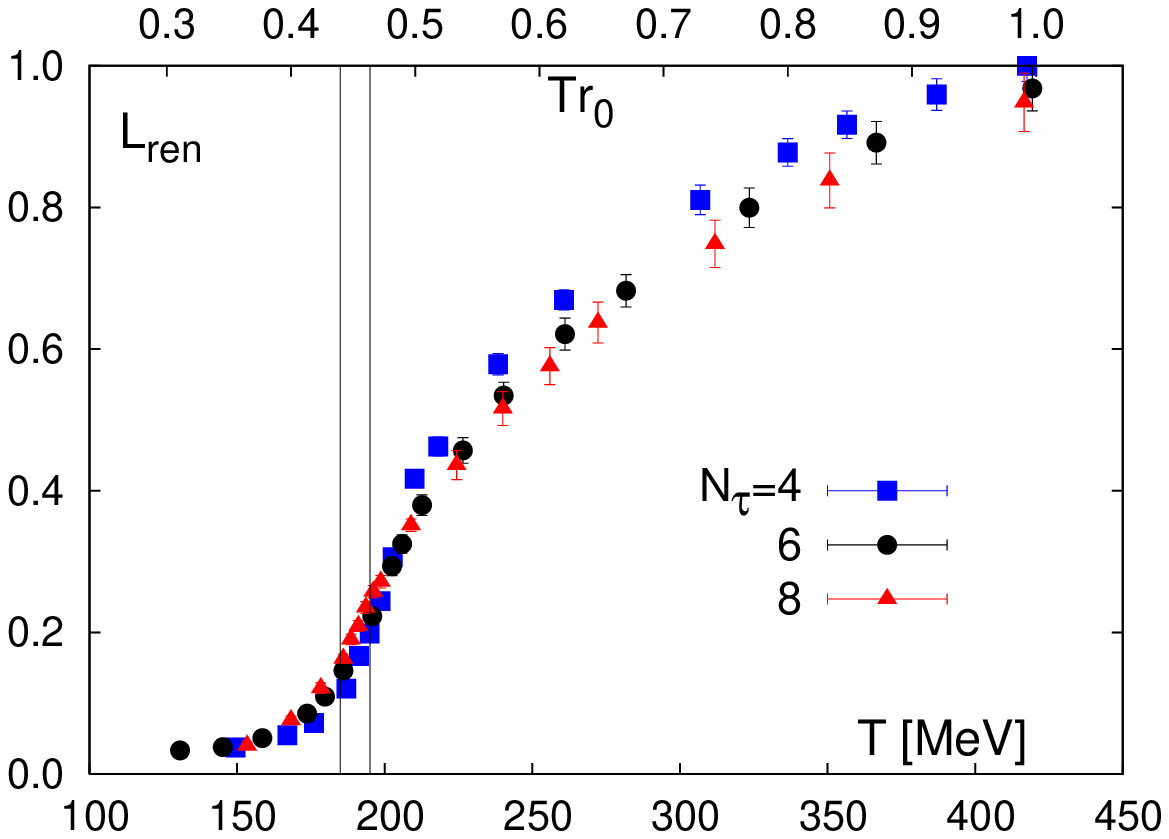}
\caption{Lattice QCD computations of the (subtracted and normalized)
  light-quark chiral condensate (left panel) and the (renormalized)
  Polyakov loop expectation value as a function of
  temperature~\cite{Cheng:2007jq,Detar:2007as}.  }
\label{fig_op-lqcd}
\end{figure}

Returning to the phase transition region, the rapid change in
$\varepsilon$ is accompanied by comparably sudden changes in the quark
condensate, and the expectation value of the so-called Polyakov loop, an
order parameter of deconfinement, cf.~Fig.~\ref{fig_op-lqcd}. The latter
is, roughly speaking, proportional to the exponent of the heavy-quark
free energy at large distance, ${\rm e}^{-F_{\bar QQ}^\infty/T}$, which
vanishes in the confined phase (or at least becomes very small since
$F_{Q\bar Q}^\infty\equiv F_{Q\bar Q}(r\to\infty)$ is large), but is
finite in the deconfined QGP.

\begin{figure}[!t]
\hspace{-0.4cm}
\includegraphics[width=0.51\linewidth]{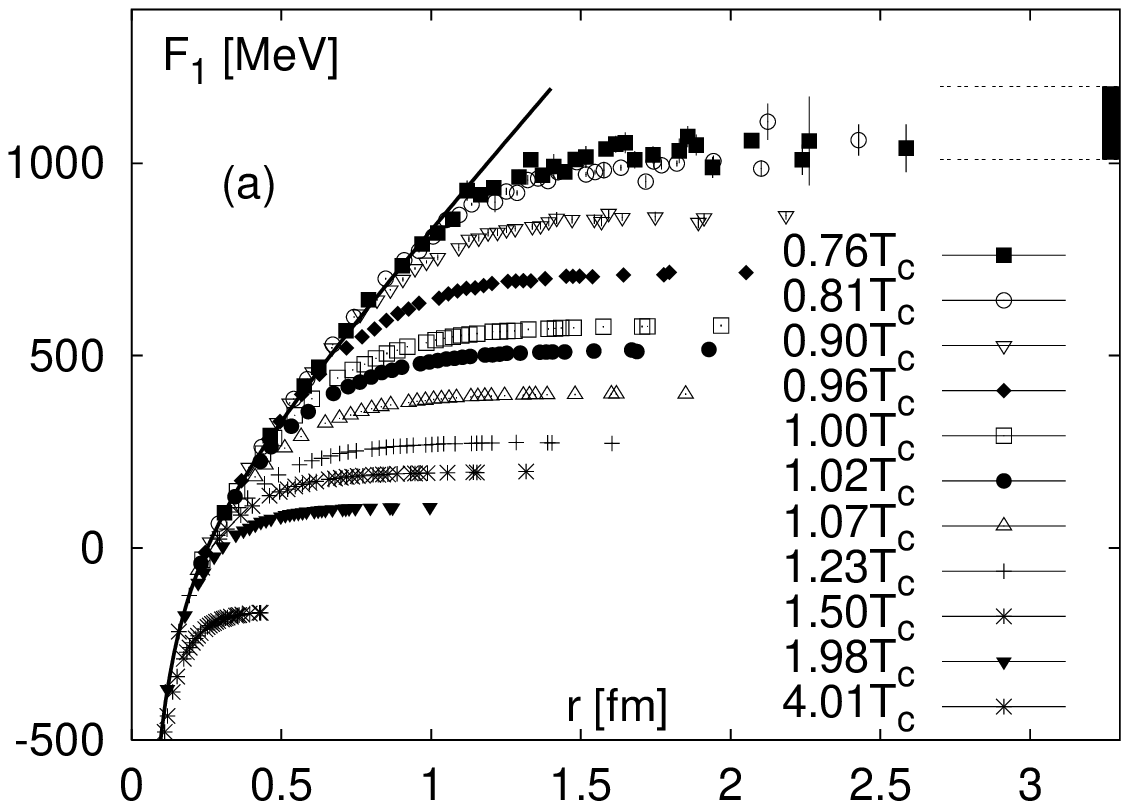}
\includegraphics[width=0.51\linewidth]{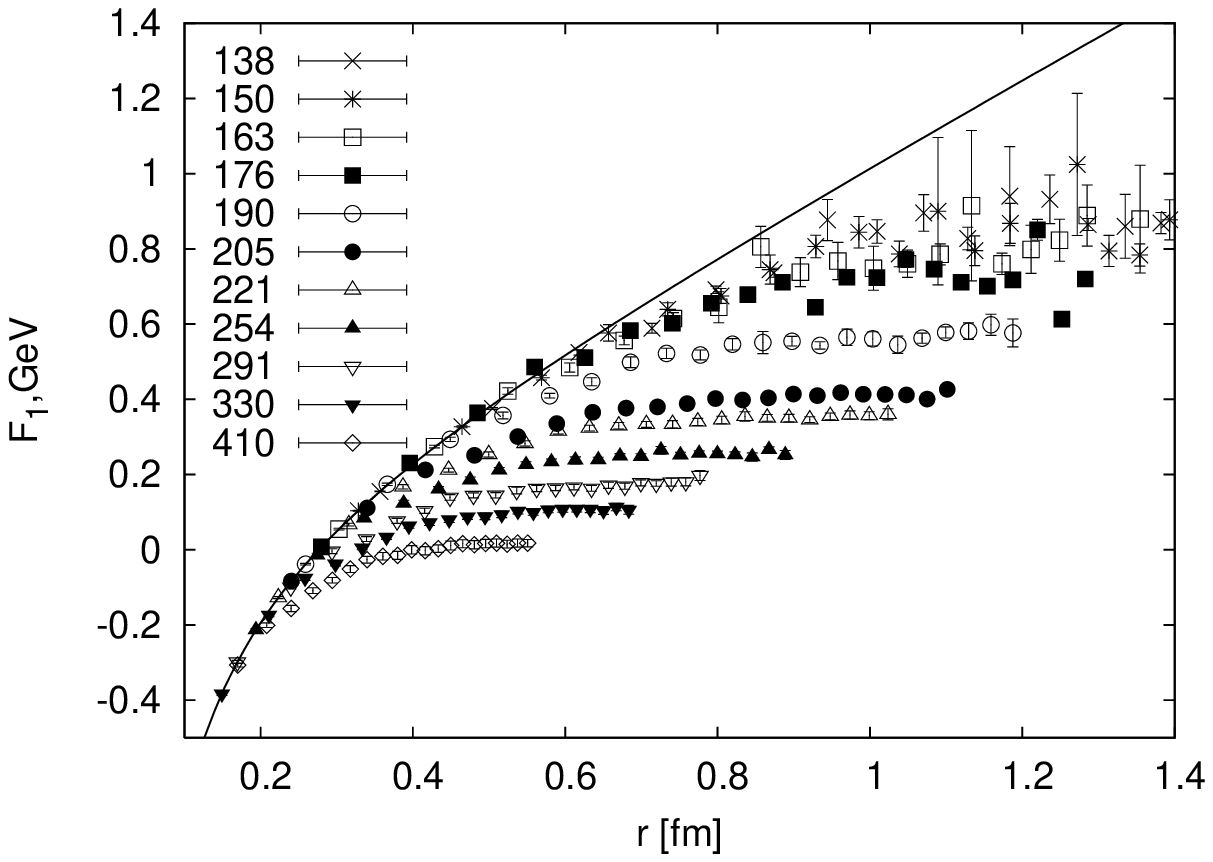}
\caption{The heavy-quark free energy as a function of size, $r$, of the
  HQ pair, for various temperatures as computed in lattice
  QCD~\cite{Kaczmarek:2005ui,Petreczky:2004pz}.}
\label{fig_F-T-lqcd}
\end{figure}

Detailed studies of the HQ free energy as a function of the relative
distance, $r$, of the $Q$-$\bar Q$ pair have also been conducted in
finite-$T$ lattice QCD, see, e.g., Fig.~\ref{fig_F-T-lqcd}.  
One finds
the qualitatively expected behavior that the interaction is increasingly
screened with increasing temperature, penetrating to smaller distances,
as is characteristic for a decreasing color-Debye screening length (or,
equivalently, increasing Debye mass, $\mu_D$).  However, the implications
of these in-medium modifications for the binding of quarkonium states
are quite subtle. The first problem is that, unlike in the vacuum case,
the identification of the free energy with an interaction potential is
no longer straightforward, due to the appearance of an entropy term, $T
S_{Q\bar Q}$,
\begin{equation}
F_{Q\bar Q}(r,T) = U_{Q\bar Q}(r,T) -T S_{Q\bar Q}(r,T) ,
\label{FQQbar}
\end{equation}
where $U_{Q\bar Q}$ denotes the internal energy. It is currently an open
problem whether $F_{Q\bar Q}$~\cite{Digal:2001iu} or $U_{Q\bar
  Q}$~\cite{Shuryak:2004tx,Mannarelli:2005pz,Mocsy:2005qw,
  Alberico:2005xw,Cabrera:2006wh} (or even a combination
thereof~\cite{Wong:2004zr}) is the most suitable quantity to be inserted
into a potential model calculation for quarkonium states in the medium
(typically carried out using a Schr\"odinger equation).
The different choices lead to considerably different results for the 
quarkonium binding in the QGP.  On the one hand, when directly using the 
free energy $F_{Q\bar Q}$, the charmonium ground state (the $S$-wave 
$J/\psi$ or $\eta_c$ states) dissolves not far above the critical 
temperature, at about 1.2~$T_c$~\cite{Digal:2001iu}. On the other hand, 
the use of $U_{Q\bar Q}$ implies deeper potentials and thus stronger 
binding, and the ground-state charmonium survives up to temperatures of
$\sim$2-2.5~$T_c$. The stronger bound bottomonia are more robust and 
may not dissolve until $\sim$4~$T_c$.

\begin{figure}[!t]
\hspace{-0.0cm}
\includegraphics[width=0.46\linewidth]{A_jpsi-lqcd.eps}
\hspace{0.2cm}
\includegraphics[width=0.48\linewidth]{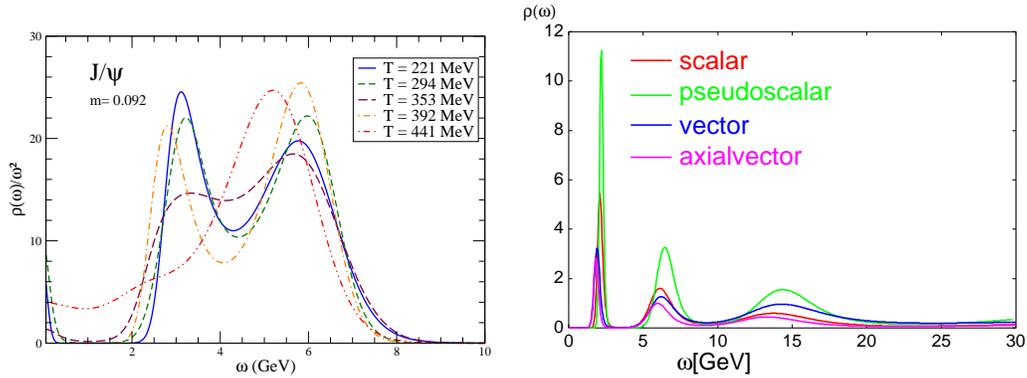}
\caption{Meson spectral functions in the Quark-Gluon Plasma, based on
  imaginary-time correlation functions computed in finite-$T$ lattice
  QCD, followed by a transformation in the physical regime using the
  Maximum Entropy Method. Left panel: $S$-wave charm-anticharm channel
  (spin-parity $J^P=1^-$, corresponding to the $J/\psi$ meson) in a QGP
  with $N_f=2$ flavors~\cite{Aarts:2007pk}; the critical temperature in
  these calculations is about $T_c\simeq 200$~MeV.  Right panel:
  quark-antiquark channels with a bare quark mass corresponding to
  strange quarks~\cite{Asakawa:2002xj}, in the scalar ($J^P=0^+$),
  pseudoscalar ($J^P=0^-$), vector ($J^P=1^-$) and axialvector
  ($J^P=1^+$) channels in a pure gluon plasma at $T$=1.4~$T_c$.  }
\label{fig_A-lqcd}
\end{figure}

An alternative way to address the problem of quarkonium dissociation in
the QGP is the computation of $Q$-$\bar Q$ correlation functions, which
are basically thermal Green's functions (or $Q$-$\bar Q$
propagators). However, the oscillating nature of the propagator for
physical energies (i.e., in real time) renders numerical lQCD
evaluations intractable. However, one can apply a trick by transforming
the problem into ``imaginary time'', where the propagators are
exponentially damped. The price one has to pay is an analytical
continuation back into the physical (real time) regime of positive
energies. This transformation is particularly problematic at finite
temperature, where the imaginary-time axis in the statistical operator
(free energy) is given by the inverse temperature, which severely limits
the accuracy in the analytic continuation (especially for a limited
discrete number of points on the imaginary-time axis). However,
probabilistic methods, in particular the so-called Maximum Entropy
Method (MEM), have proven valuable in remedying this
problem~\cite{Asakawa:2000tr}.  LQCD computations of charmonium
correlation functions in the QGP, followed by a MEM analysis to extract
pertinent spectral functions, support the notion that the ground state
($S$-wave) charmonium survives up to temperatures of
$\sim$1.5-2~$T_c$~\cite{Aarts:2007pk}, cf., e.g., the left panel of
Fig.~\ref{fig_A-lqcd}.  Even in the light-quark sector lQCD computations
indicate the possibility that mesonic resonance states persist above the
phase transition~\cite{Asakawa:2002xj,Karsch:2003jg}, as illustrated in
the right panel of Fig.~\ref{fig_A-lqcd}.

To summarize this section, first principle lattice-QCD calculations at
finite temperature have confirmed that hadronic matter undergoes a
transition into a Quark-Gluon Plasma. This transition is characterized
by rapid changes in the equation of state around a temperature of
$T_c$=0.15-0.20~GeV, which is accompanied by variations in order
parameters associated with deconfinement and the restoration of chiral
symmetry (i.e., vanishing of the chiral quark condensate).  While
thermodynamic state variables are within 15\% of the ideal gas limit at
temperatures of $\ge$3~$T_c$, the analysis of the heavy-quark potential
and mesonic spectral functions indicate substantial nonperturbative
effects at temperatures below $\sim$2~$T_c$. This suggests that up to
these temperatures the QGP is quite different from a weakly interacting
gas of quarks and gluons. Substantial progress in our understanding of
hot and dense QCD matter has emerged on a complementary front, namely
from experiments using ultrarelativistic heavy-ion collisions. The
following Section gives a short overview of the key observations and
pertinent theoretical interpretations.

\subsection{Relativistic Heavy Ion Collisions and the Quest for the QGP}
\label{ssec_urhic}
\begin{figure}[!t]
\hspace{-0.0cm}
\centerline{\includegraphics[width=0.95\linewidth]{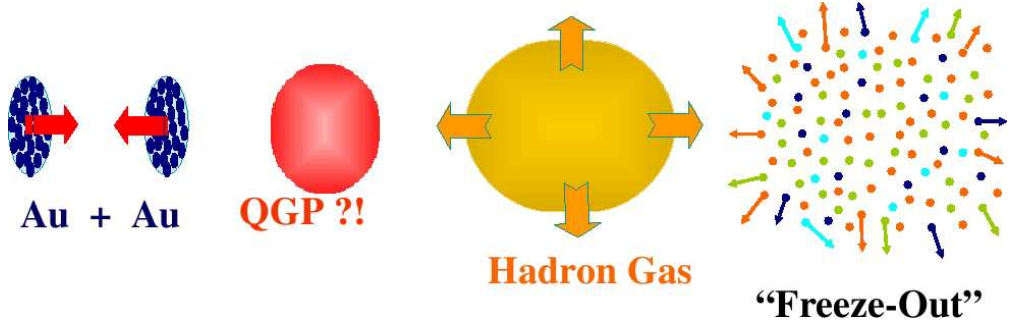}}
\caption{Schematic representation of the various stages of a heavy-ion
  collision. From left to right: incoming nuclei at highly relativistic
  energies moving at close to the speed of light (which induces a
  substantial Lorentz contraction relative to their transverse size; at
  full RHIC energy of 100+100~GeV, the Lorentz contraction is a factor
  of $\sim$1/100); upon initial impact of the nuclei, primordial
  (``hard'') nucleon-nucleon collisions occur; further reinteractions
  presumably induce the formation of a Quark-Gluon Plasma (after
  $\tau_0=0.5-1$~fm/$c$), whose pressure drives a collective expansion
  and cooling (for a duration of $\tau_{\rm QGP}\simeq 5$fm/$c$),
  followed by hadronization and further expansion in the hadronic phase
  (for a duration of $\tau_{\rm HG}\simeq 5-10$fm/$c$); at thermal
  freezeout the (short-range) strong interactions cease (after
  approximately 15~fm/$c$ total fireball lifetime).}
\label{fig_hi-evo}
\end{figure}

\begin{figure}[!t]
\hspace{-0.0cm}
\begin{center}
\begin{minipage}{0.35\linewidth}
\includegraphics[width=\textwidth]{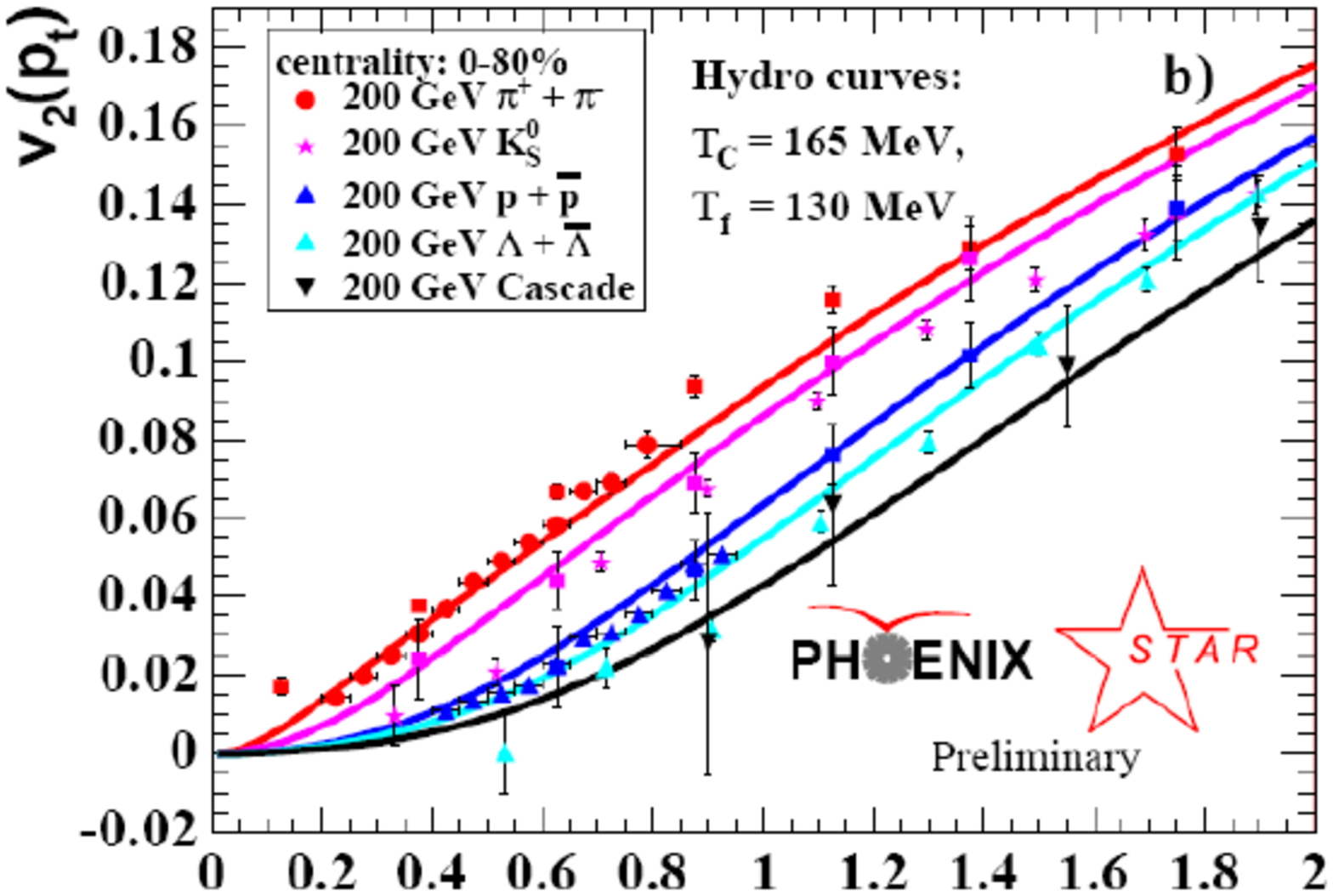}
\end{minipage}
\begin{minipage}{0.25\linewidth}
\includegraphics[width=\textwidth]{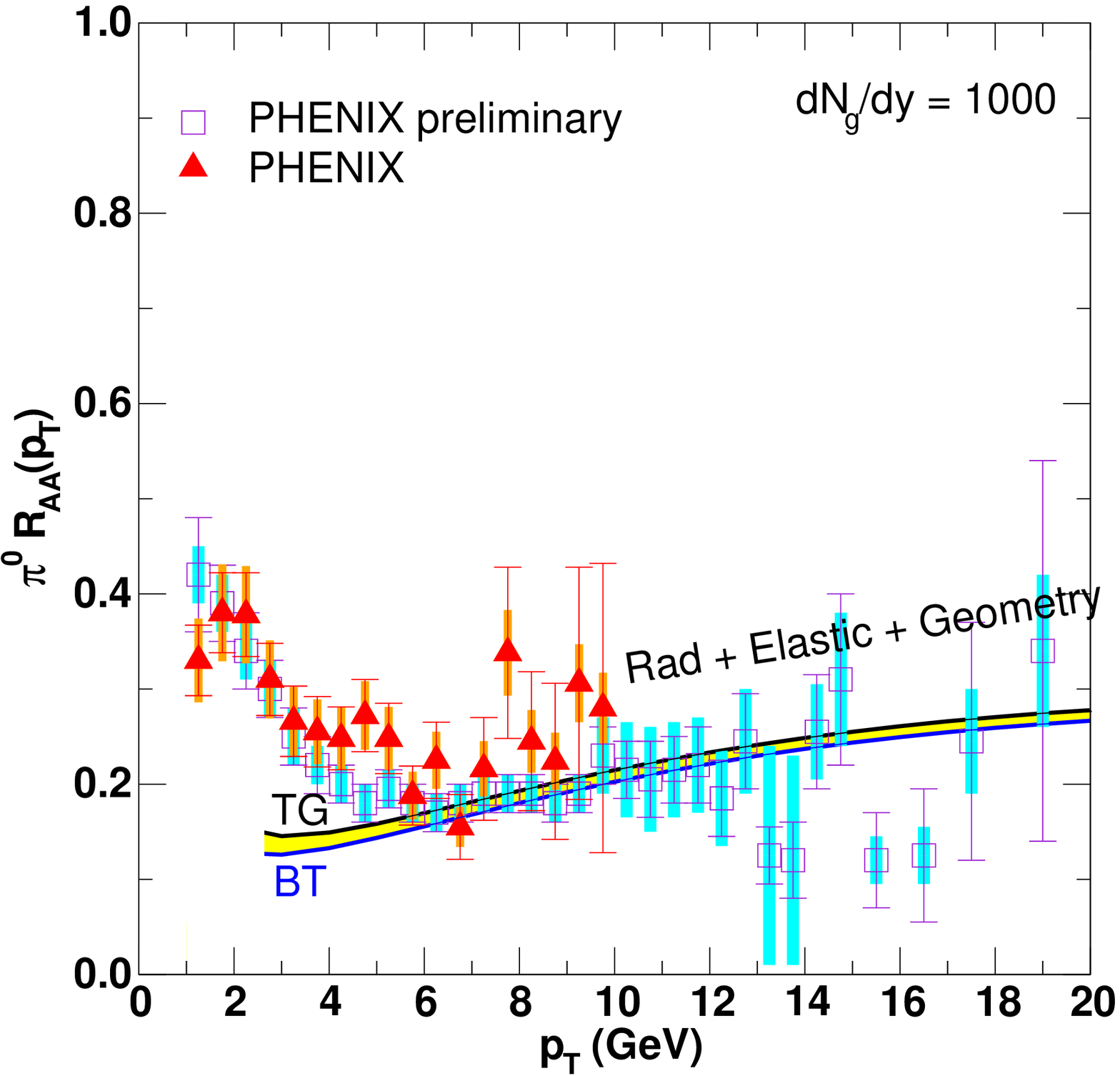}
\end{minipage}
\begin{minipage}{0.35\linewidth}
\includegraphics[width=\textwidth]{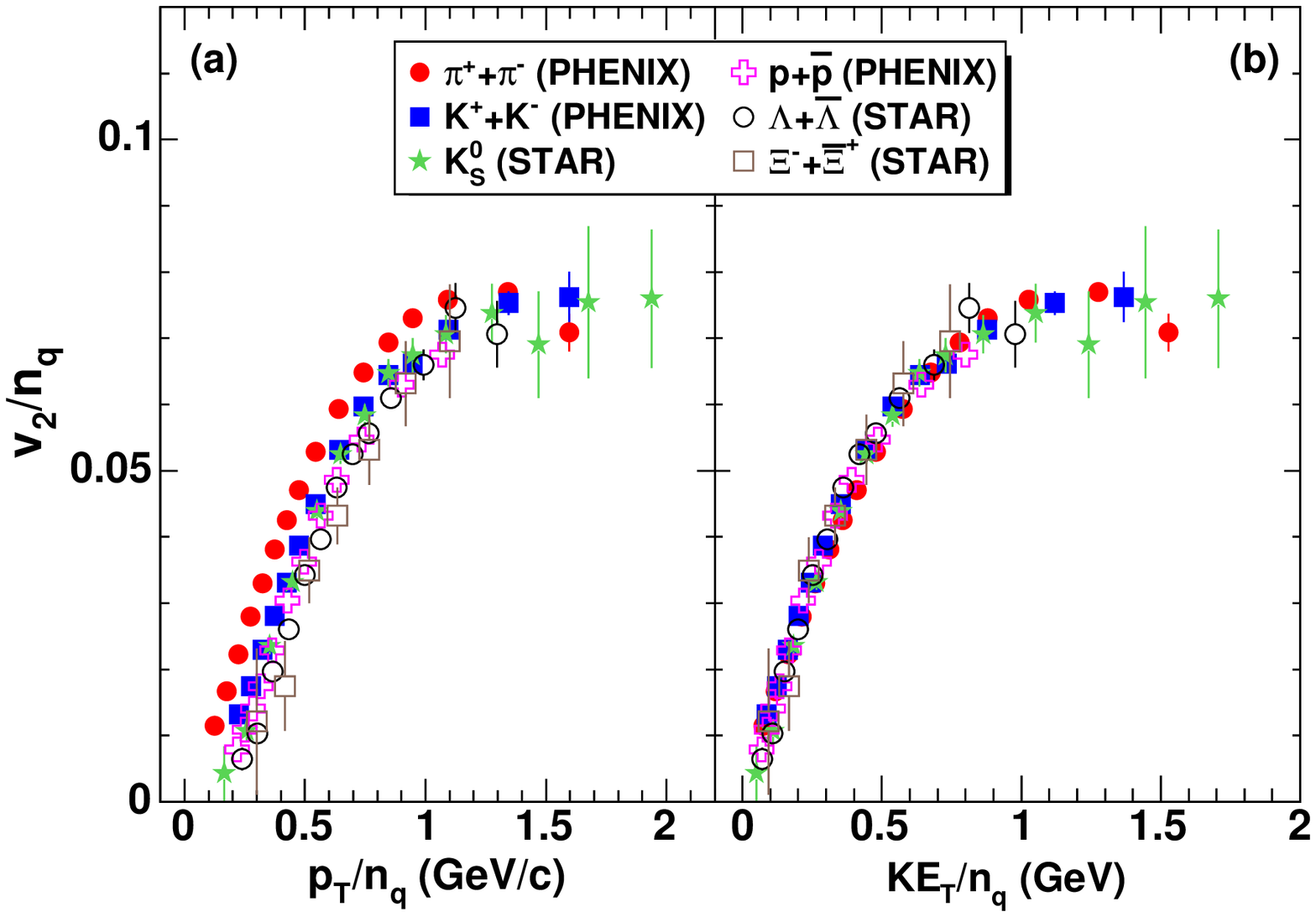}
\end{minipage}
\end{center}
\caption{Key experimental measurements in Au-Au collisions at the
  Relativistic Heavy Ion Collider in its first years of operation, as
  reflected in transverse-momentum spectra of hadrons~\cite{whitepaper}.
  Left panel: Elliptic flow compared to hydrodynamic
  calculations~\cite{Huovinen:2001cy} in the low-$p_T$ regime;
   middle panel: nuclear modification factor of neutral pions compared
  to perturbative jet-quenching calculations~\cite{Wicks:2005gt} in the
   high-$p_T$ regime;
  right panel: universal scaling of the hadronic $v_2$ interpreted as an
  underlying quark $v_2$~\cite{Adare:2006ti}.}
\label{fig_rhic-data}
\end{figure}

The first years of experiments at the Relativistic Heavy-Ion Collider
have indeed provided convincing evidence that a thermalized medium is
produced in $\sqrt{s}=200$~AGeV collisions. In this section we give a
brief summary of the basic observations and pertinent
interpretations~\cite{whitepaper}\footnote{For an assessment of the
  earlier CERN-SPS experiments at lower energies,
  cf.~Ref.~\cite{Heinz:2000xx}.}.  A schematic pictorial sketch of the
main stages of the evolution of a head-on collision of heavy nuclei is
displayed in Fig.~\ref{fig_hi-evo}.  The main observables are momentum
spectra of various hadron species.  We will concentrate on particles
with zero longitudinal momentum ($p_z$=0) in the center-of-mass frame of
a nucleus-nucleus collision, the so-called mid rapidity ($y$=0) region,
where one expects the largest energy deposition of the interpenetrating
nuclei. The main kinematic variable is thus the transverse momentum
($p_T$) of a particle. Three major findings at RHIC thus far may be
classified by their $p_T$ regime (see Fig.~\ref{fig_rhic-data} for 3
representative measurements):
\begin{itemize}
\item Thermalization and collective matter expansion in the low-$p_T$
  regime $p_T\simeq$~0-2~GeV;
\item Quark coalescence in the intermediate-$p_T$ regime,
  $p_T\simeq$~2-5~GeV;
\item Jet quenching in the high-$p_T$ regime, $p_T\ge$~5~GeV.
\end{itemize}
It turns out that all of the 3 regimes, and the associated physical
phenomena, are relevant for our discussion of heavy-quark observables
below. In the following, we will elaborate on the characteristic
features of these momentum regimes in some detail.

In the low-$p_T$ regime, the spectra of the most abundantly produced
hadrons ($\pi$, $K$, $p$, $\Lambda$, etc.) are well described by
hydrodynamic simulations of an exploding
fireball~\cite{Teaney:2000cw,Hirano:2002ds,Kolb:2003dz,Nonaka:2006yn}.
At its breakup (or thermal freezeout), where the (short-range)
interactions of the hadrons cease rather abruptly, the fireball matter
is expanding at an average collective velocity of about 60\% of the
speed of light and has cooled down to a temperature of about $T_{\rm
  fo}\simeq100$~MeV. The hadro-chemistry of the fireball, characterized
by the thermal ratios of the various hadron
species~\cite{BraunMunzinger:2003zd}, is ``frozen'' at a higher
temperature of about $T_{\rm ch}\simeq170$~MeV (as represented by the
``data'' points in Fig.~\ref{fig_phasedia}). The separation of chemical
and thermal freezeout is naturally explained by the large difference in
the inelastic and elastic reaction rates in a hadronic gas. Elastic
scattering among hadrons is dominated by strong resonances (e.g.,
$\pi\pi\to\rho\to\pi\pi$ or $\pi N\to \Delta \to \pi N$) with large
cross sections, $\sigma_{\rm res}\simeq 100$~mb, implying a thermal
relaxation time of about $\tau_{\rm therm}\simeq\langle\sigma\varrho_h
v_{\rm rel}\rangle \simeq 1$~fm/$c$ (assuming a typical hadron density
of $0.2 \, {\rm fm}^{-3}$ and a relative velocity of $v_{\rm
  rel}=0.5c$). Inelastic cross sections, on the other hand, are
typically around 1~mb, implying chemical relaxation times in a hadron
gas of $\sim$100~fm/$c$, well above its lifetime of $\sim$10~fm/$c$.

\begin{figure}[!t]
\hspace{-0.0cm}
\centerline{\includegraphics[width=0.48\linewidth]{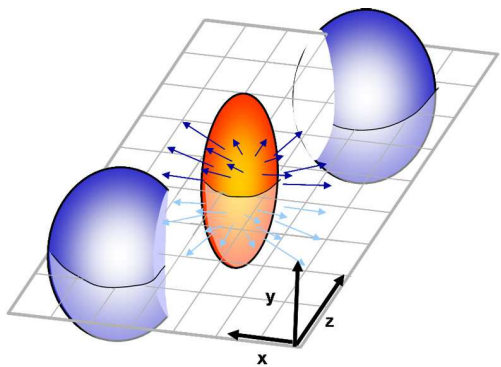}}
\caption{Schematic representation of a noncentral heavy-ion collision, 
  characterized by an almond-shaped initial overlap zone, and
  a subsequent pressure-driven build-up of elliptic flow.  }
\label{fig_elliptic}
\end{figure}

A hadronic observable which is sensitive to the early expansion phases
of the fireball is the elliptic flow, $v_2$, which characterizes the
azimuthal asymmetry in the emitted hadron spectra (i.e., the second
harmonic) according to
\begin{equation}
\frac{\dd N_h}{\dd^2p_T \dd y}=\frac{\dd N_h}{\pi \dd p_T^2 \dd y} 
\left[ 1 + 2 v_2(p_T) \cos(2\phi) + \dots \right]
\end{equation}
(at mid rapidity, the system is mirror symmetric in the transverse
$x$-$y$ plane and odd Fourier components in the azimuthal angle $\phi$
vanish). In a noncentral Au-Au collision, the initial nuclear overlap
zone in the transverse plane is ``almond''-shaped,
cf. Fig.~\ref{fig_elliptic}. Once the system thermalizes, the pressure
gradients along the short axis of the medium are larger than along the
long axis. As a result, hydrodynamic expansion will be stronger along
the $x$-axis relative to the $y$-axis, and thus build up an ``elliptic
flow'' in the collective matter expansion, which eventually reflects
itself in the final hadron spectra via a positive $v_2$ coefficient. The
important point is that a large $v_2$ can only be generated if the
thermalization of the medium is rapid enough: an initial period of
(almost) free streaming will reduce the spatial anisotropy and thus
reduce the system's ability to convert this spatial anisotropy into a
momentum anisotropy (i.e., $v_2$).  In this way, the magnitude of $v_2$
(and its $p_T$ dependence) is, in principle, a quantitative
``barometer'' of the thermalization time, $\tau_0$. Applications of
ideal relativistic hydrodynamics have shown that the experimentally
measured $v_2(p_T)$ for various hadrons ($\pi$, $K$, $p$, $\Lambda$) is
best described when implementing a thermalization time of
$\tau_0$=0.5-1~fm/$c$. The agreement with data extends from $p_T=0$ to
$\sim$2-3~GeV, which (due to the exponentially falling spectra)
comprises more than 95\% of the produced hadrons. Not only do the
interactions for thermalizing the matter rapidly have to be very strong,
but, for ideal hydrodynamics to build up the observed $v_2$, the
viscosity of the formed medium must remain very small (it is zero in
ideal hydrodynamics): the typical timescale for build-up of the observed
$v_2$ is on the order of the system size or QGP lifetime, $\tau_{\rm
  QGP}\simeq 5$~fm/$c$. These features have triggered the notion of a
``strongly-coupled'' QGP (sQGP); its initial energy density at RHIC, as
implied by the above thermalization times, amounts to
$\varepsilon_0\simeq$~10-20~GeV/fm$^3$, which is a factor of $\sim$10
above the estimated critical energy density for the phase transition
(and a factor of $\sim$100 above that for normal nuclear matter).

In the high-$p_T$ regime, the production mechanism of hadrons changes
and becomes computable in perturbative QCD, in terms of hard
parton-parton collisions upon first impact of the incoming nucleons
(cf.~left panel in Fig.~\ref{fig_hi-evo}). The produced high-energy
partons subsequently fragment into a (rather collimated) spray of
hadrons, called jet. Back-to-back jets are routinely observed in
high-energy collisions of elementary particles, but are difficult to
identify in the high-multiplicity environment of a heavy-ion
collision. However, a jet typically contains a ``leading'' particle
which carries most of the momentum of the parent parton (as described by
an empirical ``fragmentation'' function). High-$p_T$ spectra in
heavy-ion collisions thus essentially determine the modification
underlying the production of the leading hadron in a jet. This
modification is quantified via the ``nuclear modification factor'',
\begin{equation}
  R_{AA}(p_T)=\frac{\dd N_h^{AA}/\dd p_T}{N_{\rm coll} \ \dd N_h^{NN}/\dd p_T} \ ,  
\end{equation} 
where the numerator denotes the hadron spectrum in the nucleus-nucleus
collision and the denominator represents the spectrum in an elementary
nucleon-nucleon collision, weighted with the number of binary $N$-$N$
collisions in the primordial stage of the $A$-$A$ collision.  Thus, if
there is no modification of the leading hadron spectrum in the heavy-ion
collision, then $R_{AA}(p_T\ge 5~{\rm GeV})=1$.  RHIC data on $\pi$
production in central Au-Au collisions have found a large suppression by
a factor of 4-5 up to $p_T\simeq20$~GeV.  Originally, this has been
attributed to radiative energy-loss, i.e., induced gluon radiation off a
high-energy quark (or gluon) traversing a gluon
plasma~\cite{Gyulassy:2003mc,Armesto:2003jh}, as computed within
perturbative QCD (pQCD). In the approach of Ref.~\cite{Gyulassy:2003mc},
the initially extracted gluon densities turned out to be consistent with
those inferred from hydrodynamic calculations in the low-$p_T$
regime. In the approach of Ref.~\cite{Armesto:2003jh}, the interaction
strength of the energy-loss process is quantified via the transport
coefficient $\hat q =Q^2/\lambda$, which characterizes the (squared)
momentum transfer per mean free path of the fast parton; the
experimentally observed suppression requires this quantity to take on
values of 5-15~GeV$^2$/fm, which is several times larger than expected
in pQCD~\cite{Baier:2002tc}. More recently, the importance of elastic
energy loss has been realized, in particular in the context of 
heavy-quark
propagation~\cite{vanHees:2004gq,Moore:2004tg,Mustafa:2004dr,
  vanHees:2005wb,Wicks:2005gt}, as we will discuss in more detail below.

In the intermediate-$p_T$ regime, RHIC experiments have observed an
unexpectedly large ratio of baryons to mesons, e.g., $p/\pi\simeq 1$ or
$(\Lambda+\bar{\Lambda})/(4 K_s^0) \simeq 1.3$ in central Au-Au
collisions. On the one hand, within the pQCD energy loss picture, the
typical value of $p/\pi$ is close to $\sim$0.2 as observed in elementary
$p$-$p$ collisions. On the other hand, within hydrodynamic calculations,
a collective flow could, in principle, account for such an effect, but
the applicability of hydrodynamics appears to cease at momenta of
$p_T\ge 3$~GeV (as, e.g., indicated by the saturation (leveling off) of
the elliptic flow, which in hydrodynamics continues to grow).  Another
remarkable observation in intermediate-$p_T$ hadron spectra is a
constituent-quark number scaling (CQNS) of the hadronic $v_{2,h}(p_T)$,
as determined by the number, $n$, of constituent quarks in each hadron,
$h$, giving rise to a single, universal function,
\begin{equation}
v_{2,q}(p_T/n)= v_{2,h}(p_T)/n \ , 
\end{equation}   
which is interpreted as the partonic (quark) $v_2$ at the time of
hadronization. Both CQNS and the large baryon-to-meson ratios are
naturally explained in terms of quark coalescence processes of a
collectively expanding partonic source at the phase
transition~\cite{Greco:2003xt,Fries:2003vb,Hwa:2002tu}. The most recent
experimental data~\cite{Adare:2006ti,Abelev:2007qg} indicate that the
scaling persists at a surprisingly accurate level even at low $p_T$, but
only when applied as a function of the transverse kinetic energy, ${\rm
  KE}_T=m_T-m_h$, of the hadrons (where $m_T=(p_T^2+m_h^2)^{1/2}$ is the
total transverse energy of the hadron, and $m_h$ its rest mass). In
Ref.~\cite{Ravagli:2007xx}, the quark coalescence model has been
reformulated utilizing a Boltzmann transport equation where the hadron
formation process is realized via the formation of mesonic resonances
close to $T_c$. This approach overcomes the instantaneous approximation
of previous models, ensuring energy conservation in the coalescence
process, as well as the proper thermodynamic equilibrium limit. This, in
particular, allows for an extension of the coalescence idea into the
low-$p_T$ regime, and initial calculations are consistent with ${\rm
  KE}_T$ scaling for $v_{2,h}$.
   
To summarize this section, we conclude that RHIC data have provided
clear evidence for the formation of an equilibrated medium with very
small viscosity (an almost ``perfect liquid'') and large opacity with
associated energy densities well above the critical one; in addition,
indications for the presence of partonic degrees of freedom have been
observed. This medium has been named the strongly coupled QGP, or
sQGP. However, the understanding of its microscopic properties remains
an open issue at this point: What are the relevant degrees of freedom
and their interactions around and above $T_c$? Are the 3 main phenomena
described above related, and, if so, how?  Is there direct evidence for
deconfinement and/or chiral symmetry restoration?  Results from lattice
QCD, as discussed in the previous section, may already have provided
several important hints, but tighter connections to RHIC data need to be
established.

Toward this goal, heavy-quark observables are hoped to provide new
decisive and quantitative insights: Do charm and bottom quarks
participate in the flow of the medium, despite their large mass? Do they
even thermalize at low $p_T$? Do they suffer jet quenching at high
$p_T$? Do they corroborate evidence for quark coalescence processes?
Initial measurements of heavy-quark observables have been performed
providing tantalizing evidence that the answer to these questions may
indeed be largely positive: a substantial elliptic flow and suppression
of single-electron spectra associated with semileptonic heavy-meson
decays has been observed, i.e., electrons (and positrons) arising via
decays of the type $D\to e \nu X$ and $B\to e \nu X$ (the heavy-light
mesons, $D=(c\bar q)$ and $B=(b\bar q)$, are composed of a heavy quark
($c$, $b$) and a light antiquark ($\bar q=\bar u, \bar d$), and as such
are the main carriers of heavy quarks in the system).

We now turn to the main subject of this article, i.e., the theoretical
and phenomenological description of heavy-quark interactions in the QGP
and pertinent observables at RHIC.

\section{Heavy Quarks in the Quark-Gluon Plasma}
\label{sec_hq}
The special role of heavy quarks ($Q=b, c$) as a probe of the medium
created in heavy-ion collisions resides on the fact that their mass is
significantly larger than the typically attained ambient temperatures or
other nonperturbative scales, $m_Q\gg T_c, \Lambda_{\rm
  QCD}$.\footnote{Note that the production of the heaviest known quark,
  the top, is out of reach at RHIC; moreover, its lifetime,
  $\tau_t=1/\Gamma_t \simeq 0.1 $~fm/$c$, is by far too short to render
  it a useful probe.} This has several implications:
\begin{itemize}
\item [(i)] The production of heavy quarks is essentially constrained to
  the early, primordial stages of a heavy-ion collision. Thus the
  knowledge of the initial heavy-quark (HQ) spectrum (from, say, $p$-$p$
  collisions) can serve as a well defined initial state, even for
  low-momentum heavy quarks. The latter feature renders heavy-quark
  observables a prime tool to extract transport properties of the
  medium.
\item[(ii)] Thermalization of heavy quarks is ``delayed'' relative to
  light quarks by a factor of $\sim m_Q/T\simeq$~5-15. While the bulk
  thermalization time is of order $\sim$0.5~fm/$c$, the thermal
  relaxation of heavy quarks is expected to occur on a timescale
  comparable to the lifetime of the QGP at RHIC, $\tau_{\rm QGP}\simeq
  5$~fm/$c$. Based on the thus far inferred properties of the sQGP,
  charm quarks could ``thermalize'' to a certain extent, but not
  fully. Therefore, their spectra should be significantly modified, but
  still retain memory about their interaction history -- an ``optimal''
  probe. According to this estimate, bottom quarks are expected to
  exhibit notably less modification.
\item[(iii)] As is well-known from electrodynamics, Bremsstrahlung off
  an accelerated (or decelerated) charged particle is suppressed by a
  large power of its the mass, $\sim(m_q/m_{c})^{4}$ for heavy relative
  to light quarks. Therefore, induced gluon radiation off heavy quarks
  (i.e., radiative energy loss) is much suppressed relative to elastic
  scattering.
\item[(iv)] The typical momentum transfer from a thermal medium to a
  heavy quark is small compared to the HQ momentum, $p_{\rm th}^2 \sim
  m_Q T \gg T^2$ (where $p_{\rm th}$ is the thermal momentum of a heavy
  quark in nonrelativistic approximation, $p_{\rm th}^2/2m_Q \simeq
  3~T/2$; for a relativistically moving quark, it is parametrically even
  larger, $p_{\rm th}\sim m_Q v$). This renders a Brownian Motion
  approach (Fokker-Planck equation) a suitable and controlled
  theoretical framework for describing the diffusion of a heavy quark in
  the (s)QGP.
\item[(v)] The HQ mass can furthermore be utilized as a large scale in
  developing effective (nonperturbative) interactions, such as HQ
  effective theory or potential model approaches. In principle, this
  allows to make contact with static interaction potentials from
  finite-temperature lattice QCD, although a number of issues have to be
  resolved before reliable quantitative predictions can be made.
\item[(vi)] At low momenta, nonperturbative (resummation) effects become
  relatively more important with increasing mass. E.g., for bound state
  formation, the binding energy is known to increase with the (reduced)
  mass of the constituents. For the concrete problem of the charm
  diffusion constant, it has recently been shown that the perturbation
  series is badly convergent even for values of the strong coupling
  constant as small as $\alpha_s=0.1$~\cite{CaronHuot:2007gq}.
\end{itemize}

Before first RHIC data on HQ observables became available, the
expectation based on pQCD radiative energy loss was that the $D$-meson
spectra are much less suppressed than light hadron
spectra~\cite{Djordjevic:2004nq}, with a small elliptic flow of up to
$v_2^D\simeq4\%$~\cite{Armesto:2005mz}. At the same time, the importance
of elastic collisions was emphasized in
Refs.~\cite{vanHees:2004gq,Moore:2004tg,Mustafa:2004dr}. In particular,
in Ref.~\cite{vanHees:2004gq} nonperturbative HQ resonance interactions
in the QGP (motivated by lattice QCD results, cf.~Fig.~\ref{fig_A-lqcd})
were introduced and found to reduce the HQ thermalization times by a
factor of $\sim$3 compared to elastic pQCD scattering. Predictions for
the $D$-meson $v_2$ and $p_T$ spectra in the limiting case in which the
degree of HQ thermalization is similar to light quarks can be found in
Ref.~\cite{Greco:2003vf}; including the effects of coalescence with
light quarks, the $D$-meson elliptic flow reaches up to around
$v_2^D\simeq15\%$, about a factor of $\sim$4 larger than in the pQCD
energy-loss calculations. This analysis also demonstrated the important
feature that the single-electron ($e^\pm$) spectra arising from the
semileptonic decays, $D\to e \nu X$, closely trace the suppression
($R_{AA}$) and elliptic flow of the parent $D$ meson (see also
Ref.~\cite{Dong:2004ve}). In the following, we elaborate on the
underlying approaches and further developments with applications to HQ
transport properties and RHIC data.

\subsection{Heavy-Quark Interactions in the QGP}
\label{ssec_hq-int}
The scattering of charm and bottom quarks in a Quark-Gluon Plasma is
dominated by interactions with the most abundant particles in the
medium, i.e., gluons as well as light ($u$ and $d$) and strange
quarks. The basic quantity to be evaluated is therefore the scattering
matrix (or cross section), which can be further utilized to compute
in-medium HQ self-energies and transport coefficients. For reasons given
in the previous Section we focus on elastic $2\to 2$ scattering
processes, $Q+i\to Q+i$ with $i=g,u,d,s$. Our discussion is organized
into perturbative (Sec.~\ref{sssec_pqcd}) and nonperturbative approaches
(Secs.~\ref{sssec_reso} and \ref{sssec_pot}). While the effective 
resonance model (Sec.~\ref{sssec_reso}) involves a priori undetermined 
parameters in terms of the resonance masses and coupling constants, 
the lattice-QCD based potential scattering (Sec.~\ref{sssec_pot})
is an attempt to generate the heavy-light quark interactions 
microscopically without free parameters (albeit with significant
uncertainties in the extraction of the interaction potentials).

\subsubsection{Perturbative Scattering}
\label{sssec_pqcd}
\begin{figure}[!t]
\begin{center}
\begin{minipage}{0.28\textwidth}{\includegraphics[width=\textwidth]{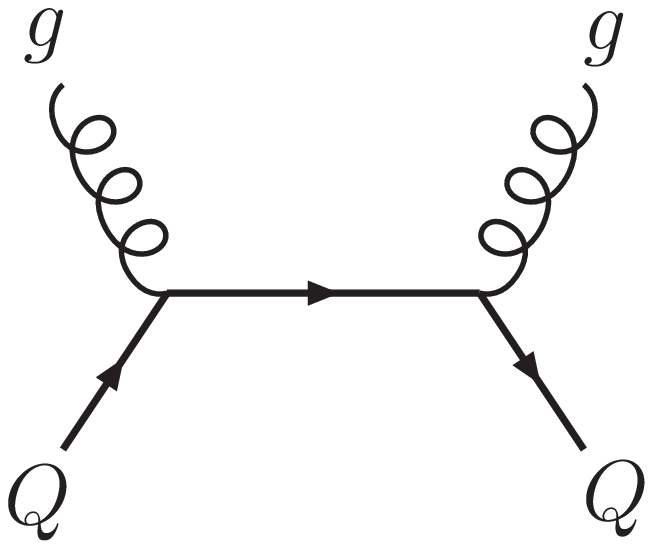}}
\end{minipage}
\begin{minipage}{0.28\textwidth}{\includegraphics[width=\textwidth]{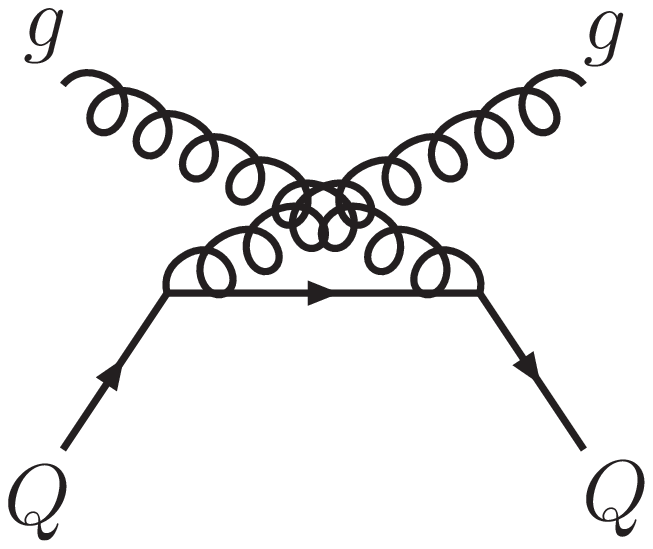}}
\end{minipage}
\begin{minipage}{0.21\textwidth}{\includegraphics[width=\textwidth]{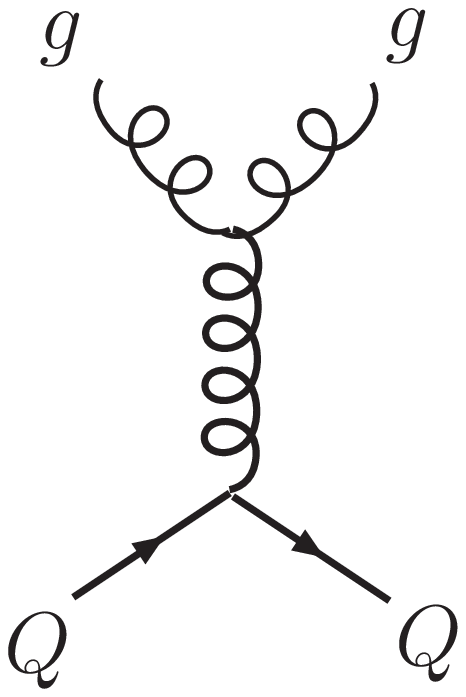}}
\end{minipage}
\begin{minipage}{0.21\textwidth}{\includegraphics[width=\textwidth]{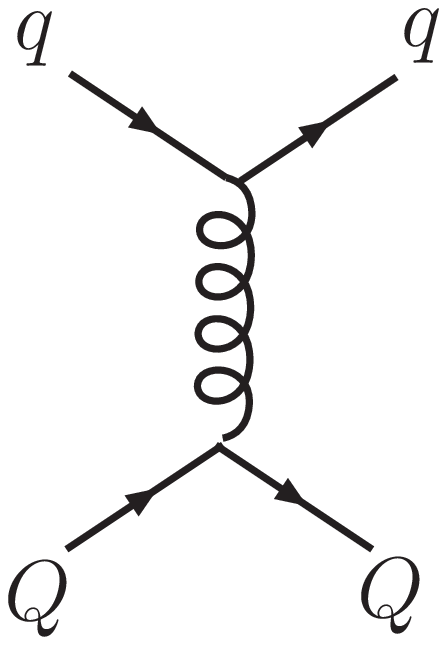}}
\end{minipage}
\end{center}
\caption{Feynman diagrams for leading-order perturbative HQ scattering
  off light partons.}
\label{fig_HQ-pqcd}
\end{figure}

\begin{figure}[!t]
\begin{center}
\includegraphics[width=0.5\linewidth]{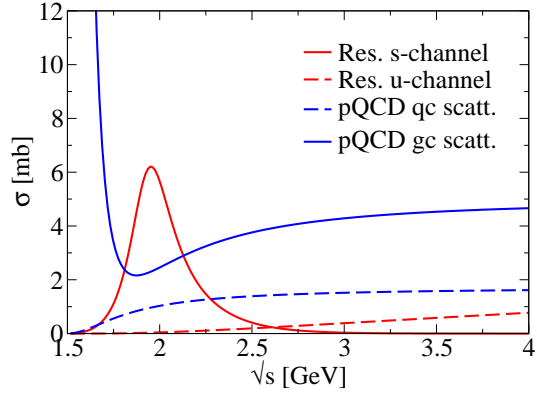}
\end{center}
\caption{Total HQ scattering cross sections off light partons in pQCD
  (blue lines) and within the effective resonance model (red 
  lines)~\cite{Rapp:2006ta}.}
\label{fig_xsec}
\end{figure}

In QCD, the simplest possible diagrams for HQ interactions with light
partons are given by leading order (LO) perturbation theory. The
pertinent Feynman diagrams are very similar to the ones depicted in
Fig.~\ref{fig_gluo-ex}, and are summarized in Fig.~\ref{fig_HQ-pqcd}.
As discussed in Sec.~\ref{ssec_qcd}, these processes can be expected to
constitute a realistic description of HQ scattering in regimes where the
strong coupling constant, $\alpha_s$, is small (higher order diagrams
will contribute with higher powers in $\alpha_s$). In principle, this
can be realized either for large HQ momenta (implying the relevant
momentum transfers to be large) or at high temperatures where the
interaction is screened and/or the typical momentum transfer of order
$Q^2\sim T^2$ is large. Since the color charge of gluons is larger than
that of quarks (by a factor of 9/4 to order $\alpha_s$), the dominant
contribution to pQCD scattering arises from interactions with gluons,
more precisely the $t$-channel gluon exchange in
$Q(p_1^{(4)})+g(p_2^{(4)})\to Q(p_3^{(4)})+g(p_4^{(4)})$, where
$p_i^{(4)}=(E_{i},\vec p_i)$ denotes the energy-momentum vector (or
4-momentum) of particle $i$ (we use $p_i= |\vec{p}_i|$ for the magnitude
of the 3-momentum).  The corresponding differential cross section is
given by
\begin{equation}
  \frac{\dd \sigma_{gQ}}{\dd t} = \frac{1}{16 \pi (s-M^2)^2}
  \overline{|\mathcal{M}|^2} \ , 
\label{xsec-pqcd}
\end{equation}
where $t=(p_3^{(4)}-p_1^{(4)})^2=2(m_Q^2-E_1E_2+\vec p_1\cdot \vec p_3)$
is the energy-momentum transfer on the heavy quark with $\vec p_1\cdot
\vec p_3= p_1 p_3 \cos\Theta$, $s$ is the squared center-of-mass energy
$s=(p_1^{(4)}+p_2^{(4)})^2$, and $\Theta$ the scattering angle of the
heavy quark. The squared scattering amplitude, averaged over initial and
summed over final spin polarizations,
\begin{equation}
  \overline{|\mathcal{M}|^2} = \pi^2 \alpha_s^2 \left [
  \frac{32(s-M^2)(s+t-M^2)}{t^2} + \ldots \right ] \ , 
\label{Mscatt}
\end{equation}
turns out to be dominated by the $t$-channel gluon exchange diagram
(third panel in Fig.~\ref{fig_HQ-pqcd}), corresponding to the expression
explicitly written in the brackets. The factor $1/t^2$ represents the
(squared) propagator of the exchanged gluon which leads to a large cross
section for small $t$ implying a small scattering angle $\Theta$, i.e.,
forward scattering. However, as discussed in Sec.~\ref{ssec_phasedia},
the interaction is screened in the medium which, in a simplistic form,
can be implemented as an in-medium gluon propagator $1/(t-\mu_D^2)$,
leading to a reduction of the cross section (note that for $t$-channel
gluon exchange, the 4-momentum transfer, $t$, is negative). To lowest
order in perturbation theory, the gluon Debye mass is given by
$\mu_D\sim gT$. The total cross section follows from integrating the
expression in Eq.~(\ref{xsec-pqcd}) over $t$; the results including all
LO diagrams for charm-quark scattering off quarks and gluons are
displayed in Fig.~\ref{fig_xsec}.
The total pQCD cross sections, computed for an optimistically large
$\alpha_s=0.4$, are, in fact, quite sizable, at around $\sigma_{\rm
  tot}=4(1.5)$~mb for gluons (quarks). In a schematic estimate these
cross sections may be converted into a reaction rate by using the
``pocket formula'' $\Gamma_c = \sigma_{ci} n_i v_{\rm rel}$. To obtain
an upper limit, one may use ideal-gas massless parton densities,
$n_i=d_i\pi^2 T^3$, with quark, antiquark and gluon degeneracies of
$d_{q\bar{q}} =10.5 N_f$ with $N_f=2.5$ and $d_{g}=16$ (recall
Sec.~\ref{ssec_phasedia} and Fig.~\ref{fig_eos-lqcd}), in connection
with a relative velocity $v_{\rm rel}=c$. One finds
$\Gamma_c\simeq0.4$~GeV=$(0.5{\rm fm}/c)^{-1}$, a rather large elastic
scattering rate.  However, as we will see below, the relevant quantity
for determining the thermalization time scale is the transport cross
section, which involves an extra angular weight proportional to
$\sin^2\Theta$ when integrating over the differential cross section.
This renders pQCD scattering rather ineffective in isotropizing HQ
distributions, due to its predominantly forward scattering angles; the
resulting thermal relaxation times, $\tau_Q^{\rm therm}$, are therefore
much larger than what one would naively expect from the large scattering
rate estimated above.

\subsubsection{Resonance Model}
\label{sssec_reso}
\begin{figure}[!t]
\begin{center}
\includegraphics[width=0.5\linewidth]{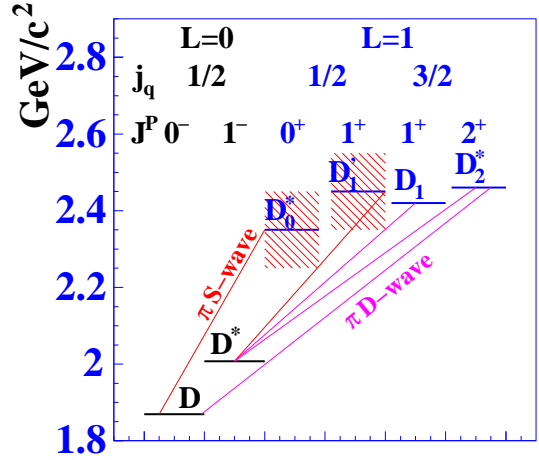}
\end{center}
\caption{Level spectrum of $D$-mesons, $D=(c\bar q)$, in the vacuum
  according to recent measurements reported in Ref.~\cite{Abe:2003zm}.
  The masses of the states are shown as a function of their quantum
  numbers indicated in the upper portion of the figure. Note, on the one
  hand, the splitting (non-degeneracy) of the chiral partners $D$ and
  $D_0^*$, as well as of $D^*$ and $D_1$, by about $\Delta M=0.4$~GeV
  (believed to be a consequence of spontaneous breaking of chiral
  symmetry in the vacuum).  On the other hand, HQ symmetry, implying
  degeneracy of $D$-$D^*$ and of $D_0^*$-$D_1$, is satisfied within
  $\Delta M=0.15$~GeV.  }
\label{fig_Dmes-vac}
\end{figure}

As was discussed in Sec.~\ref{ssec_phasedia}, lattice calculations
suggest the existence of hadronic resonances (or bound states) in the
QGP for temperatures of 1-2~$T_c$. In Ref.~\cite{vanHees:2004gq} the
idea has been introduced that such resonances are present in the
heavy-light quark sector and are operative in a significant increase of
the interaction strength of heavy quarks in the QGP. The starting point
of this investigation is an effective Lagrangian, {\em assuming} the
presence of heavy-light fields, $\Phi$, which couple to a heavy quark
and light antiquark according to
\begin{equation}
{\cal L}_{\Phi Qq} = {\cal L}^0_{\Phi} +  {\cal L}^0_{Q} + {\cal L}^0_{q} 
+ \sum\limits_m G_m Q~\frac{1+\not{\!v}}{2}~\Phi_m~\Gamma_m~\bar q + 
{\rm h.c.} \ . 
\label{Leff}
\end{equation}
The first 3 terms on the right-hand-side represent the kinetic energy
and mass terms giving rise to free particle propagation while the last
term (h.c. = hermitean conjugate), roughly speaking, indicates all terms
with particles and antiparticles interchanged (i.e., $Q\to \bar Q$,
$\bar q \to q$, etc.). The key term is the explicitly written
interaction term generating $\Phi$-$Q$-$\bar q$ (3-point) vertices whose
strength is controlled by pertinent coupling constants, $G_m$, which in
this approach are free parameters. The summation over $m$ accounts for
the different quantum numbers of the $\Phi$-fields (which in turn are
related to the structure of the coupling matrices, $\Gamma_m$). The
couplings can be largely inferred from symmetry considerations. As
alluded to toward the end of Sec.~\ref{ssec_qcd}, the spontaneous
breaking of chiral symmetry (SBCS) in the vacuum implies hadronic chiral
partners to split in mass; in the $D$-meson sector, this applies to the
chiral partners in the scalar ($J^P=0^+$) - pseudoscalar ($J^P=0^-$)
multiplet, $D$ and $D_0^*$, as well as to the vector ($J^P=1^-$) and
axialvector ($J^P=1^+$) chiral multiplet, $D^*$ and $D_1$. The pertinent
chiral breaking is nicely borne out of recent measurements of the
$D$-meson spectrum in vacuum~\cite{Abe:2003zm}, where the chiral
splitting has been established to amount to about $\Delta M=0.4$~GeV,
cf.~Fig.~\ref{fig_Dmes-vac}. In the QGP, however, the spontaneously
broken chiral symmetry will be restored, recall, e.g., the left panel in
Fig.~\ref{fig_op-lqcd}. Chiral restoration is necessarily accompanied by
the degeneracy of chiral partners, as is indeed observed in lQCD
computations of meson spectral functions above $T_c$, cf.~right panel of
Fig.~\ref{fig_A-lqcd}. We thus infer that the chiral partners in the
$D$-meson spectrum, i.e., scalar and pseudoscalar (as well as vector and
axialvector) should have the same mass and width in the QGP.

In addition, QCD possesses heavy-quark symmetries, in particular a spin
symmetry which states that hadrons containing heavy quarks are
degenerate if the HQ spin is flipped inside the hadron. Within the
constituent quark model, one therefore expects a degeneracy of
pseudoscalar and vector mesons, where the heavy and light quark are
coupled in a relative $S$-wave (orbital angular momentum $l=0$) and the
total spin, $J$, of the meson is solely determined by the coupling of
the two quark spins. The asserted symmetry is accurate within
$\sim$0.1-0.15~GeV in the $D$-meson spectrum ($D$-$D^*$ and
$D_0^*$-$D_1^*$, cf.~Fig.~\ref{fig_Dmes-vac}), and within $\sim$0.05~GeV
in the $B$-meson spectrum (as given by the $B(5280)$-$B^*(5325)$ mass
difference; note that the accuracy of the HQ symmetry indeed appears to
scale with the inverse HQ mass, $m_c/m_b\simeq 1/3$).  Since the
heavy-light resonance mass in the QGP itself is subject to uncertainties
on the order of possibly up to a few hundred MeV, there is little point
in accounting for the relatively small violations of HQ spin
symmetry. In Ref.~\cite{vanHees:2004gq} it was therefore assumed that
also the pseudoscalar-vector (as well as scalar-axialvector) states are
degenerate.

With both chiral and HQ symmetry, the effective Lagrangian,
Eq.~(\ref{Leff}), essentially contains 2 parameters in both the charm
and bottom sector, which are the (universal) resonance mass, $m_{D,B}$,
and coupling constant, $G_{D,B}$. The former has been varied over a
rather broad window above the $Q$-$\bar q$ threshold. Note that bound
states cannot be accessed in two-body scattering due to energy
conservation, i.e., the bound state mass is by definition below the
2-particle threshold, $E_{\rm thr}$=$m_Q$+$m_{\bar q}$. Once the mass is
fixed, the (energy-dependent) width for the two-body decay of the
resonance, $\Phi \to Q+\bar q$, is determined by the coupling constant
as
\begin{equation} 
\Gamma_\Phi(M) = \frac{3G_\Phi^2}{8\pi} \ \frac{(M^2-m_Q^2)^2}{M^3} 
\ \Theta(M-m_Q) \ ,
\end{equation}
where the mass of the light antiquark has been put to zero. To determine
the coupling constant, some guidance can be obtained from effective
quark models at finite
temperature\cite{Blaschke:2002ws,Blaschke:2003ji}, where an in-medium
$D$-meson width of several hundred MeV was found (see also
Ref.~\cite{vanHees:2007me} and Fig.~\ref{fig_ImT} below).

\begin{figure}[!t]
\hspace{1.5cm}
\begin{minipage}{7cm}
\includegraphics[width=0.8\linewidth]{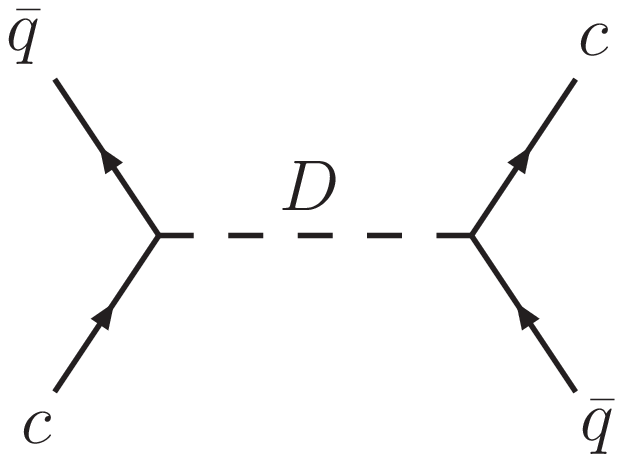}
\end{minipage}
\hspace{0cm}
\begin{minipage}{7cm}
\includegraphics[width=0.55\linewidth]{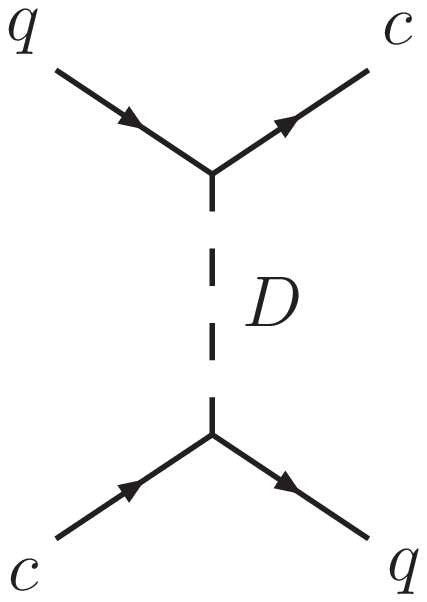}
\end{minipage}
\caption{Feynman diagrams for nonperturbative charm-quark interactions
  via effective $D$-meson exchange in $s$-channel scattering off
  antiquarks (left panel) and $u$ channel scattering off quarks (right
  panel)~\cite{vanHees:2004gq}.}
\label{fig_HQ-reso}
\end{figure}

The effective Lagrangian, Eq.~(\ref{Leff}), generates 2 diagrams for
resonance exchange in heavy-light quark scattering, so-called $s$- and
$u$-channels, as shown in Fig.~\ref{fig_HQ-reso}. The pertinent total
cross sections for charm-quark scattering are displayed by the red lines
in Fig.~\ref{fig_xsec}, assuming $D$-meson masses and widths of
$m_D=2$~GeV and $\Gamma_D=0.4$~GeV (the charm- and bottom-quark masses
have been set to $m_{c,b}=1.5, 4.5$~GeV).
In the energy regime relevant for scattering of thermal partons in the
QGP, $E_{\rm cm}\simeq m_Q + E_q^{\rm th}\simeq 2.2$~GeV ($E_q^{\rm
  th}\simeq 3T \simeq 0.75$~GeV at $T=0.25$~GeV), the cross section is
largely dominated by the $s$-channel resonance formation, and is
substantially larger than the LO pQCD result for $Q$-$q$
scattering. This is not so compared to pQCD scattering off
gluons. However, the angular distribution of the differential
$s$-channel resonance cross section (not shown) is essentially isotropic
(in the rest system of the resonance), which renders it significantly
more efficient for thermalizing the HQ distributions (as we will see in
Sec.~\ref{ssec_transport} below). The interaction introduced via the
Lagrangian, Eq.~(\ref{Leff}), cannot predict at what temperatures the
effective resonance fields dissolve. This, however, can be overcome if
the interaction underlying the resonance formation is treated on a more
microscopic level. Even more importantly, a microscopic treatment could,
in principle, eliminate the coupling-constant and mass parameters. First
steps in this direction will be discussed in the next section.

\subsubsection{Potential Scattering Based on Lattice QCD} 
\label{sssec_pot}
The advances in finite-temperature lQCD to compute the in-medium free
energy of a $Q$-$\bar Q$ pair as a function of its size have, in
principle, opened the possibility to extract their static
(chromo-electric) interaction potential, see the discussion around
Eq.~(\ref{FQQbar}). If this problem can be well defined and solved, the
next desirable step is to check the quantitative consequences of such an
interaction in the light quark sector (where lQCD has also found
indications for resonance formation), eventually including transport and
bulk properties. In Refs.~\cite{Brown:2003km,Shuryak:2004tx} a
relativistic correction has been suggested in terms of a
velocity-velocity interaction (known as the Breit interaction in
electrodynamics). In those works the focus has been on the bound state
problem, by solving an underlying Schr\"odinger equation.  In
Ref.~\cite{Mannarelli:2005pz}, a $T$-matrix approach to the $q$-$\bar q$
interaction has been set up, which allows for a simultaneous treatment
of bound and scattering states. This framework has been applied for
heavy-light quark scattering in Ref.~\cite{vanHees:2007me}. Thus far the
discussion was constrained to the color neutral (singlet) $Q$-$\bar q$
channel (i.e., a blue-antiblue, green-antigreen or red-antired
color-charge combination), but a quark and an antiquark can also combine
into a color-octet (a combination of red-antiblue, blue-antigreen
etc.). In addition, one can extend the approach to the diquark ($Q$-$q$)
channel, where color-antitriplet and sextet combinations are possible
(see also Ref.~\cite{Shuryak:2004tx}). Finite-temperature lattice
computations of the free energy of a heavy diquark~\cite{Doring:2007uh}
suggest that the relative interaction strength in the meson and diquark
systems follows the expectations of perturbative QCD (so-called Casimir
scaling, which essentially reflects the color-charges of the partons),
namely $V_{Q\bar Q}^{(1)} = 2 V_{QQ}^{(\bar 3)} = - 4 V_{QQ}^{(6)}$;
that is, the interaction in the color-triplet diquark channel is half as
attractive as in the color-neutral diquark channel while it is weakly
repulsive for a color-sextet diquark.

\begin{figure}[!t]
\begin{center}
\includegraphics[width=0.95\linewidth]{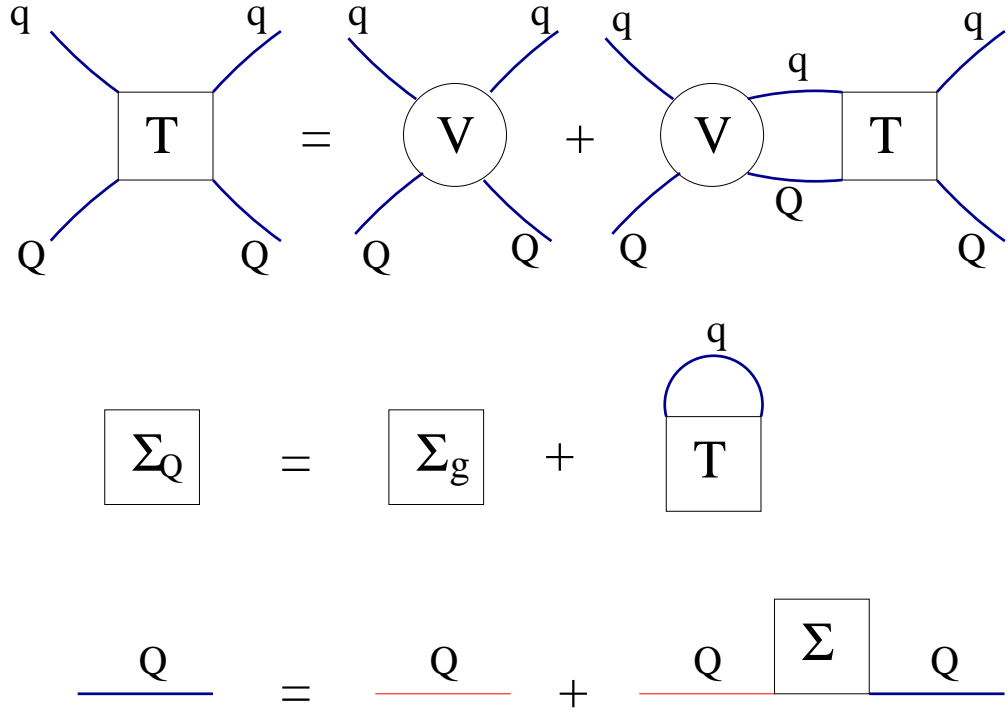}
\end{center}
\caption{Diagrammatic representation of the Brueckner problem for HQ
  interactions in the QGP~\cite{vanHees:2007me,vanHees:2008xx}; upper
  panel: $T$-matrix equation for HQ scattering off thermal light quarks
  or antiquarks; middle panel: HQ self-energy due to interactions with
  gluons and quarks or antiquarks; lower panel: Dyson equation for the
  in-medium HQ propagator.  }
\label{fig_brueck}
\end{figure}

The starting point for the calculations of the heavy-light quark
$T$-matrix is the relativistic Bethe-Salpeter equation for elastic
$1+2\to 3+4$ scattering,
\begin{equation}
  T_m(1,2;3,4) = K_m(1,2,3,4) + \int \frac{d^4k}{(2\pi)^4} \ K_m(1,2;5,6) 
  \ G^{(2)}(5,6) \ T_m(5,6;3,4) \ ,  
\label{TBS}
\end{equation}
which accounts for the full 4-dimensional energy-momentum dependence of
the scattering process (including ``off-shell'' particles for which
energy and 3-momentum are independent variables); $m$ characterizes all
quantum numbers of the composite meson or diquark (color, total spin and
flavor), $K_m$ is the interaction kernel and $G^{(2)}(5,6)$ is the
two-particle propagator in the intermediate state of the scattering
process. The integration in the second term of Eq.~(\ref{TBS}) accounts
for all possible momentum transfers in compliance with energy-momentum
conservation. The labels $j=1, \dots, 6$ denote the quantum numbers
(including 4-momentum) of the scattered single particles.
Eq.~(\ref{TBS}) represents a rather involved integral equation for the
full scattering $T$-matrix.  However, the (static) potentials extracted
from lQCD do not contain the rich information required for the 4-D
interaction kernel $K_m$.  It is therefore in order to adopt suitable
reduction schemes~\cite{Blankenbecler:1965gx,Thompson:1970wt}, which are
well-known in nuclear physics, e.g., in the context of nucleon-nucleon
scattering~\cite{Machleidt:1989tm}. A reduction amounts to neglecting
additional particle-antiparticle fluctuations for the intermediate
2-particle propagator and puts the latter on the energy shell, which
allows one to perform the energy integration (over $k_0$) in
Eq.~(\ref{TBS}). Importantly, it furthermore enables to identify the
reduced interaction kernel with a 2-body potential, $V_m$, and therefore
to establish the connection to the potentials extracted from lattice
QCD. The now 3-dimensional scattering equation, known as
Lippmann-Schwinger (LS) equation, is amenable to an expansion in partial
waves characterized by the angular momentum quantum number, $l=0, 1,
\dots$, corresponding to relative $S$- and $P$-waves, etc. Assuming HQ
spin symmetry, the spin quantum number does not explicitly enter, and
one arrives at
\begin{equation}
  T_{a,l}(E;p^\prime, p) = V_{a,l}(p^\prime, p) + 
  \frac{2}{\pi} \int \dd k \, k^2 \, V_{a,l}(p^\prime, k) 
  \, G_{Qq}(E; k) \, T_{a,l}(E;k,p) \, ,  
\label{Tmat}
\end{equation}
where $a=1,\bar 3,6,8$ labels the color channel. In the Thompson
reduction scheme, the intermediate heavy-light quark-quark (or
quark-antiquark) propagator takes the form
\begin{equation}
  G_{Qq}(E;k) = \frac{1}{4} 
  \frac{1-f(\omega^Q_k)-f(\omega^q_k)}{E-(\omega^q_k
    +{\rm i}\Sigma^q_I(\omega^q_k,k)) -(\omega^Q_k 
    + {\rm i} \Sigma^Q_I(\omega^Q_k,k))} \  
\label{GTh}
\end{equation}
with $\omega_k^{q,Q}=(m_{q,Q}^2+k^2)^{1/2}$ the on-shell quark energies,
and $f(\omega_k^{q,Q})$ the thermal Fermi distribution functions (the
numerator in Eq.~(\ref{GTh}) accounts for Pauli blocking, which,
however, is essentially irrelevant at the temperatures considered here).
A pictorial representation of the $T$-matrix scattering equation is
given in the upper panel of Fig.~\ref{fig_brueck}.
Due to the interactions with the medium, the quark propagators
themselves are modified which is encoded in a single-quark self-energy,
$\Sigma_{q,Q}$, figuring into the 2-particle propagator,
Eq.~(\ref{GTh}). It can be related to the heavy-light $T$-matrix, as
well as due to interactions with gluons, as
\begin{equation}
  \Sigma_Q(k;T) = \Sigma_{Qg} + 
  \int \frac{\dd^3p}{2\omega_p^q} \, f(\omega_p^q) \, T_{Qq}(E;p,k) \ , 
\label{Sigma}
\end{equation}
cf.~the middle panel in Fig.~\ref{fig_brueck}. The self-energy can be
resummed in a Dyson equation for the single-quark propagator according
to
\begin{equation}
D_Q = D_Q^0  + D_Q^0 \, \Sigma_Q \, D_Q \ , 
\label{dyson}
\end{equation}  
where the free propagator is given by $D_Q^0(\omega,k)=1/(\omega -
\omega_k^0)$ (omitting any quantum number structure and denoting the
free particle on-shell energy by $\omega_k^0$).  The Dyson equation is
graphically displayed in the lower panel of Fig.~\ref{fig_brueck}. It
can be solved algebraically, resulting in the full in-medium
single-particle propagator
\begin{equation}
D_Q = \frac{1}{\omega - \omega_k^0 - \Sigma_Q} \ .  
\label{DQ}
\end{equation} 
The convolution of the full propagators, $D_q$ and $D_Q$, within the
Thompson reduction scheme leads to the 2-particle propagator,
Eq.~(\ref{GTh}). In general, the self-energy is a complex-valued quantity
(as is the $T$-matrix), with its real part affecting the in-medium
quasiparticle mass while the imaginary part characterizes the
attenuation (absorption) of the propagating particle (wave).

\begin{figure}[!t]
\begin{minipage}{0.5 \linewidth}
\vspace{0.0cm}
\includegraphics[width=0.94\linewidth]{UQQ-latKZ.eps}
\end{minipage}
\begin{minipage}{0.5 \linewidth}
\vspace{-0.5cm}
\includegraphics[width=0.83\linewidth,angle=-90]{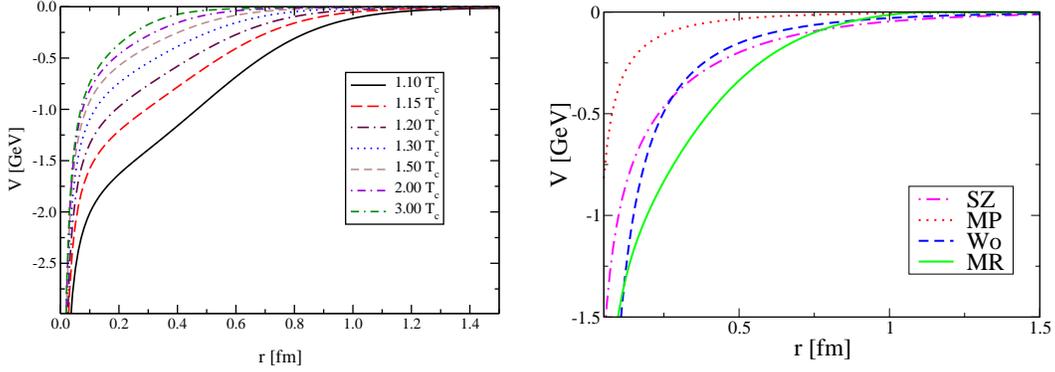}
\end{minipage}
\caption{Left panel: subtracted color-singlet heavy-quark internal energy,
$U_{Q\bar Q}(r)-U_{Q\bar Q}^\infty$, numerically evaluated from $N_f$=2 
lattice QCD computations in Ref.~\cite{Kaczmarek:2005gi}, as a function 
of $Q$-$\bar Q$ distance for various temperatures (figure taken from 
Ref.~\cite{Cabrera:2006wh}).
Right panel: Comparison~\cite{Mannarelli:2005pz} of different extractions
of the color-singlet $Q$-$\bar Q$ potential, Eq.~(\ref{Vqqb-lat}), at 
$T$=1.5~$T_c$
(SZ=\cite{Shuryak:2004tx}, MP=\cite{Mocsy:2005qw}, Wo=\cite{Wong:2004zr},
MR=\cite{Mannarelli:2005pz}).}
\label{fig_UQQ}
\end{figure}

Since the HQ self-energy depends on the heavy-light $T$-matrix, and the
latter, in turn, depends on the self-energy, one is facing a
Brueckner-type self-consistency problem, as illustrated in
Fig.~\ref{fig_brueck}. In the light-quark sector, the system of
Eqs.~(\ref{Tmat}) and (\ref{Sigma}) has been solved by numerical
iteration and moderate effects due to self-consistency have been found.
For a heavy quark, the impact of self-consistency (which to a large
extent is governed by the real parts of $\Sigma$) is weaker (since the
relative corrections to the HQ mass are not as large as for light
quarks). Therefore, in the calculations of the $T$-matrix in
Ref.~\cite{vanHees:2007me}, the real and imaginary parts of the
self-energies of light and heavy quarks have been approximated by
(constant) thermal mass corrections (largely being attributed to the
gluon-induced self-energy term, $\Sigma_{Qg}$, in Eq.~(\ref{Sigma}) and
resulting in $m_q^{\rm th}=0.25$~GeV and $m_{c,b}^{\rm th}=1.5,4.5~$GeV)
and quasiparticle widths of $\Gamma_{Q,q}=0.2$~GeV.  The underlying
heavy-light interaction potentials, $V_{1,\bar 3,6,8}$, have been
identified with the internal energy, $U_{\bar QQ}$, extracted from the
lattice results of the in-medium singlet free energy, 
$F^{(1)}_{\bar QQ}$, in combination with Casimir scaling for the 
strengths in the non-singlet color channels. Since the long-distance 
limit of the internal energy, 
$U_{\bar QQ}^\infty\equiv U_{\bar QQ}(r\to \infty)$,
does not go to zero, it needs to be subtracted to ensure the convergence
of the scattering equation,
\begin{equation}
  V_{\bar QQ} (r) = U_{\bar QQ}(r) - U_{\bar QQ}^\infty \ .
\label{Vqqb-lat}
\end{equation}
Clearly, for infinite separation in a deconfined medium, the force
between 2 quarks should vanish, which is consistent with the leveling
off of the free (and internal) energy at large $r$ in
Fig.~\ref{fig_F-T-lqcd}\footnote{Recall that the force is the gradient
  of the potential, $\vec {\cal F}=-\vec\nabla V(r)$}. This energy
contribution is naturally associated with an in-medium mass correction
(in fact, in the vacuum, it can be identified with the difference
between the bare charm-quark and the $D$-meson mass).  However, an
open problem is the entropy contribution in the internal energy,
$U_{\bar QQ}=F_{\bar QQ}+T S_{\bar QQ}$, which does not vanish for
$r\to\infty$; especially close to the critical temperature, the $T
S_{\bar QQ}^\infty$ term becomes uncomfortably large, and its
subtraction consequently leads to a rather strong effective potential,
cf.~Eq.~(\ref{Vqqb-lat}). This is currently the largest uncertainty in
the extraction of potentials from the lQCD free energy, which may be as
large as 50\%, as illustrated in Fig.~\ref{fig_UQQ}.
The use of the internal energy may be regarded as an upper estimate for 
the strength of the thus constructed potentials.

\begin{figure}[!t]
\begin{center}
\includegraphics[width=0.48\linewidth,height=0.35\linewidth]{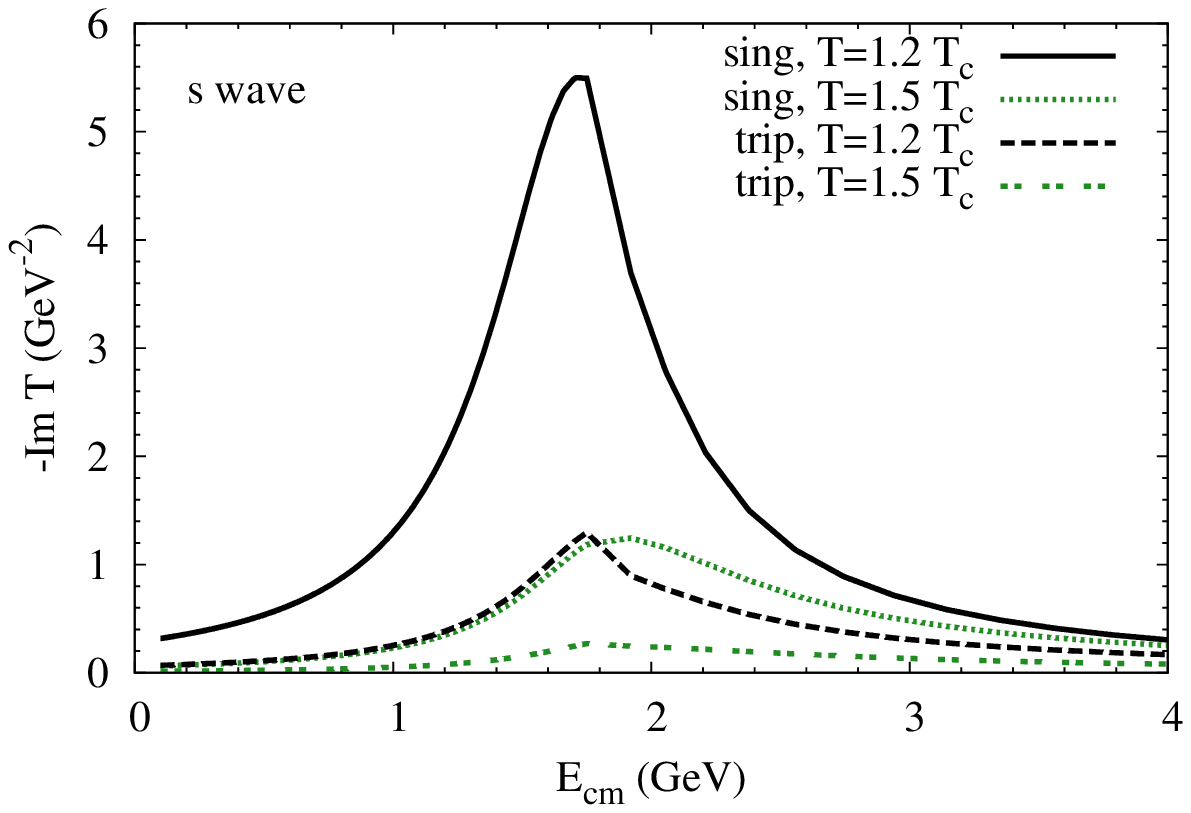}
\includegraphics[width=0.48\linewidth]{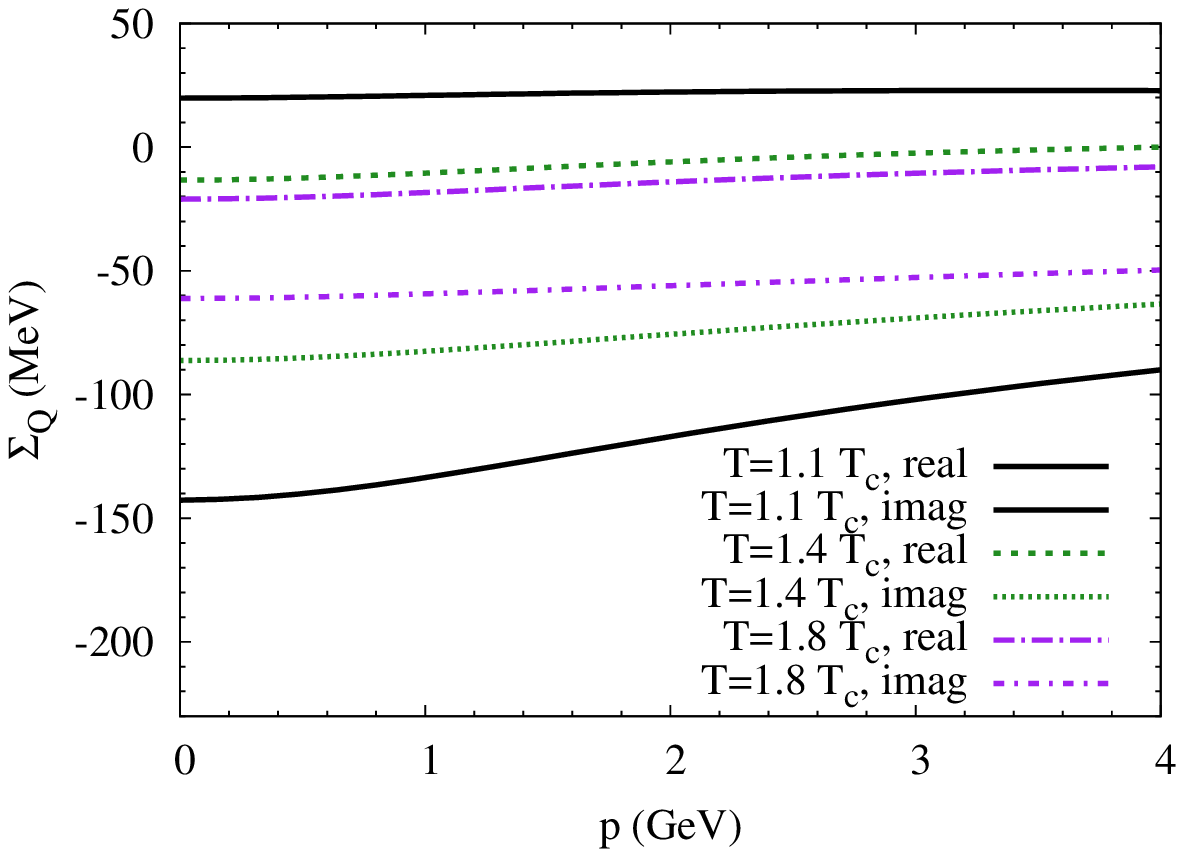}
\end{center}
\caption{Left panel: in-medium $T$-matrix for HQ potential scattering
  off light quarks and antiquarks in the QGP in the color-antitriplet
  diquark and color-singlet meson channels, respectively, at 2 different
  temperatures~\cite{vanHees:2007me}; right panel: real and imaginary
  parts of charm-quark self-energies at different temperatures resulting
  from a sum over all $T$-matrix channels; the real parts are generally
  small, while the imaginary parts are related to the scattering rate by
  $\Gamma_c=-2~{\rm Im}\Sigma_c$.}
\label{fig_ImT}
\end{figure}

The numerical calculations of the in-medium heavy-light $T$-matrices in
the scheme outlined above~\cite{vanHees:2007me} have found that the
dominant interactions are operative in the attractive color-singlet
(mesons) and color-antitriplet (diquark) channels, while they are
strongly suppressed in the repulsive sextet and octet channels. This can
be understood from the iterative structure of the $T$-matrix
Eq.~(\ref{Tmat}), implying that higher order terms in the Born series,
$T=V+VGV+VGVGV +\cdots $, are of alternating sign for repulsive
potentials, while they add for attractive potentials. Moreover, from the
left panel of Fig.~\ref{fig_ImT} one sees that the color-singlet meson
channel supports charm-light quark resonance structures in the vicinity
of the 2-body threshold up to temperatures of possibly $\sim$1.5~$T_c$,
and up to $\sim$1.2~$T_c$ in the antitriplet diquark channel (all
numerical results related to the $T$-matrix approach as shown here and
below are based on the potential labelled ``Wo'' in the right panel of
Fig~\ref{fig_UQQ}).

The interaction strength may be better quantified via the HQ
self-energies, displayed in the right panel of Fig.~\ref{fig_ImT} for
charm quarks. While the real parts are small at all temperatures (not
exceeding 0.02~GeV), the imaginary parts are substantial, translating
into scattering rates (or quasiparticle widths) of up to
$\Gamma_c=-2~{\rm Im}\Sigma_c \simeq$~0.2-0.3~GeV not too far above
$T_c$. Part of the reason for the different magnitudes in real and
imaginary parts is that the real part of the $T$-matrix assumes both
positive (repulsive) and negative (attractive) values, which compensate
each other when integrated over (or from different channels). The
imaginary part of the $T$-matrix is strictly negative (absorptive) and
always adds in the self-energy.

In the following Section we will set up a Brownian-Motion framework for
HQ diffusion in the QGP and address the question of how the different
interactions we have discussed above reflect themselves in the thermal
relaxation for charm and bottom quarks.
                                                                                    
\subsection{Heavy-Quark Transport}
\label{ssec_transport}
The description of the motion of a heavy particle in a fluid (or heat
bath) has a long history and a wide range of applications, including
problems as mundane as the diffusion of a drop of ink in a glass of
water. A suitable approach is given in terms of a diffusion equation for
the probability density, $\rho_1$, of the ``test particle'' 1. More
rigorously, one starts from the full Boltzmann equation for the phase
space density, $f_1(\vec r,\vec p,t)$,
\begin{equation}
  \left(\frac{\partial}{\partial t} + \frac{\vec p}{E} \cdot
    \vec\nabla_r - (\vec\nabla_r V)  \cdot \vec\nabla_p \right)
  f_1(\vec r,\vec p,t) = I_{\rm coll}(f_1)     \ , 
\label{boltz}
\end{equation}
where $\vec{\cal F}= -\vec\nabla_r V$ represents the force on the test
particle due to an external (in-medium) potential, $V$, and $I_{\rm
  coll}(f_1)$ is the collision integral induced by scattering off
particles in the heat bath.  The application to heavy-quark motion has
first been advocated in Ref.~\cite{Svetitsky:1987gq}. Neglecting the
mean-field term in Eq.~(\ref{boltz}), and assuming a uniform medium, one
can integrate the Boltzmann equation over the spatial coordinates to
obtain an equation for the distribution function, $f_Q$, of the heavy
quark in momentum space,
\begin{equation}
  f_Q(\vec p;t) \equiv \int d^3r f_Q(\vec r,\vec p,t) \ , 
\end{equation}  
which is solely determined by the collision term,
\begin{equation}
  \frac{\partial f_Q(\vec p;t)}{\partial t} =  I_{\rm coll}(f_Q) \ , 
\label{boltz-hom}
\end{equation}
The latter can be written as an integral over all momentum transfers,
$k$,
\begin{equation}
I_{\rm coll}(f_Q)= \int \dd^3 k [w(p+k,k) f_Q(p+k)-w(p,k) f_Q(p)] \  ,
\label{Icoll}
\end{equation}
where the key ingredient is a transition rate, $w(p,k)$, for the HQ
momentum to change from $p$ to $p-k$. The two terms in the integral
represent the scattering of the heavy quark into (``gain term'') and out
(``loss term'') of the momentum state $p$. The transport equation
(\ref{boltz-hom}) still constitutes a differential-integral equation for
$f_Q$ which in general is not easily solved. At this point, one can take
advantage of the HQ quark mass providing a large scale, so that the
momentum $p$ of the heavy quark can be considered to be much larger than
the typical momentum transfer, $k\sim T$, imparted on it from the
surrounding medium. Under these conditions, the transition rate in
Eq.~(\ref{Icoll}) can be expanded for small $k$. Keeping the first two
terms of this expansion, Eq.~(\ref{boltz-hom}) is approximated by the
Fokker-Planck equation,
\begin{equation}
  \frac{\partial f_Q(p,t)}{\partial t}=\frac{\partial}{\partial p_i} 
  \left[ A_i(p) + \frac{\partial}{\partial p_j} B_{ij}(p) \right] 
  f_Q(p,t) \ .
\label{FP}
\end{equation}
where the transport coefficients, $A_i$ and $B_{ij}$, are given by
\begin{equation}
  A_i(p)=\int \dd^3k \, w(p,k) \, k_i, 
  \quad B_{ij}(p)=\frac{1}{2} \int \dd^3k \, w(p,k) \, k_i k_j \ 
\label{ABw}
\end{equation}
in terms of the transition rate $w$; $A_i$ encodes the average momentum
change of the heavy quark per unit time and thus describes the friction
in the medium, while $B_{ij}$ represents the average momentum broadening
per unit time, i.e., the diffusion in momentum-space.  For an isotropic
medium (in particular a medium in thermal equilibrium), the transport
coefficients can be reduced to
\begin{equation}
  A_i(p)=\gamma(p^2) p_i, \quad B_{ij}(p)=\left [\delta_{ij} - \frac{p_i
      p_j}{p^2} \right ]B_0(p^2) + \frac{p_i p_j}{p^2} B_1(p^2) \ , 
\label{ABij}
\end{equation}
where the friction coefficient $\gamma =\tau_{\rm therm}^{-1}$ is
equivalent to a thermal relaxation time, and $B_0$ and $B_1$ are
diffusion coefficients. It is very instructive to examine the limit of
momentum independent coefficients (which in general is not the case and
will not be assumed below), in which case the Fokker-Planck equation
reduces to a particularly simple form,
\begin{equation}
\frac{\partial f_Q}{\partial t}=\gamma \frac{\partial}{\partial p_i}
(p_i f_Q) + D \frac{\partial}{\partial p_i} \frac{\partial}{\partial
  p_i} f_Q \ , 
\label{FP0}
\end{equation}
which clearly illustrates the form of the diffusion term, proportional
to a single diffusion constant $D$. The diffusion and friction constants
are, in fact, related via Einstein's famous fluctuation-dissipation
relation,
\begin{equation}
\gamma= \frac{D}{T m_Q} \ ,  
\label{Einstein}
\end{equation}
reducing the problem to a single transport coefficient. Note the
intimate connection between the HQ transport coefficients and the
temperature of the surrounding medium. This demonstrates that the
Fokker-Planck equation is a consistent approximation to the Boltzmann
equation in the sense that it recovers the proper equilibrium limit:
both friction and diffusion terms are essential to implement the
principle of detailed balance; a ``pure"  diffusion equation (i.e., 
$\gamma$=0) is only applicable in the limit $T,m_Q\to\infty$. 
Even in the presence of 
momentum dependent transport coefficients, the Einstein relation
(\ref{Einstein}), remains valid in the zero-momentum limit ($p\to 0$)
and provides for a valuable check whether the computed HQ diffusion
constants $D$ and $\gamma$ recover the temperature of the ambient
medium.

In phenomenological applications to heavy-ion collisions, the evolution
of high-$p_T$ particles is often approximated within an energy-loss
treatment, which amounts to neglecting the diffusion term, i.e., $D\to
0$. This means that only momentum- (or energy-) degrading processes are
taken into account, which is reflected in the Einstein equation as the
$T\to 0$ limit (with $D/T$ finite). 
The lack of momentum diffusion implies that both momentum randomization
and energy-gain processes are neglected. Thus, the particles can neither 
equilibrate nor become part of the collectively expanding medium (which
is, of course, crucial for the transfer of transverse and elliptic flow
from the medium to the particles propagating through it). 
Nevertheless, at high momentum this approximation may be in
order if the microscopic processes underlying energy loss are rare and
at high momentum transfer. In this case the Fokker-Planck equation is
not reliably applicable.

Let us now turn to the microscopic input to the calculation of the
transport coefficients. Throughout the remainder of this paper, we
employ fixed HQ masses at $m_c=1.5$~GeV and $m_b=4.5$~GeV. Using Fermi's
Golden Rule of quantum mechanics (or quantum field theory), the
transition rate $w$ can be expressed via the quantum mechanical
scattering amplitude, ${\cal M}$, underlying the pertinent scattering
process in the medium.  For elastic $Q+i\to Q+i$ scattering ($i=q,\bar
q, g$), the rate can be written as
\begin{eqnarray}
  w(p,k)= \int \frac{d_i \ \dd^3 q}
  {16(2 \pi)^9 \omega_p \omega_q \omega_{q+k} \omega_{p-k}}
  f_i(q) \, \overline{|\mathcal{M}_{iQ}|^2} \, 
  (2 \pi)^4 \delta(\omega_p+\omega_q-\omega_{p-k}-\omega_{q+k}) \ . 
\label{trans-rate}
\end{eqnarray}
In perturbative QCD, the amplitude is explicitly given by
Eq.~(\ref{Mscatt}) representing the Feynman diagrams depicted in
Fig.~\ref{fig_HQ-pqcd}. Likewise, the $T$-matrix discussed in the
previous Section can be directly related to the ${\cal M}$ amplitude,
see Ref.~\cite{vanHees:2007me}.  In Eq.~(\ref{trans-rate}), $d_i$
denotes the spin-color degeneracy of the parton, $f_i(q)$ is its
phase-space distribution in the medium, while $p$ and $q$ ($p-k$ and
$q+k$) are the initial (final) momenta of the heavy quark and the
parton, respectively. All in- and outgoing particles are on their mass
shell, i.e., $\omega_p=\sqrt{m^2+p^2}$ with their respective masses,
$m=m_{Q,i}$; the $\delta$-function enforces energy conservation in the
process.

\begin{figure}[!t]
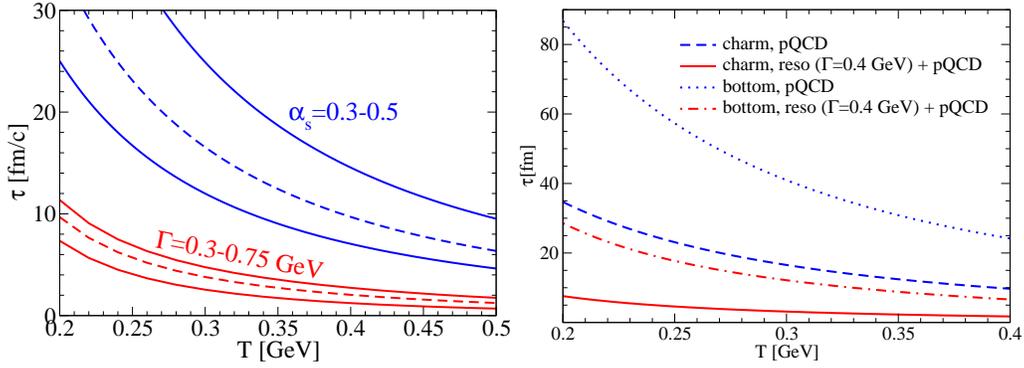

\begin{center}
\includegraphics[width=0.48\linewidth]{tau-charm-vsT.eps}
\includegraphics[width=0.48\linewidth]{tau-charm-bottom.eps}
\end{center}
\caption{Thermal relaxation times of heavy quarks at zero 3-momentum as 
  a function of temperature in the QGP~\cite{vanHees:2004gq}; left panel:
  charm quarks in the resonance model (with a $D$-meson width 
  $\Gamma_D$=0.3-0.8~GeV; lower band) compared to LO-pQCD (with a strong  
  coupling constant $\alpha_s$=0.3-0.5; upper band)~\cite{Rapp:2005at}.
  Right panel: comparison of resonance+pQCD interactions (red lines)
  and pQCD only (blue lines) for charm and bottom quarks.}
\label{fig_tau-T}
\end{figure}
In Fig.~\ref{fig_tau-T}, we compare the temperature dependence of the
inverse friction coefficient (thermal relaxation time) for charm quarks
at zero momentum for elastic scattering in pQCD (as discussed in
Sec.~\ref{sssec_pqcd}) with the effective resonance model
(Sec.~\ref{sssec_reso}). The latter leads to substantially lower
thermalization times than pQCD scattering, by around a factor of $\sim$3
at temperatures $T\simeq$~1-2~$T_c$. In contrast to pQCD, the values of
$\tau_{c,\rm therm}\simeq$~2-10~fm/$c$ within the resonance model are
comparable to the expected QGP lifetime at RHIC, $\tau_{\rm QGP}\simeq
5$~fm/$c$; thus, if resonances are operative, significant modifications
of charm spectra at RHIC are anticipated due to (the approach to)
thermalization. The uncertainty band covered by varying the effective
coupling constant, $G$, over a wide range is comparatively moderate. We
recall that the magnitude of the underlying total cross sections for
pQCD and the resonance model (cf.~left panel of Fig.~\ref{fig_xsec}) are
not largely different; an important effect thus arises due to the angular
dependence of the differential cross section (or scattering amplitudes),
which is forward dominated in pQCD (corresponding to a small 3- or
4-momentum transfer, $t$ or $k$) while isotropic in the rest frame of a
$D$-meson resonance implying larger momentum transfers, $k$. Since the
expression for $A_i$ (and thus for $\gamma$), Eqs.~(\ref{ABw}) and
(\ref{ABij}), directly involves $k$, large-angle scattering is more
efficient in thermalizing the $c$-quark distributions.  
Charm-quark relaxation in the resonance model appears to become less 
efficient at high temperatures. This is not due to the disappearance of 
the resonances (as will be the case in the lattice-QCD based potential
approach), but due to a mismatch between the excitation energy of the
resonances and the increasing average thermal energy of the partons in
the heat bath. E.g., at $T=0.5$~GeV, the latter amounts to 
$\omega_{\rm thermal} = 3~T =1.5$~GeV, which is well above the optimal 
energy for forming a $D$-meson resonance in collisions with a 
zero-momentum charm
quark. Thus, with increasing temperatures the efficiency of the
resonances in $c$-quark scattering is diminished since the partons in
the surrounding medium become too energetic.

\begin{figure}[!t]
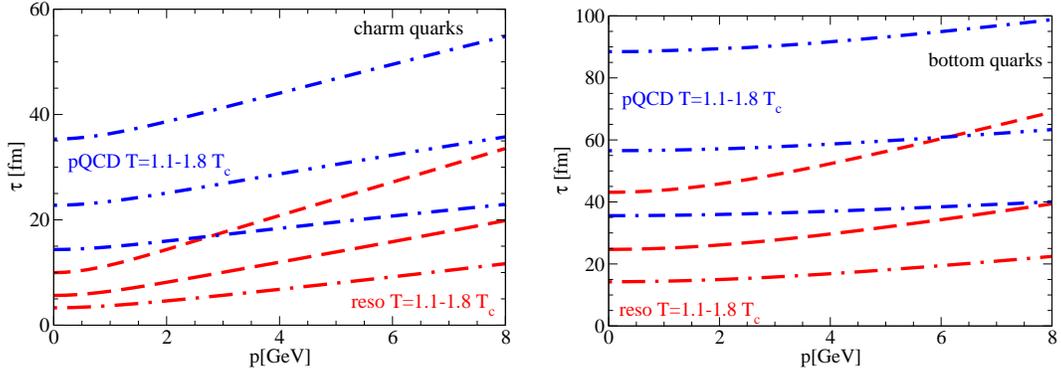

\begin{center}
\includegraphics[width=0.48\linewidth]{reso-tau-vs-p.eps} \hfill
\includegraphics[width=0.48\linewidth]{reso-tau-vs-p-bottom.eps}
\end{center}
\caption{Thermal relaxation times of charm (left panel) and bottom
  quarks (right panel) in the QGP as a function of 3-momentum at 3
  different temperatures, using either pQCD ($\alpha_s$=0.4; upper
  3 lines at $p$=0 for charm, and lines 1, 2 and 4 from above at $p$=0
  for bottom) or resonance scattering ($\Gamma_\Phi$=0.4~GeV; lower
  3 lines at $p$=0 for charm, and lines 1, 2 and 4 from below at $p$=0
  for bottom)~\cite{vanHees:2004gq}. The temperatures are $T$=1.1~$T_c$, 
  1.4~$T_c$ and 1.8~$T_c$ from top to bottom within each set of 
  interaction.}
\label{fig_tau-p}
\end{figure}

The effect of $B$-meson resonances on bottom quarks is relatively
similar to the charm sector, i.e., a factor of $\sim$3 reduction in the
thermal relaxation time compared to pQCD, cf.~right panel of
Fig.~\ref{fig_tau-T}. However, the magnitude of $\tau_{b,\rm therm}$
stays above typical QGP lifetimes at RHIC, especially below 2~$T_c$
where almost all of the QGP evolution is expected to occur (based, e.g.,
on hydrodynamic simulations). The momentum dependence of the relaxation
times for LO-pQCD and resonance interactions is illustrated in
Fig.~\ref{fig_tau-p}. The latter show a more pronounced increase of the
relaxation time with increasing $p$, since their interaction strength is
concentrated at low energies. It has been found that even at high
momenta, the main interaction is still via resonance formation with a
``comoving'' parton from the heat bath (rather than from tails of the
resonance in interactions with partons of typical thermal energies).

\begin{figure}[!t]
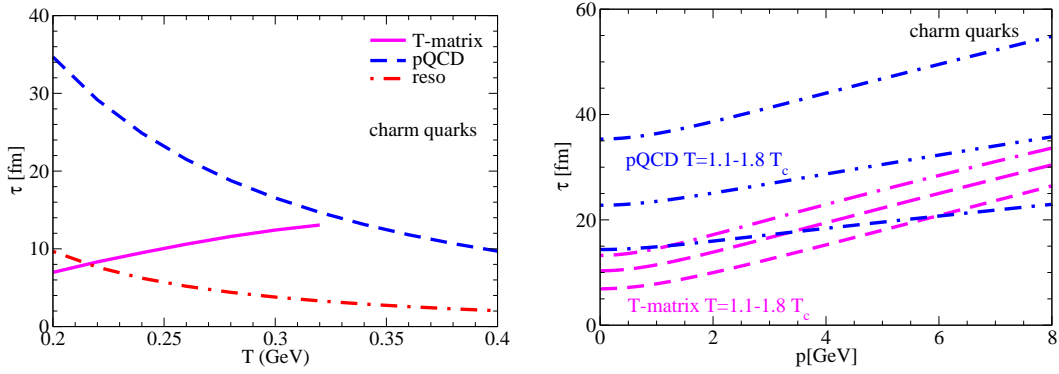

\begin{center}
\includegraphics[width=0.48\linewidth]{tau-vs-T.eps} \hfill
\includegraphics[width=0.48\linewidth]{t-matrix-tau-vs-p.eps} 
\end{center}
\caption{Thermal relaxation times for charm quarks in the QGP as
  computed from the heavy-light quark $T$-matrices utilizing potentials
  estimated from lattice QCD, including all 4 color combinations in
  $c\bar q$ and $cq$ channels~\cite{vanHees:2007me,vanHees:2008xx}. 
  The temperature dependence at zero 3-momentum (middle line in the left 
  panel) and the 3-momentum dependence at 3 temperatures (lower 3 lines
  at $p$=0 in the right panel) of $\tau_{c,\rm therm}$ are compared to 
  LO-pQCD scattering (upper blue curves at $p$=0). Note that in the 
  right panel, the curves are for temperatures $T$=1.1~$T_c$, 1.4~$T_c$, 
  1.8~$T_c$ from bottom top (top to bottom) for the $T$-matrix (pQCD) 
  interactions.}
\label{fig_tau-lqcd}
\end{figure}

\begin{figure}[!t]
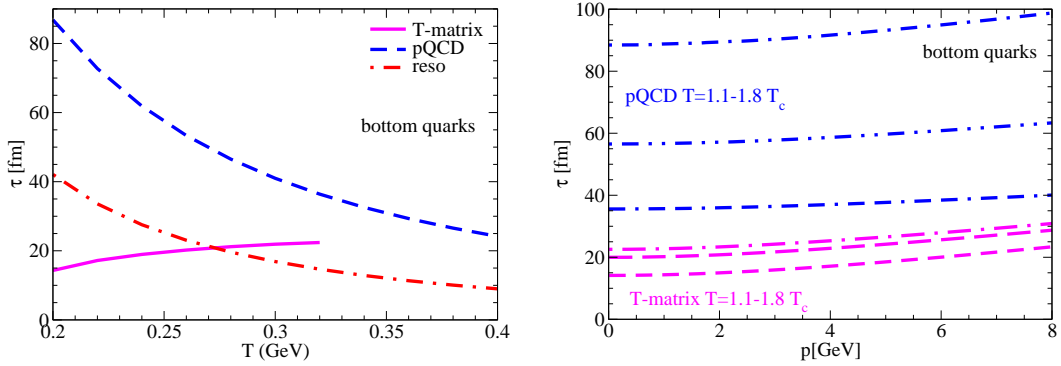

\begin{center}
\includegraphics[width=0.48\linewidth]{tau-vs-T-bottom.eps} \hfill
\includegraphics[width=0.48\linewidth]{t-matrix-tau-vs-p-bottom.eps}
\end{center}
\vspace{-0.4cm}
\caption{Same as Fig.~\ref{fig_tau-lqcd} but for bottom 
 quarks~\cite{vanHees:2007me,vanHees:2008xx}.}
\label{fig_tau-lqcd-bot}
\end{figure}

Next, we examine the transport coefficients as computed from the
$T$-matrix approach using lQCD-based potentials (as elaborated in in
Sec.~\ref{sssec_pot}). In Fig.~\ref{fig_tau-lqcd} we display the
temperature dependence of the thermal relaxation time for charm quarks.
Close to $T_c$, the strength of the $T$-matrix based
interactions (including all color channels) is very comparable to the
effective resonance model. It turns out that the color-singlet meson
channel and the color-antitriplet diquark channel contribute to the
friction coefficient by about equal parts; the somewhat smaller
$T$-matrix in the diquark channel is compensated by the 3-fold color
degeneracy in the intermediate scattering states. The contribution of
the repulsive sextet and octet channels is essentially
negligible. Contrary to both pQCD and the resonance model, the
lQCD-based $T$-matrix approach leads to an increase of the $c$-quark
relaxation time with increasing temperature, despite the substantial
increase of the parton densities with approximately $T^3$. The increase
in scattering partners is overcompensated by the loss of interaction
strength as determined by weakening of the lQCD-based potentials, which
is largely attributed to color screening. As a consequence,
thermalization due to LO-pQCD scattering becomes more efficient than the
nonperturbative $T$-matrix for temperatures above $T\simeq
1.8~T_c$. This feature is very much in line with the generally expected
behavior that the QGP becomes a more weakly coupled gas at sufficiently
large temperatures. More quantitatively, it is reflected by lattice QCD 
computations of bulk matter properties, in particular by the 
so-called ``interaction measure", $\varepsilon-3P$, which may be 
interpreted as an indicator of nonperturbative effects: as apparent
from the left panel of Fig.~\ref{fig_eos-lqcd}, this quantity becomes
close to zero at temperatures above $\sim$2$T_c\simeq$~0.4~GeV. 
Similarly, up to $\sim$2$T_c$, the (renormalized) Polyakov loop 
exhibits substantial deviations from one (i.e., the value corresponding 
to a ``fully" deconfined QGP), cf.~right panel of Fig.~\ref{fig_op-lqcd}.  
The 3-momentum dependence of the $T$-matrix based relaxation times, shown 
in the right panel of Fig.~\ref{fig_tau-lqcd}, reconfirms the loss of 
interaction strength at high momenta as found in the resonance model. The 
transport properties of the bottom quarks as evaluated in the $T$-matrix 
approach are summarized in Fig.~\ref{fig_tau-lqcd-bot}; one finds very 
similar features as in the charm sector, quantitatively differing by a 
factor of $\sim$$m_b/m_c=3$.

\begin{figure}[!t]
\vspace{0.4cm}
\begin{center}
\includegraphics[width=0.49\linewidth]{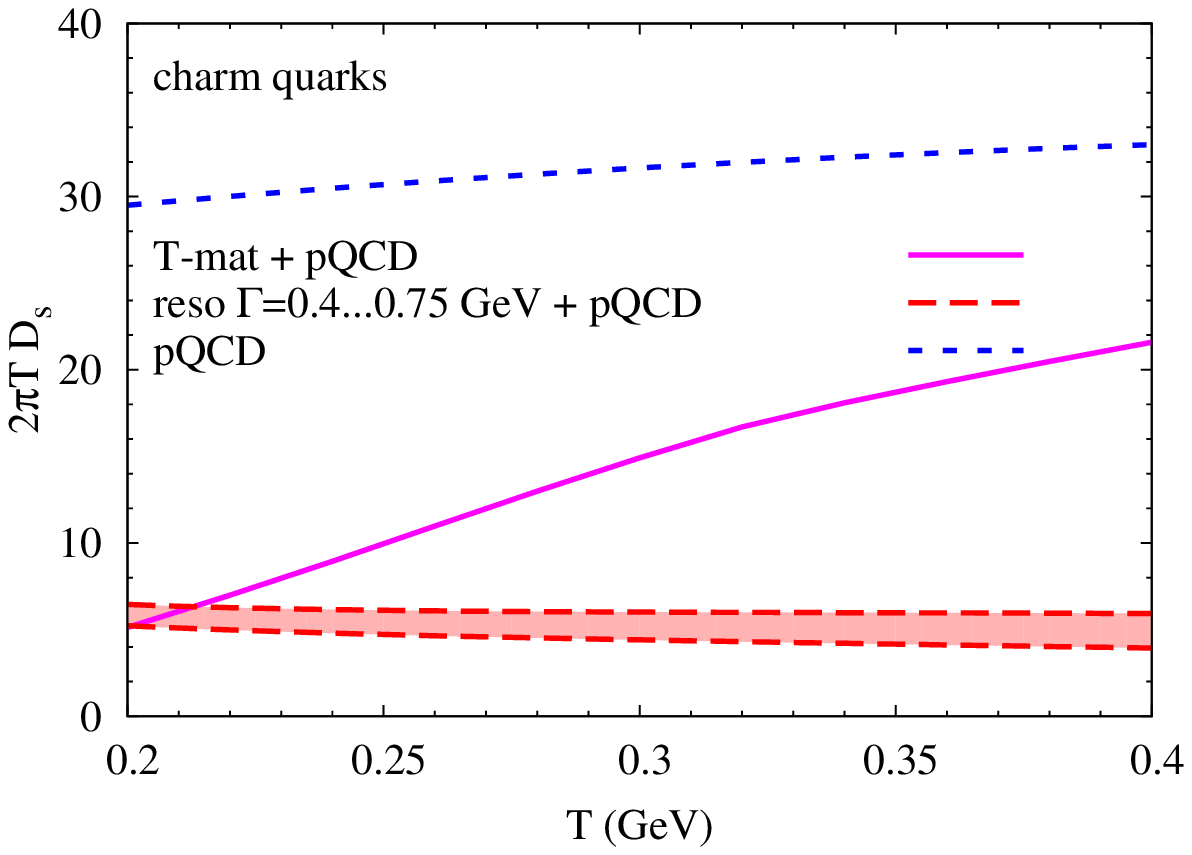}
\includegraphics[width=0.49\linewidth]{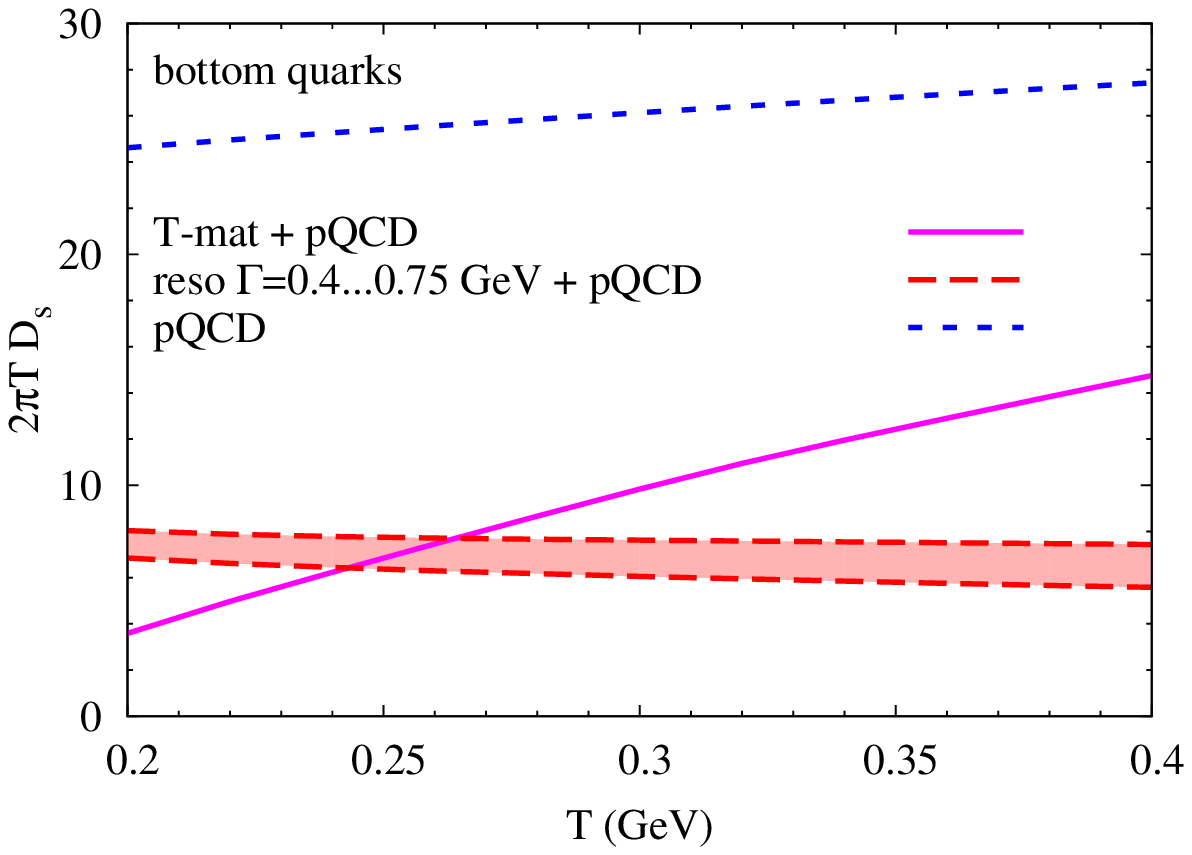}
\end{center}
\vspace{-0.4cm}
\caption{Spatial diffusion constant, $D_s$, for charm (left panel) and
  bottom (right panel) quarks in units of the thermal wave length,
  $1/(2\pi T)$, as a function of temperature. Compared are results from
  elastic pQCD scattering, resonance+pQCD model and $T$-matrix+pQCD model.}
\label{fig_Ds}
\end{figure}

The momentum-space diffusion (or friction) coefficient can be converted
into a spatial diffusion constant, defined in the standard way via the
variance of the time ($t$) evolution of the particle's position,
\begin{equation}
\langle x^2 \rangle - \langle x \rangle^2 = 2 D_x t \ .  
\end{equation}
with
\begin{equation}
D_x =\frac{dT}{m_Q \gamma}
\label{Dx}
\end{equation}
and $D_s=D_x/d$ where $d$ denotes the number of spatial dimensions.  In
Fig.~\ref{fig_Ds}, $D_s$ for charm and bottom quarks is summarized for
the 3 different interactions (LO-pQCD, resonance+pQCD model and
nonperturbative $T$-matrix+pQCD).
Nonperturbative interactions lead to a substantial reduction of
diffusion in coordinate space compared to LO pQCD, especially for charm
quarks.  E.g., for a central Au-Au collision at RHIC, taking as an
illustration $T\simeq0.2$~GeV, $t=\tau_{\rm QGP}\simeq 5$~fm/$c$,
$D_s=6/2\pi T$ and $d=2$ (in the transverse plane), one finds $\Delta x
= \sqrt{2 D_x t} \simeq 4.3$~fm/$c$ compared to $\sim 8.6$~fm/$c$ for LO
pQCD only. This indicates that spatial diffusion is significantly
inhibited in the presence of nonperturbative interactions, and is
smaller than the typical transverse size of the fireball, $R\simeq
8$~fm/$c$, indicating a strong coupling of the charm quark to the medium
over the duration of its evolution.

\subsection{Heavy-Quark Observables at RHIC}
\label{ssec_obs}
In this Section we elaborate on how the above developed Brownian Motion
approach can be implemented into a description of HQ observables in
ultrarelativistic heavy-ion collisions
(URHICs)~\cite{Moore:2004tg,vanHees:2005wb,Gossiaux:2006yu,
Liu:2006vi}\footnote{For implementations into parton transport models, 
see, e.g., Refs.~\cite{Molnar:2004ph,Zhang:2005ni}.}. This requires the
following ingredients: (i) a realistic evolution of the expanding QGP
fireball; (ii) Langevin simulations of the heavy quarks in the fireball
background with realistic input spectra; (iii) hadronization of the HQ
spectra into $D$- and $B$-meson spectra at the end of the QGP fireball
evolution, and (iv) semileptonic decays of the $D$- and $B$-mesons to
compare to experimental single-electron ($e^\pm$) spectra.  We will
focus on Au-Au~($\sqrt{s}=200$~AGeV) collisions at RHIC where first
measurements of the nuclear modification factor, $R_{AA}(p_T)$, and
elliptic flow, $v_2(p_T$), have become available over the last
$\sim$3-4~years.

\subsubsection{Langevin Simulations}
\label{sssec_langevin}
The Fokker-Planck equation for the time evolution of the phase-space
distribution of a heavy particle moving through a fluid can be solved
stochastically utilizing a Langevin process. The change in position and
momentum of the heavy quark over a discrete but small time interval,
$\delta t$, are evaluated in the rest frame of the medium (QGP)
according to
\begin{equation}
\label{langevin}
\delta \vec{x}=\frac{\vec{p}}{\omega_p}~\delta t \ , \quad
\delta \vec{p}=-A(t,\vec{p}+\delta \vec{p})~\vec{p}~\delta t + \delta
\vec{W}(t,\vec{p}+\delta \vec{p}) \ , 
\end{equation}
where $\omega_p$ denotes the on-shell HQ energy and $\vec p /\omega_p$
is its relativistic velocity. The drag and diffusion terms of the
Fokker-Planck equation determine the change of momentum, $\delta \vec
p$.  The momentum diffusion is realized by a random change of momentum
$\delta \vec{W}$ which is assumed to be distributed according to
Gaussian noise~\cite{Dunkel:2005},
\begin{equation}
\label{fluct-force}
P(\delta \vec{W}) \propto \exp \left [-\frac{\hat{B}_{jk} \delta W^j \delta
    W^{k}}{4 \delta t} \right] \ , \quad \hat{B}_{ij}=(B^{-1})_{ij} \ ,
\end{equation}
where $(B^{-1})_{ij}$ denotes the inverse of the matrix $B_{ij}$, the
momentum-diffusion coefficients of Eq.~(\ref{ABij}). The Gaussian form
of this force is inherently consistent with the underlying Fokker-Planck
equation which was derived from the Boltzmann equation in the limit of
many small momentum transfers (central limit theorem).

A realistic application to URHICs hinges on a proper description of the
evolving medium. As discussed in Sec.~\ref{ssec_urhic}, ideal
hydrodynamic simulations, especially for the QGP phase, reproduce the
observed collective expansion properties in central and semicentral
Au-Au collisions at RHIC very well, and have been employed for HQ
Langevin simulations in connection with LO-pQCD transport coefficients
in Ref.~\cite{Moore:2004tg}. Alternatively, the basic features of
hydrodynamic evolutions (collective flow and expansion timescales) may
be parametrized using expanding fireball models. In
Ref.~\cite{vanHees:2005wb}, an earlier developed fireball model for
central Au-Au collisions~\cite{Rapp:2000pe} has been extended to account
for the azimuthally asymmetric (elliptic) expansion dynamics closely
reminiscent of the hydrodynamic simulations of
Ref.~\cite{Kolb:2000sd}. Let us briefly outline the main components of
such a description. The starting point is the time dependent volume
expansion of a fire cylinder, $V_{\rm FB}=z(t) \pi a(t) b(t) $ where
$a(t)$ and $b(t)$ characterize the elliptical expansion in the
transverse plane and $z(t)$ the longitudinal size (typically covering
$\Delta y = 1.8$ units in rapidity, corresponding to the width of a
thermal distribution).
\begin{figure}[!t]
\begin{center}
\includegraphics[width=0.48\linewidth]{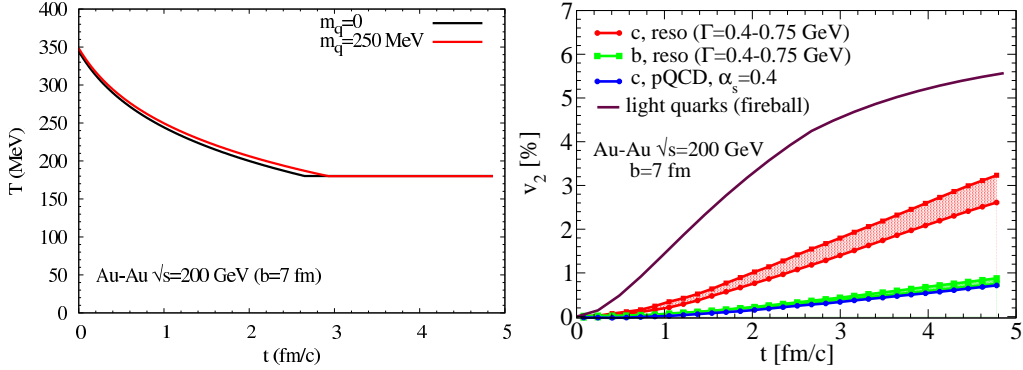}
\includegraphics[width=0.48\linewidth]{v2-vs-time.eps}
\end{center}
\caption{Thermal fireball expansion for semicentral Au-Au collisions at
  RHIC. Left panel: temperature evolution for either a massless gas with
  $d_{\rm eff}$=42 or a massive quasiparticle gas with $d_{\rm QGP}$=48
  and thermal parton masses of $m_i$=0.25~GeV; right panel: time
  evolution of the inclusive elliptic flow, $v_2$, of the bulk medium
  (upper solid curve) and for heavy quarks as following from
  relativistic Langevin calculations: charm quarks with LO-pQCD
  interactions only ($\alpha_s$=0.4, lowest line), as well as charm
  (upper band) and bottom quarks (lower band) for the resonance model
  ($\Gamma_D$=0.4~GeV) + pQCD.}
\label{fig_fireball}
\end{figure}
As in ideal hydrodynamics, the evolution is assumed to be isentropic,
i.e., to proceed at a fixed total entropy, $S$, which is matched to the
number of observed hadrons at the empirically inferred chemical
freezeout, cf. Fig.~\ref{fig_phasedia}. The time dependence of the
entropy density, $s(t)=S/V_{FB}(t)=s(T)$, then determines the
temperature evolution, $T(t)$ of the medium using $s_{\rm QGP}(T)=d_{\rm
  eff} \frac{4\pi^2}{90}T^3$ in the QGP (with $d_{\rm eff}\simeq40$ to
account for deviations from the ideal gas, cf.~Fig.~\ref{fig_eos-lqcd})
and a (numerical) hadron resonance gas equation of state, $s_{\rm
  HG}(T)$, in the hadronic phase. At $T_c$, which at RHIC energies is
assumed to coincide with chemical freezeout at $T_{\rm chem}=180~{\rm
  MeV}$, the hadron gas is connected to the QGP phase in a standard
mixed-phase construction. Assuming a formation time of the thermal
medium of $\tau_0=1/3$~fm/$c$ (translating into a longitudinal size of
$z_0=\tau_0 \Delta y=0.6$~fm) after the initial overlap of the colliding
nuclei, the initial temperature for semicentral (central)
Au-Au($\sqrt{s}=200$~AGeV) collisions amounts to
$T_0=0.34(0.37)$~GeV. The subsequent cooling curve and elliptic flow of
the bulk medium are displayed in Fig.~\ref{fig_fireball}.  The last
ingredient needed for the HQ Langevin simulations are the initial charm
and bottom-quark spectra. They have been constructed to reproduce
available experimental information on $D$-meson and $e^\pm$ spectra in
elementary $p$-$p$ and $d$-Au collisions at RHIC~\cite{Rapp:2005at},
where no significant medium formation (and thus modification of their
production spectra) is expected.

\begin{figure}[!t]
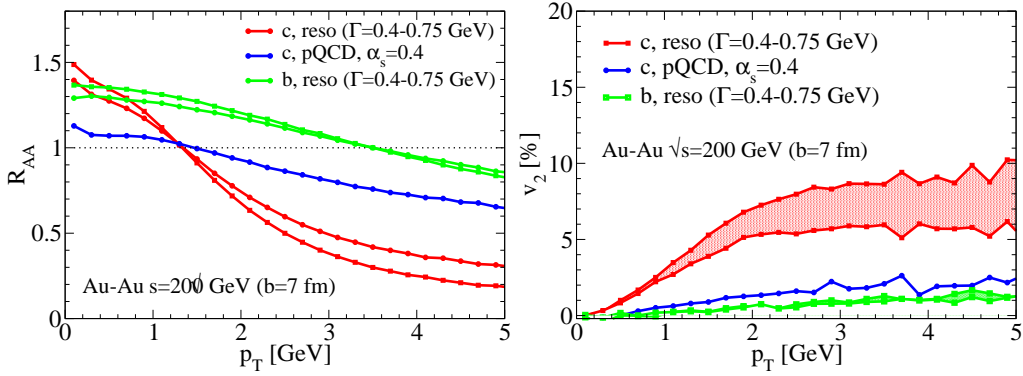

\begin{center}
\includegraphics[width=0.48\linewidth]{RAA-quark-minbias.eps}
\includegraphics[width=0.48\linewidth]{v2-quark-minbias.eps}
\end{center}
\caption{Nuclear modification factor (left panel) and elliptic flow
  (right panel) of heavy quarks as a function of their transverse
  momentum in semicentral ($b$=7~fm) Au-Au 
  collisions~\cite{vanHees:2005wb}.}
\label{fig_v2-Raa-quark}
\end{figure}

In Fig.~\ref{fig_v2-Raa-quark} we compare the results for HQ $p_T$
spectra and elliptic flow of Langevin simulations in the above described
fireball expansion using either LO-pQCD scattering with $\alpha_s=0.4$
(which may be considered as an upper estimate) or a combination of the
$Q+\bar q\to \Phi$ resonance interaction with
LO-pQCD~\cite{vanHees:2005wb}. This combination is motivated by the fact
that in the resonance model the interaction of a heavy quark is
restricted to (light) antiquarks from the medium, while LO-pQCD is
dominated by interactions with thermal gluons (the contribution from
antiquarks is small, at the $\sim$10-15\% level of the total pQCD part).
One finds that both the suppression at intermediate $p_T$ and the
elliptic flow of charm quarks are augmented by a factor 3-5 over LO-pQCD
interactions, quite reminiscent to what has been found at the level of
the transport coefficients. The uncertainty due to variations in the
effective resonance parameters is moderate, around $\pm$30\%. It
is remarkable that the Langevin simulations naturally provide for a
leveling off of the elliptic flow as a characteristic signature of the
transition for a quasi-thermal regime at low $p_T$ to a kinetic regime
for $p_T\ge2$~GeV, very reminiscent to the empirical ${\rm KE}_T$
scaling shown in Fig.~\ref{fig_rhic-data}.
\begin{figure}[!b]
\begin{center}
\includegraphics[width=0.42\linewidth,height=0.45\linewidth,angle=-90]
{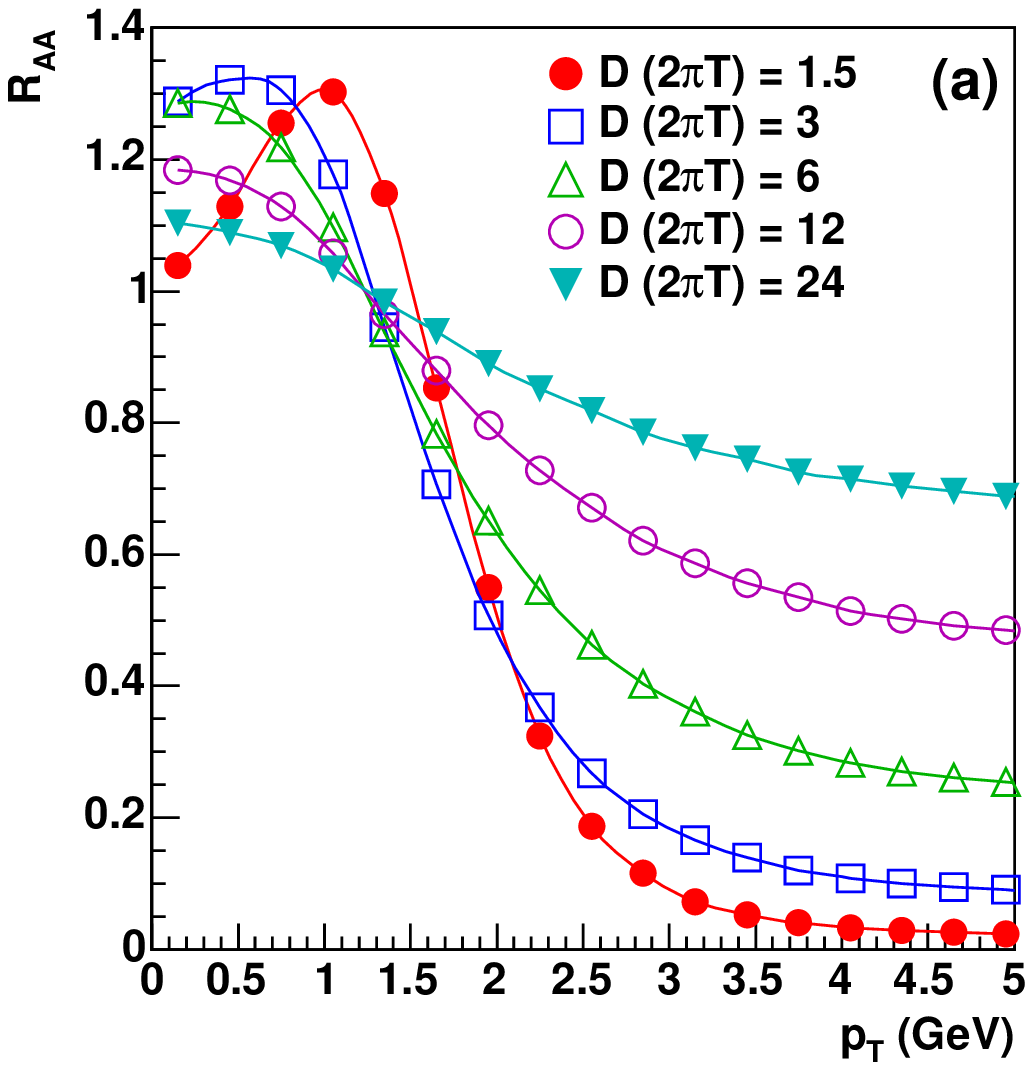}
\hspace{1cm}
\includegraphics[width=0.42\linewidth,height=0.45\linewidth,angle=-90]
{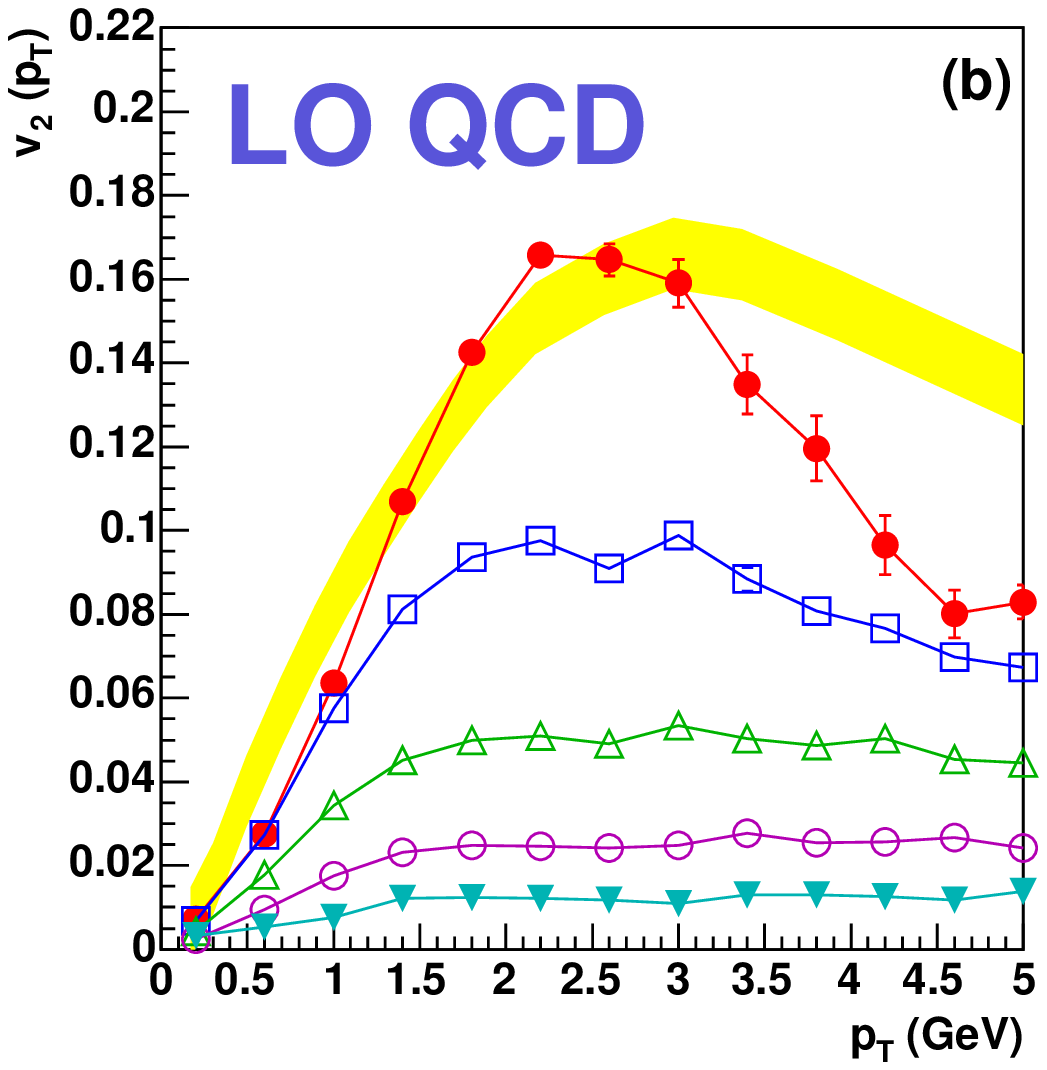}
\end{center}
\caption{Nuclear modification factor and elliptic flow of charm quarks
  as a function of their transverse momentum in semicentral ($b$=6.5~fm)
  Au-Au collisions using a hydrodynamic evolution of the bulk medium at
  RHIC~\cite{Moore:2004tg}.}
\label{fig_v2-Raa-teaney}
\label{fig_MT04}
\end{figure}
Even the quantitative plateau value of 7-8\% is recovered, indicating
that with the resonance+pQCD model the $c$-quarks are largely
participating in the collective expansion of the medium. On the other
hand, bottom quarks strongly deviate from the universal behavior, which
is of course due to their much larger mass, $m_b=4.5$~GeV.

The results of the HQ Langevin calculations in a hydrodynamic simulation
for semicentral Au-Au collisions are summarized in
Fig.~\ref{fig_MT04}~\cite{Moore:2004tg}. In these calculations, the pQCD
HQ scattering amplitudes have been augmented by a full perturbative
in-medium gluon exchange propagator with a fixed Debye mass of
$\mu_D$=1.5~$T$, while the strong coupling constant $\alpha_s$ has been
varied to produce a rather large range of diffusion constants.
The hydrodynamic evolution is performed for Au-Au collisions at impact
parameter $b$=6.5~fm with a thermalization time of $\tau_0$=1~fm/$c$
(corresponding to an initial temperature of $T_0=0.265$~GeV) and a
critical temperature of $T_c$=0.165~GeV. The basic trends of the HQ
spectra and elliptic flow are consistent with the fireball simulations
of Ref.~\cite{vanHees:2005wb}, indicating a strong correlation between
large $v_2$ and small $R_{AA}$ (strong suppression). Comparing more
quantitatively the simulations with a ``realistic'' pQCD HQ diffusion
constant of $D_s=24/2\pi T$ to LO-pQCD in the fireball evolution
(corresponding to $D_s\simeq30/2\pi T$), reasonable agreement is found,
with $R_{AA}$($p_T$=5~GeV)$\simeq$0.7 and
$v_2$($p_T$=5~GeV)$\simeq$1.5-2~\%, especially when accounting for the
lower $T_0$ in the hydro evolution (implying less suppression) and the
smaller impact parameter (implying less $v_2$).

\begin{figure}[!t]
\begin{center}
\includegraphics[width=0.48\linewidth]{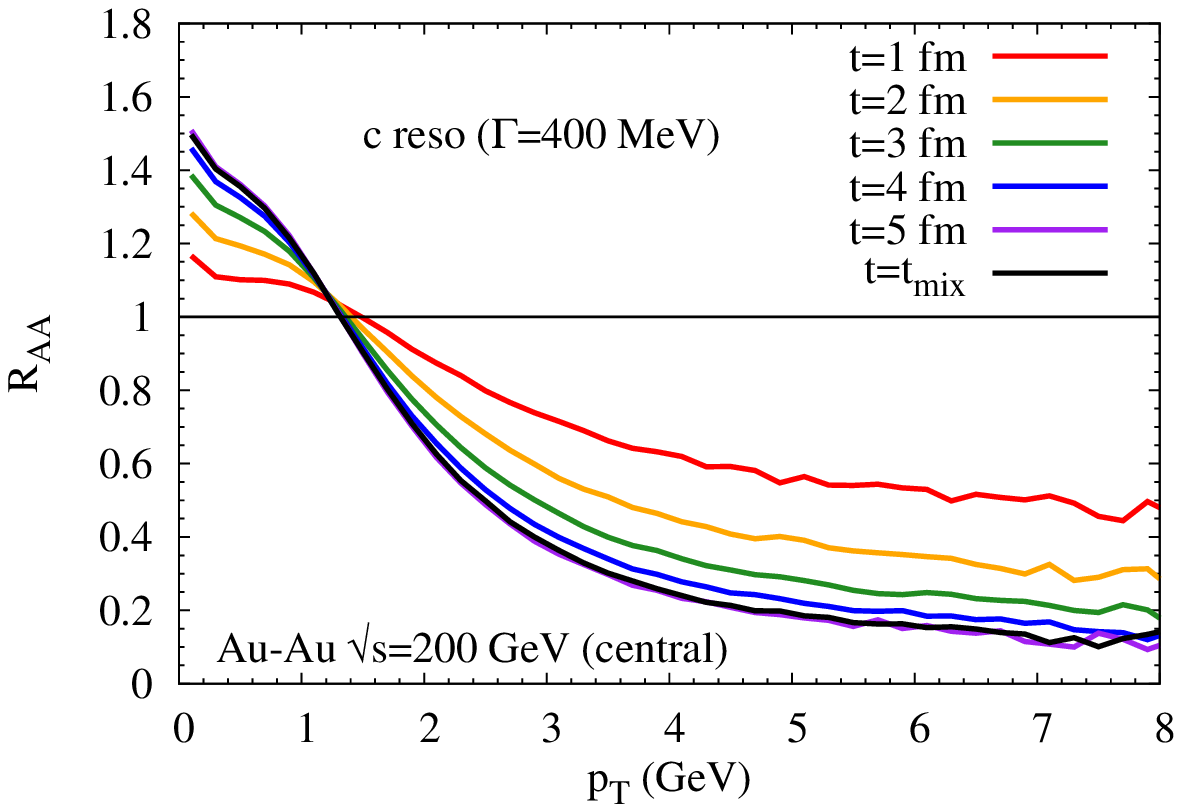}
\includegraphics[width=0.48\linewidth]{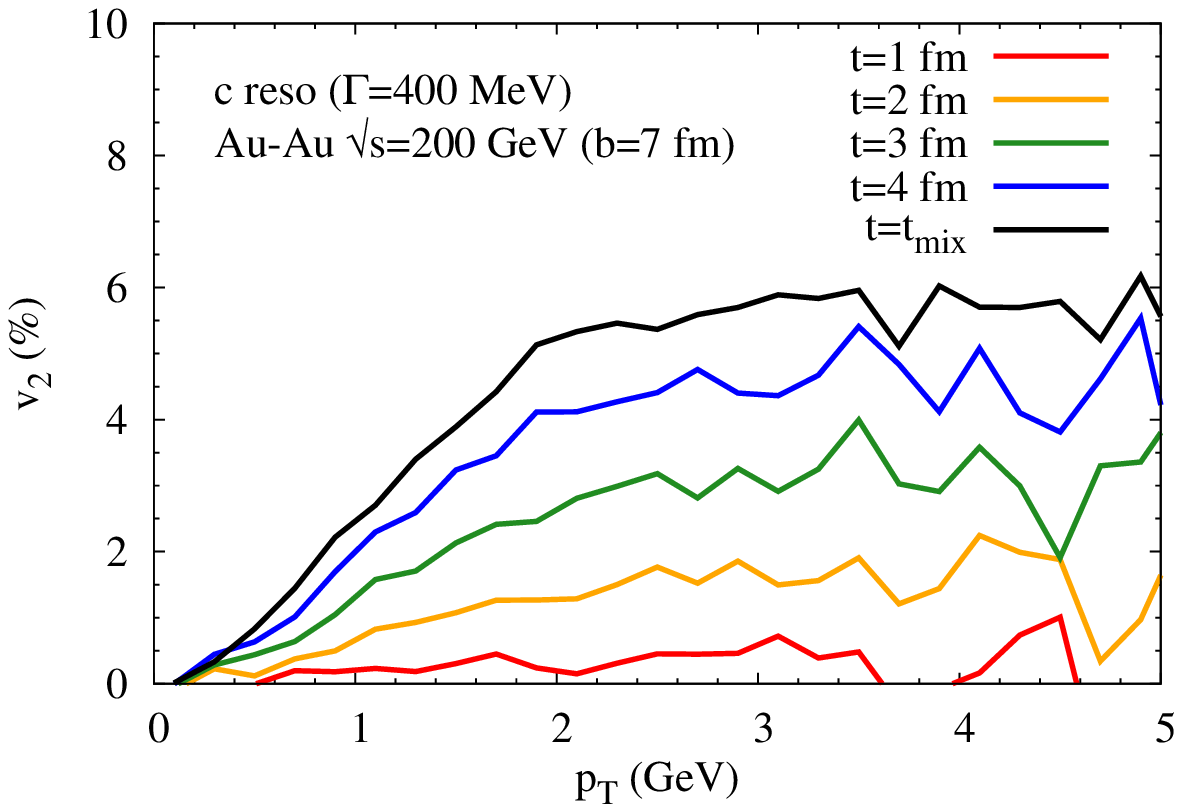}
\end{center}
\vspace{-0.4cm}
\caption{Time evolution of the nuclear modification factor (left panel,
  central collisions) and elliptic flow (right panel, semicentral
  collisions) of charm quarks in the resonance+pQCD model in Au-Au
  collisions at RHIC~\cite{vanHees:2005wb,Rapp:2006ta}.}
\label{fig_v2-Raa-time}
\end{figure}

It is very instructive to investigate the time evolution of the
suppression and elliptic flow, displayed in Fig.~\ref{fig_v2-Raa-time}
for the resonance+pQCD model in the fireball evolution. It turns out
that the suppression in $R_{AA}$ is actually largely built up in the
early stages of the medium expansion, where the latter is the hottest
and densest, i.e., characterized by a large (local) opacity. On the
other hand, the elliptic flow, being a collective phenomenon, requires
about $\sim$4~fm/$c$ to build up in the ambient fireball matter,
implying that the charm-quark $v_2$ starts building up only 1-1.5~fm/$c$
after thermalization, and then rises rather gradually, see left panel of
Fig.~\ref{fig_fireball}. Thus, the time evolution of $R_{AA}$ and $v_2$
is not much correlated, quite contrary to the tight correlation
suggested by the final results. This will have important consequences
for the interpretation of the experimental data in
Sec.~\ref{sssec_spectra}.

\begin{figure}[!t]
\vspace{0.4cm}
\begin{center}
\includegraphics[width=0.48\linewidth]{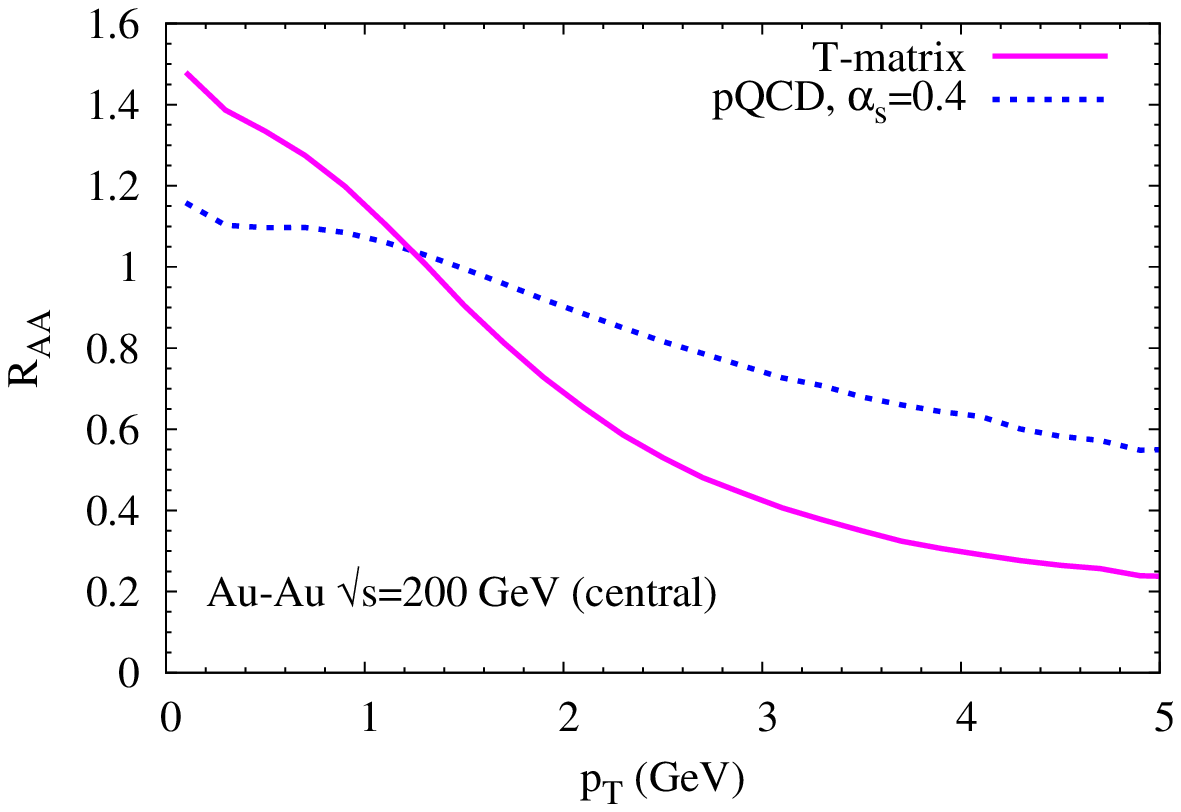}
\includegraphics[width=0.48\linewidth]{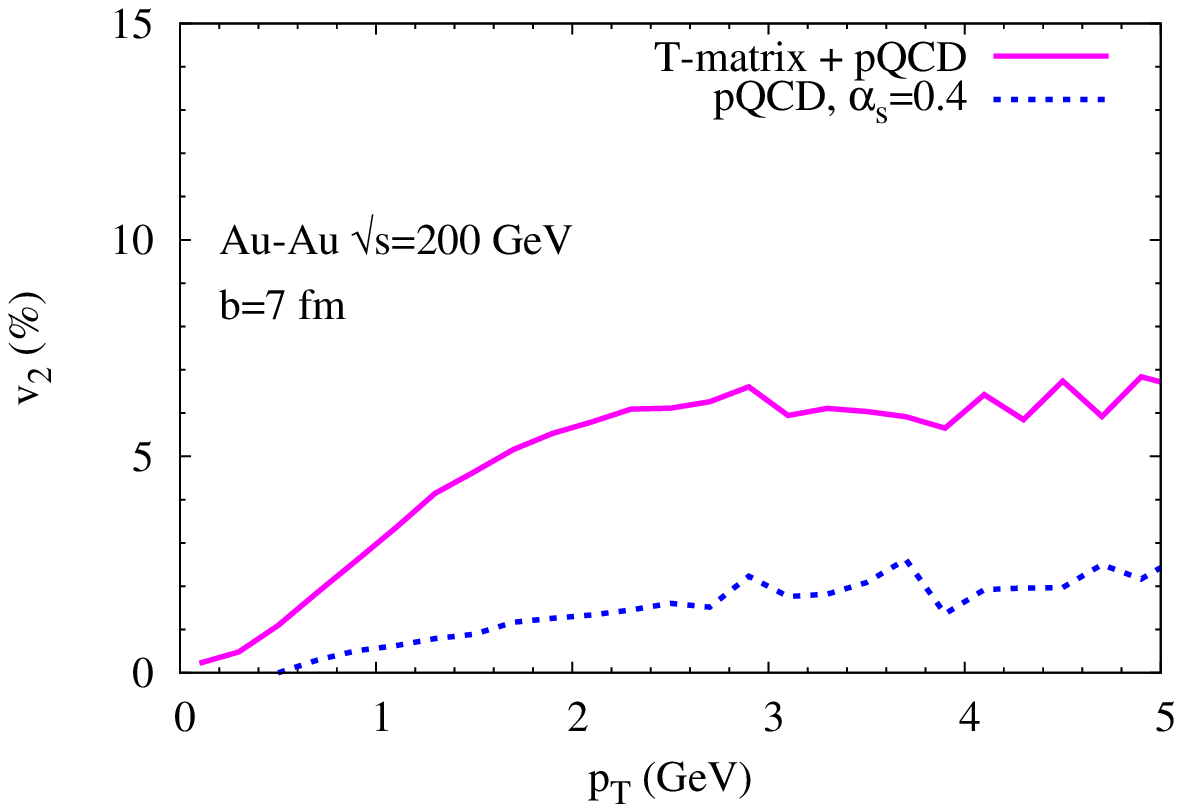}
\end{center}
\vspace{-0.4cm}
\caption{Nuclear modification factor (left panel) and elliptic flow
  (right panel) of charm quarks in central/semicentral Au-Au collisions
  at RHIC as computed within HQ Langevin simulations in a thermal
  fireball background with nonperturbative potential scattering off
  thermal quarks and antiquarks plus LO-pQCD scattering off
  gluons~\cite{vanHees:2007me,vanHees:2008xx}.}
\label{fig_v2-Raa-tmat}
\end{figure}

Finally, we show in Fig.~\ref{fig_v2-Raa-tmat} the fireball simulation
results for heavy quarks in the nonperturbative $T$-matrix approach
augmented by LO-pQCD scattering. Here, the latter only includes gluonic
interactions to strictly avoid any double counting with the perturbative
(Born) term of the $T$-matrix (which involves scattering off both quarks
and antiquarks from the heat bath). The results are very similar to the 
effective resonance+pQCD model.

\subsubsection{Heavy-Meson and Single-Electron Spectra}
\label{sssec_spectra}

To make contact with experiment, the quark spectra as computed in the
previous section require further processing. First, the quarks need to
be converted into hadrons ($D$ and $B$ mesons, their excited states, and
possibly baryons containing heavy quarks). Second, since the current
RHIC data are primarily for single electrons, the pertinent semileptonic
decays, $D,B \to e \nu_e X$, need to be evaluated.

The hadronization of quarks produced in energetic collisions of
elementary particles (e.g., $e^+e^-$ annihilation or hadronic
collisions) is a notoriously difficult problem that has so far evaded a
strict treatment within QCD and thus requires phenomenological input. A
commonly employed empirical procedure to describe hadronization of
quarks produced at large transverse momentum is to define a
fragmentation function, $D_{h/i}(z)$, which represents a probability
distribution that a parton, $i$, of momentum $p_i$ hadronizes into a
hadron, $h$, carrying a momentum fraction $z=p_h/p_i$ of the parent
parton (with $0<z\le1$, reflecting the fact that color-neutralization in
the fragmentation process requires the production of extra ``soft''
partons, which, in general, do not end up in $h$). At large enough
$p_T$, the parton production occurs at a very short time scale,
$\tau_{\rm prod}\simeq 1/p_T$, and thus hadronization, characterized by
a typical hadronic scale, $\tau_{\rm had}\simeq 1/\Lambda_{\rm QCD}$,
becomes independent of the production process (this is roughly the
essence of the ``factorization theorem'' of
QCD~\cite{qcdfac}). Therefore, the distribution $D(z)$ is supposed to be
universal, i.e., can be determined (or fit) in, e.g., $e^+e^-\to
hadrons$ and then be applied to hadronic collisions.  For light quarks
and gluons, $D(z)$ is typically a rather broad distribution centered
around $z\simeq 0.5$, while for heavy quarks it is increasingly peaked
toward $z=1$, sometimes even approximated by a Dirac $\delta$-function,
$D(z)=\delta(z-1)$ (so called $\delta$-function fragmentation).  Toward
lower $p_T$, other hadronization processes are expected to come into
play. In hadronic collisions, a possibility is that a produced quark
recombines with another quark or antiquark from its environment, e.g.,
valence quarks of the colliding hadrons~\cite{DH77}. There is ample
empirical evidence for the presence (and even dominance) of the
recombination mechanism in both $p$-$p$ and $\pi$-$p$ collisions, in
terms of flavor asymmetries of hadrons (including charmed
hadrons~\cite{KLS81,BJM02,Rapp:2003wn}) produced at forward/backward
rapidities, where recombination with valence quarks is favored). E.g.,
for charm production in $\pi^- N$ collisions, the $D^-/D^+$ ratio is
enhanced at large rapidity, $y$, indicating the presence of $\bar{c}d\to D^-$
recombination with a $d$-quark from the $\pi^-=d\bar{u}$ (but not $c
\bar{d}\to D^+$).

\begin{figure}[!t]
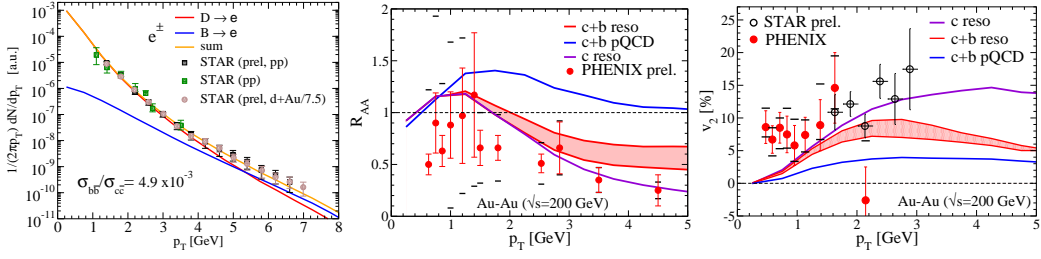

\includegraphics[width=0.32\linewidth]{elec-pp-dAu.eps}
\includegraphics[width=0.325\linewidth]{raa_e_mb-prc.eps}
\includegraphics[width=0.325\linewidth]{v2_e_mb-prc.eps}
\caption{Left panel: single-electron spectra from heavy-flavor decays in
  $p$-$p$ and $d$-Au collisions; the empirically inferred
  decomposition~\cite{vanHees:2005wb,Rapp:2005at} is compared to STAR
  data~\cite{star-dAu-pp,star-dAu}. Middle and right panel: $e^\pm$
  elliptic flow and nuclear modification factor in semicentral Au-Au
  collisions ($b$=7fm/$c$)~\cite{vanHees:2005wb,Rapp:2005at} compared to
  first RHIC data~\cite{v2-phenix,v2pre-star,Jacak:2005af}.}
\label{fig_v2-Raa-elec0}
\end{figure}

As discussed in Sec.~\ref{ssec_urhic}, the quark recombination (or
coalescence) model has received renewed interest in the context of RHIC
data, by providing a successful explanation of 2 phenomena observed in
intermediate-$p_T$ hadron spectra, namely the constituent quark-number
scaling of the elliptic flow and the large baryon-to-meson ratios. It is
therefore natural to also apply it to the hadronization of heavy
quarks~\cite{Lin:2003jy,Greco:2003vf}, where it appears to be even more
suited since the HQ mass provides a large scale relative to which
corrections to the coalescence model are relatively suppressed even at
low momentum. We here follow the approach of Ref.~\cite{Greco:2003vf}
where the $p_T$ spectrum of a $D$-meson is given in terms of the light
and charm quark or antiquark phase space distributions, $f_{\bar q,c}$,
as
\begin{equation}
  \frac{\dd N_D^{\rm coal}}{\dd y\,\dd^2p_T} = g_D \int \frac{p\cdot \dd
    \sigma}{(2\pi)^3}\int \dd^3q \ f_D(q,x) \ f_{\bar q}(\vec p_{\bar q},\vec r_{\bar q}) 
  \ f_{c}(\vec p_{c},\vec r_{c}) \ , 
\end{equation}
where $\vec{p} = \vec{p}_{\bar q} + \vec{p}_{c}$ denotes the momentum of
the $D$ meson, $g_D$ a combinatorial factor (ensuring color-neutrality
and spin-isospin averaging), $f_D(q,x)$ the Wigner function of the
$D$-meson which is usually assumed to be a double Gaussian in relative
momentum, $\vec q= \vec p_c - \vec p_{\bar q}$ and size, $\vec r=\vec
r_c-\vec r_{\bar q}$, and $\dd \sigma$ represents an integration over
the hadronization volume. The charm-quark distribution function is
directly taken from the output of the Langevin simulations discussed in
the previous section, while the light-quark distributions are taken as
determined from the successful application of the coalescence model of
Ref.~\cite{Greco:2003xt} to light hadron observables at RHIC. The
coalescence mechanism, however, does not exhaust all charm quarks for
hadronization, especially at high $p_T$ where the light-quark
phase-space density becomes very small; in Ref.~\cite{vanHees:2005wb},
the ``left-over'' charm quarks, $\hat{N}_c$, have been hadronized with
$\delta$-function fragmentation (as was done when constructing the input
spectrum in connection with $p$-$p$ and $d$-Au data). The total
$D$-meson spectrum thus takes the form
\begin{equation}
  \frac{\dd N_D^{\rm tot}}{\dd y \, \dd^2p_T} = \frac{\dd N_D^{\rm
      coal}}{\dd y\,\dd^2p_T} + \frac{\dd \hat{N}_c^{\rm frac}}{\dd y\,\dd^2p_T} \ . 
\end{equation} 
In the approach of Ref.~\cite{vanHees:2005wb} the formation of baryons
containing charm quarks (most notably $\Lambda_c=udc$) has been
estimated to be rather small, with a $\Lambda_c/D$ ratio significantly
smaller than 1, and is therefore neglected. The same procedure as for
charm quarks is also applied for bottom-quark hadronization. In
principle, the QGP and mixed phase is followed by an interacting
hadronic phase (cf.~Fig.~\ref{fig_hi-evo}), where $D$ and $B$ mesons are
subject to further reinteractions. In Ref.~\cite{Fuchs:2004fh} $D$-meson
reaction rates, $\Gamma_D^{\rm HG}$, have been estimated in a hot pion
gas. Even at temperatures close to $T_c\simeq0.18$~GeV, $\Gamma_D^{\rm
  HG}\le 0.05$~GeV, which is significantly smaller than in the QGP at
all considered temperatures, cf.~Fig.~\ref{fig_ImT}, and are therefore
neglected in the calculations of Ref.~\cite{vanHees:2005wb}.  Finally,
the $D$- and $B$-meson spectra are decayed with their (weighted average)
semileptonic decay branching ($\sim$10\% for $D$ mesons), assuming a
dominance of 3-body decays (e.g., $D\to K e \nu$). Before discussing the
pertinent single-$e^\pm$ spectra in more detail, two important features
should be recalled: (i) the shape and magnitude of the decay electrons
closely follows those from the parent $D$
mesons~\cite{Greco:2003vf,Dong:2004ve}; (ii) the electron spectra are a
combination of charm and bottom decays which experimentally have not yet
been separated. Since bottom-quark spectra (and thus their decay
electrons) are predicted to be much less modified than charm spectra, a
reliable interpretation of the $e^\pm$ spectra mandates a realistic
partitioning of the 2 contributions. Following the strategy of
Ref.~\cite{vanHees:2005wb}, the input charm and electron spectra are
constructed as follows: one first reproduces available $D$-meson spectra
in $d$-Au collisions~\cite{star-D}, calculates the pertinent electron
decays and then adjusts the bottom contribution to reproduce the $e^\pm$
spectra in $p$-$p$ and $d$-Au reactions. As a result of this procedure,
the bottom contribution to the $e^\pm$ spectra in the elementary system
exceeds the charm contribution at momenta $p_T$$\simeq$5-5.5~GeV, see
left panel of Fig.~\ref{fig_v2-Raa-elec0}; this is consistent with the
(rather large) margin predicted by perturbative
QCD~\cite{Cacciari:2005rk}.

The first comparison of the $e^\pm$ spectra obtained within the Langevin
resonance+pQCD+coalescence model~\cite{vanHees:2005wb,Rapp:2005at}
to RHIC data available at the time is shown in the middle and right
panel of Fig.~\ref{fig_v2-Raa-elec0}. While no quantitative conclusions
could be drawn, a calculation with pQCD elastic scattering alone was
disfavored. One also notices that the bottom contribution leads to a
significant reduction of the $v_2^e$ at $p_T\ge 3$~GeV, as well as a
reduced suppression in $R_{AA}^e$, where the $c$ and $b$ contributions
become comparable for $p_T\ge 4.5$~GeV in semicentral collisions.

\begin{figure}[!t]
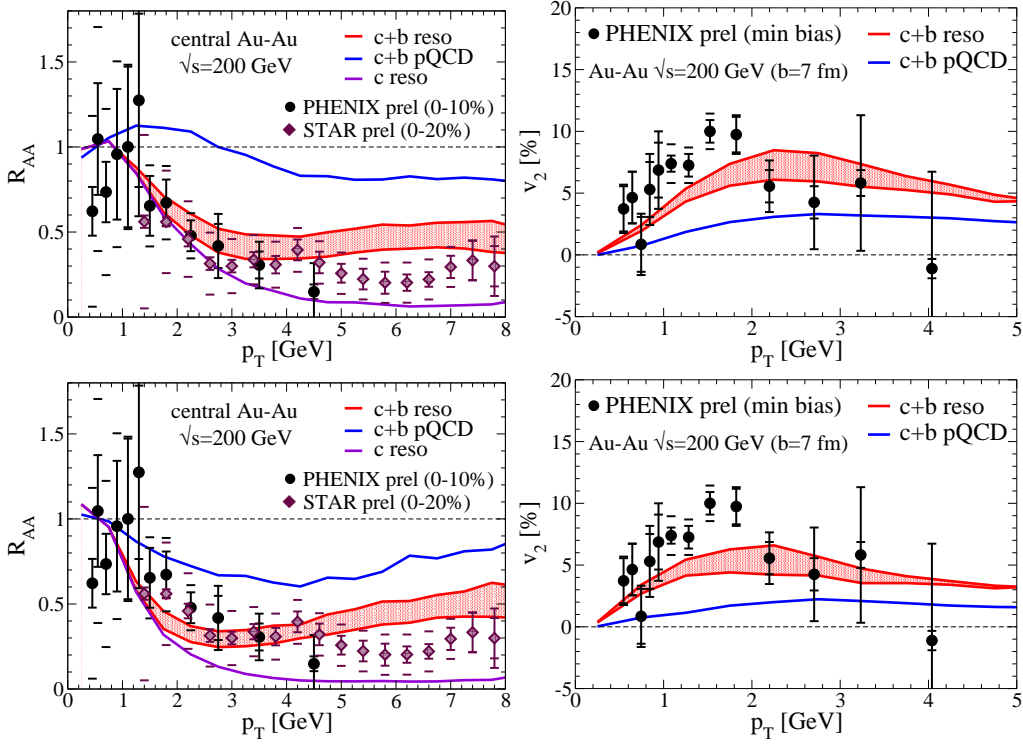

\begin{center}
\includegraphics[width=0.48\linewidth]{raa_e_cent-therm.eps}
\includegraphics[width=0.48\linewidth]{v2_e_MB_therm.eps}
\includegraphics[width=0.48\linewidth]{raa_e_cent_frag.eps}
\includegraphics[width=0.48\linewidth]{v2_e_MB_frag.eps}
\end{center}
\caption{Elliptic flow and nuclear modification factor of electrons from
  heavy-flavor decays in Au-Au collisions at RHIC as computed within HQ
  Langevin simulations in a thermal fireball background employing the
  resonance+pQCD model for HQ interactions in the
  QGP~\cite{vanHees:2005wb,Rapp:2006ta}. In the lower panels, quark
  coalescence processes at hadronization are switched off. The data are
  from Refs.~\cite{Adler:2005xv,Butsyk:2005qn,Bielcik:2005wu}.}
\label{fig_v2-Raa-elec1}
\end{figure}

\begin{figure}[!t]
\begin{minipage}{7cm}
\vspace{-0.3cm}
\includegraphics[width=0.93\linewidth]{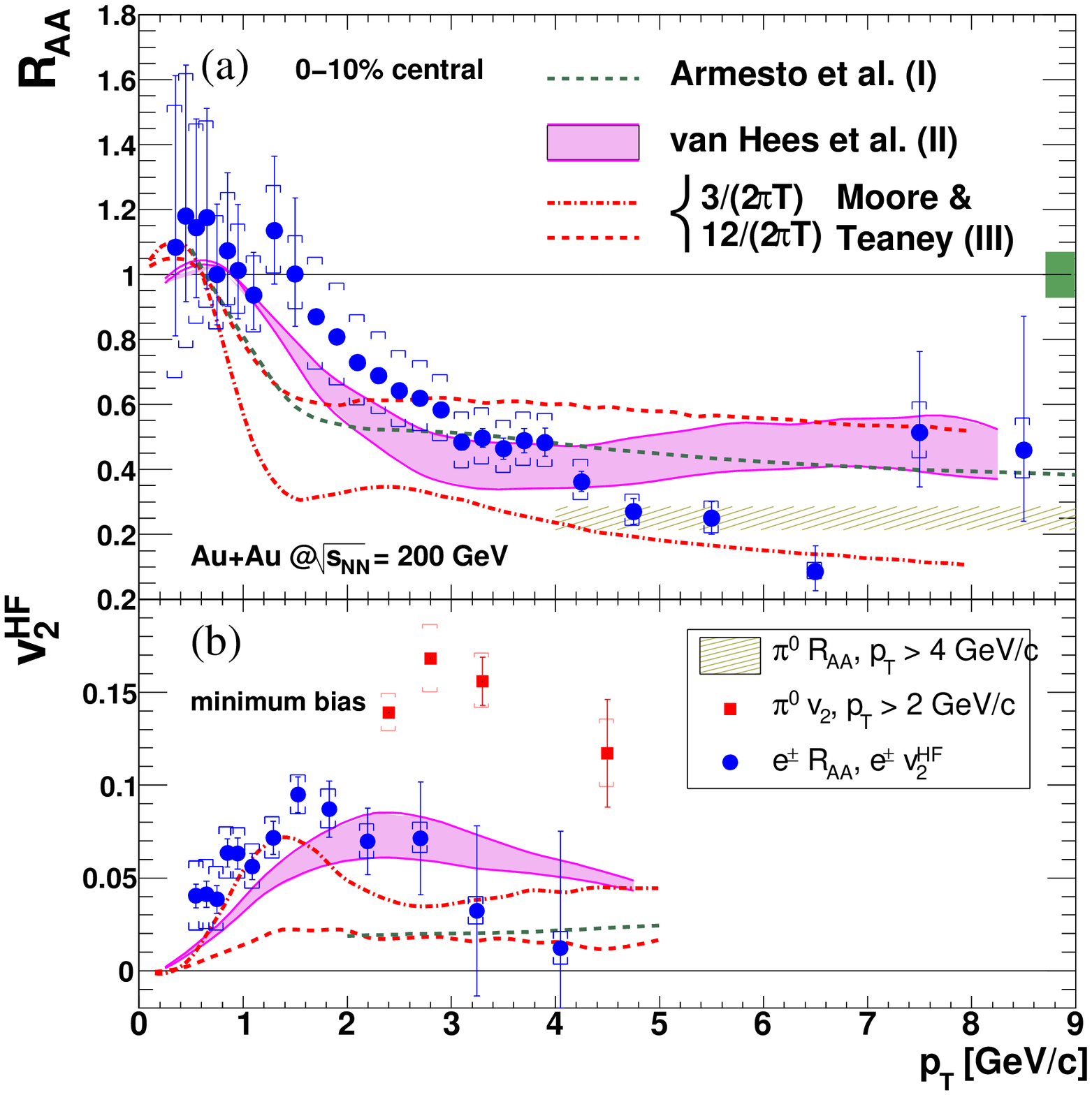}
\end{minipage}
\begin{minipage}{7cm}
\includegraphics[width=0.93\linewidth]{raa-central-v2mb.eps}
\end{minipage}
\caption{Elliptic flow and nuclear modification factor of electrons from
  heavy-flavor decays in Au-Au collisions at RHIC. Left panel: PHENIX
  data~\cite{Adare:2006nq} compared to theoretical predictions based on the
  Langevin simulations in the resonance+pQCD+coalescence model 
  (bands)~\cite{vanHees:2005wb} or with upscaled pQCD interactions 
  (dash-dotted and dotted lines)~\cite{Moore:2004tg}, as well as with 
  radiative energy-loss
  calculations (dashed lines)~\cite{Armesto:2005mz}.  Right panel:
  PHENIX~\cite{Adare:2006nq} and STAR~\cite{Abelev:2006db} are 
  compared to Langevin simulations employing HQ
  $T$-matrix interactions plus coalescence~\cite{vanHees:2007me}. }
\label{fig_v2-Raa-elec2}
\end{figure}

The resonance+pQCD+coalescence model is compared to improved $e^\pm$ 
data~\cite{Adler:2005xv,Butsyk:2005qn,Bielcik:2005wu} in the
upper panels of Fig.~\ref{fig_v2-Raa-elec1}. For central Au-Au
collisions, the $e^\pm$ suppression appears to be underpredicted
starting at $p_T\simeq$4-5~GeV (with the $b$ contribution exceeding the
charm at $p_T\ge3.7$~GeV), indicating the presence of additional
suppression mechanisms at higher momenta.
The lower panels in Fig.~\ref{fig_v2-Raa-elec1} show the results of
calculations without quark coalescence, i.e., all $c$ and $b$ quarks are
hadronized with $\delta$-function fragmentation. The shape of the
$R_{AA}^e(p_T)$, as well as the magnitude of $v_2^e$, are not properly
reproduced. This conclusion has been consolidated by another improvement
of the experimental results, displayed in the left panel of
Fig.~\ref{fig_v2-Raa-elec2} together with theoretical predictions within
the resonance+pQCD+coalescence model~\cite{vanHees:2005wb}, the 
hydrodynamic Langevin simulations with pQCD-inspired transport
coefficients~\cite{Moore:2004tg}, as well as the radiative energy-loss
approach with a transport coefficient
$\hat{q}$=14~GeV$^2$/fm~\cite{Armesto:2005mz}. Both Langevin simulations
point at a HQ diffusion coefficient of around $D_s\simeq 5/(2\pi T)$,
cf.~Fig.~\ref{fig_Ds}. The comparison of the 2 Langevin approaches
reiterates the importance of the coalescence contribution: the latter is
absent in the hydro calculations of Ref.~\cite{Moore:2004tg} which
cannot simultaneously describe the measured elliptic flow and
suppression with a single value of the diffusion coefficient. Quark
coalescence, on the other hand, introduces an ``anti-correlation'' of
$v_2$ and $R_{AA}$ into the spectra which increases the $v_2$ but
decreases the suppression (larger $R_{AA}$), which clearly improves on a
consistent description of the data. In the radiative energy loss
approach~\cite{Armesto:2005mz}, the suppression is approximately
reproduced but the azimuthal asymmetry is too small, especially at low
$p_T$. As discussed in connection with the Fokker-Planck
Eq.~(\ref{FP0}), energy-loss calculations do not account for momentum
diffusion; a non-zero $v_2$ is therefore solely due to the geometric
path length difference across the long and the short axes of the
almond-shaped transverse fireball area (a shorter path length inducing
less suppression). The lack of $v_2$ thus corroborates the
interpretation that the charm (and maybe bottom) quarks become part of
the collectively expanding medium, while the large transport coefficient
supports the strongly coupled nature of the medium, even without
diffusion and coalescence.

Finally, the right panel of Fig.~\ref{fig_v2-Raa-elec2} shows the
predictions of fireball Langevin simulations employing the
nonperturbative $T$-matrix+pQCD+coalescence approach for HQ
interactions~\cite{vanHees:2007me}; the agreement with PHENIX data is
fair. One should keep in mind that the inherent
uncertainties in this approach, e.g., in the extraction and definition
of an in-medium two-body potential from the lattice-QCD free energy, are
still appreciable, estimated at around $\pm$30\% at the level of $e^\pm$
observables~\cite{vanHees:2007me}. A conceptually attractive feature of
the $T$-matrix+coalescence approach is that it directly connects two
thus far disconnected phenomena observed at RHIC, namely the strongly
coupled QGP and quark coalescence: the very same interaction that
induces the strong coupling of the heavy quarks to the QGP leads to the
formation of ``pre-hadronic'' resonance structures close to $T_c$.  The
latter are naturally identified with the meson ($M$) and diquark ($dq$)
states building up hadrons in a $q+\bar q \to M$ and $q+q\to dq, dq+q\to
B$ processes ($B$: baryon), as suggested by the universal
constituent-quark number scaling (CQNS), see also
Ref.~\cite{Ravagli:2007xx}.

\subsection{SQGP at RHIC?}
\label{ssec_sqgp}

Let us now try to elaborate on the possible broader impact of the
current status of the HQ observables and their interpretation. It is
gratifying to see that the available electron data thus far confirm the
strongly coupled nature of the QGP produced at RHIC; relativistic
Langevin simulations have quantified this notion in terms of extracted
transport coefficients which are a factor of 3-5 stronger than
expectations based on elastic perturbative QCD interactions. At least in
the low-momentum regime, this should be a reliable statement since the
heavy-quark mass warrants the main underlying assumptions, namely: (i)
the applicability of the Brownian motion approach, and (ii) the
dominance of elastic interactions. Clearly, an important next step is to
augment these calculations by a controlled implementation of radiative
energy-loss mechanisms, see, e.g., Ref.~\cite{Vitev:2007jj} for a first
estimate.  As for the microscopic understanding of the relevant
interactions underlying the HQ rescattering, it should be noted that
large couplings necessarily require resummations of some sort,
especially for the problem at hand, i.e., HQ diffusion, for which the
perturbation series converges especially
poorly~\cite{CaronHuot:2007gq}. An example of such a (partial)
resummation is given by the $T$-matrix
approach~\cite{Mannarelli:2005pz,vanHees:2007me} discussed above, which
is particularly valuable when it can be combined with model-independent
input from lattice QCD. Here the objective must be to reduce the
uncertainties in the definition and extraction of a suitable potential.
This task can be facilitated by computing ``Euclidean'' (imaginary time)
correlation functions for heavy-light mesons and check them against
direct lattice computations which can be carried out with good accuracy.
These kinds of constraints are currently pursued in the heavy quarkonium
sector~\cite{Mocsy:2005qw,Alberico:2005xw,Cabrera:2006wh}, i.e., for
$Q$-$\bar Q$ correlation functions, indicating that potential models are
a viable framework to describe in-medium HQ interactions.

\begin{figure}[!t]
\begin{minipage}{7cm}
\vspace{-0.15cm}
\includegraphics[width=0.97\linewidth]{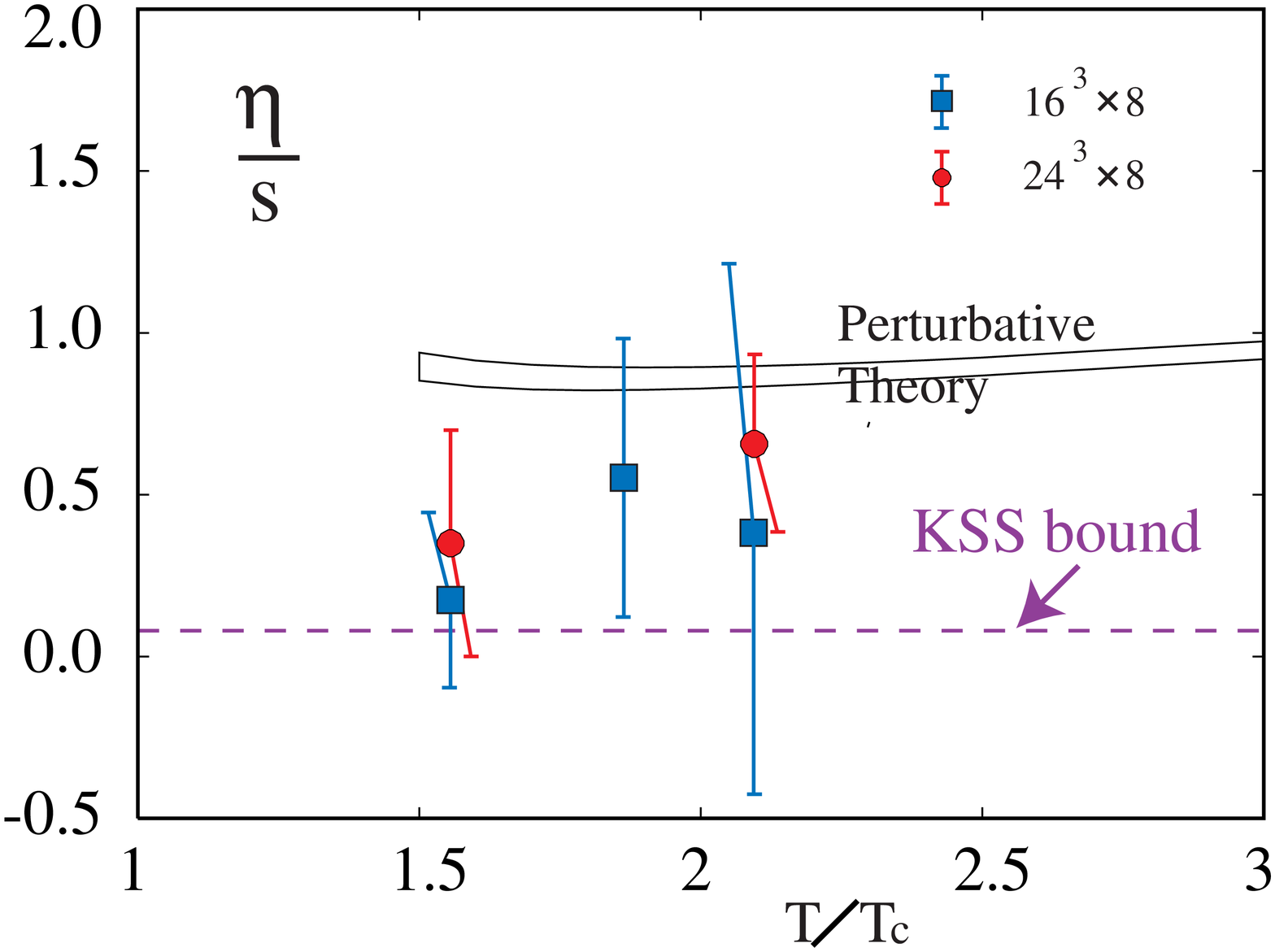}
\end{minipage}
\begin{minipage}{7cm}
\includegraphics[width=0.95\linewidth]{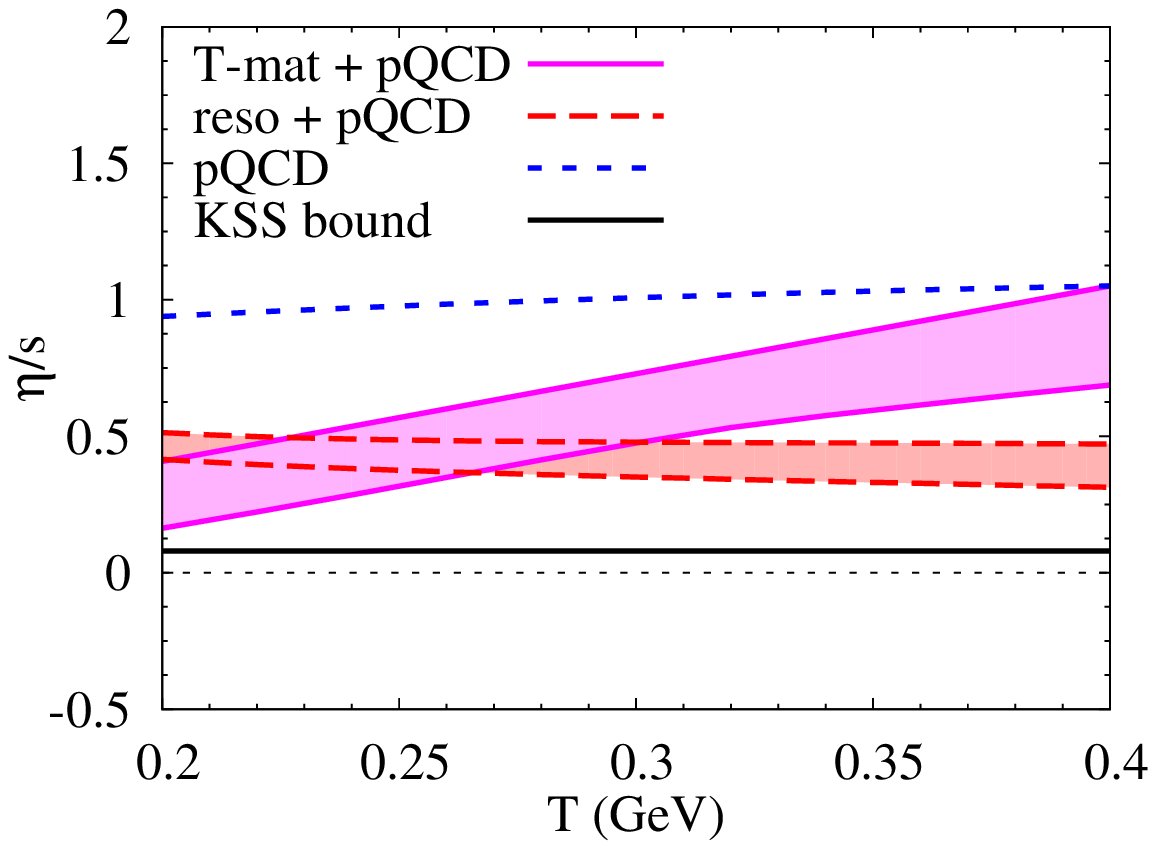}
\end{minipage}
\caption{The ratio of shear viscosity to entropy density, $\eta/s$. Left
  panel: lattice QCD computations in a gluon
  plasma~\cite{Nakamura:2004sy} compared to results inferred from
  perturbation theory~\cite{Blaizot99,AMY03}. Right panel: schematic
  estimates utilizing the HQ diffusion constant employing (a) LO pQCD
  elastic scattering ($\alpha_s$=0.4) in the weakly interacting limit
  (\ref{etas-pqgp}) (dashed line), (b) the effective resonance model 
  in the strong-coupling limit (\ref{etas-cft}) (band enclosed by dashed
  lines), and (c) the lattice-QCD potential based $T$-matrix
  approach (augmented by pQCD scattering off gluons) represented by 
  the band enclosed by solid lines constructed from the weak and strong 
  coupling limits.}
\label{fig_eta-s}
\end{figure}

An alternative nonperturbative approach to describe the medium of a
strongly-coupled gauge theory has recently been put forward by
exploiting connections between string theory and Conformal Field Theory
(CFT), the so-called AdS/CFT correspondence. The key point here is a
conjectured duality between the weak-coupling limit of a certain string
theory (defined in Anti-de-Sitter (AdS) space) and the strong-coupling
limit of a supersymmetric gauge theory (``conformal'' indicates that the
theory does not carry any intrinsic scale, such as $\Lambda_{QCD}$ in
QCD; this difference may, in fact, be the weakest link in the
identification of the CFT plasma with the QGP; it implies, e.g., the
absence of a critical temperature in CFT). A remarkable result of such a
correspondence is the derivation of a universal value for the ratio of
shear viscosity to entropy density pertaining to a large class of
strongly coupled quantum field theories, $\eta/s=1/(4\pi)$, which was
furthermore conjectured to be an absolute lower bound for any quantum
liquid~\cite{Kovtun:2004de}. Within the same framework, the HQ diffusion
has been computed, with the result $D_s\simeq 1/(2\pi
T)$~\cite{Herzog:2006gh,CasalderreySolana:2006rq}. The assumption that
the QGP is indeed in a strongly coupled regime (sQGP) can then be
utilized to establish a relation between $\eta/s$ and the heavy-quark
diffusion constant, and thus obtain a quantitative estimate of
$\eta/s$. Based on AdS/CFT, and exploiting the proportionality
$\eta/s\propto D_s$ (as, e.g., borne out of kinetic theory) one has
\begin{equation}
  \frac{\eta}{s} \approx  \frac{1}{4\pi} D_s (2\pi T) = 
  \frac{1}{2} \ T \ D_s \qquad
  {\rm (AdS/CFT)}  
\label{etas-cft}
\end{equation}
using the empirically inferred $D_s(2\pi T)$=4-6, one arrives at
$\eta/s$=(4-6)/(4$\pi$)=(1-1.5)/$\pi$. This may be compared to a rather
recent lattice QCD computation displayed in the left panel of
Fig.~\ref{fig_eta-s}~\cite{Nakamura:2004sy}, which also contains the
result of a perturbative calculation. A caveat here is that the lattice
computations are for a pure gluon plasma (GP).

An alternative estimate for $\eta/s$ might be obtained in the
weak-coupling regime. Starting point is a kinetic theory estimate of
$\eta$ for an ultrarelativistic gas~\cite{Danielewicz:1984ww,IV70}
\begin{equation}
\eta \approx \frac{4}{15} n \ \langle p \rangle \ \lambda_{\rm tr} \ ,  
\end{equation} 
where $n$ is the particle density, $\lambda_{\rm tr}$ the transport mean
free path over which a particle's momentum is degraded by an average
momentum $\langle p \rangle$. Assuming the latter to be of order of the
average thermal energy of a massless parton, one has $n \langle p
\rangle \approx \varepsilon$ (the energy density of the gas). Further
using $Ts = \varepsilon + P = \frac{4}{3} \varepsilon$ and $\lambda_{\rm
  tr}= \tau_{\rm tr}$, one obtains
\begin{equation}
\frac{\eta}{s} \approx \frac{1}{5} \ T \ \tau_{\rm tr} \ .  
\end{equation} 
Finally, accounting for the delayed thermal relaxation time of a heavy
quark via $\tau_{Q}\approx (T/m_Q) \tau_{\rm tr}$, and using the
expression, Eq.~(\ref{Dx}), for the HQ diffusion constant, one arrives
at the following rough estimate for a weakly coupled (perturbative) QGP
(wQGP),
\begin{equation}
\frac{\eta}{s} \approx \frac{1}{5} \ T \ D_s  \qquad {\rm (wQGP)} \ .  
\label{etas-pqgp}
\end{equation}
Note the significantly smaller coefficient in this estimate compared to
the one in expression (\ref{etas-cft}), which reflects the expected
underestimation of the shear viscosity if a gas estimate is applied in a
liquid-like regime (as emphasized in Ref.~\cite{Danielewicz:1984ww}).
As a rough application we may use the LO pQCD results for the HQ
diffusion coefficient. With $D (2\pi T)\simeq 40$ for a gluon plasma
(GP) at $T$=0.4~GeV (see Fig.~\ref{fig_Ds}, with a $\sim$25\% increase
for removing the contributions from thermal quarks and antiquarks), one
finds $\eta/s\simeq1.25$, which is surprisingly close to the
perturbative estimate constructed in Ref.~\cite{Nakamura:2004sy} which
is based on a next-to-leading logarithm of the shear
viscosity~\cite{AMY03} and a hard-thermal-loop calculation of the
entropy density~\cite{Blaizot99} (both of which represent pQCD
calculations beyond the LO estimate of the HQ diffusion constant in
Fig.~\ref{fig_Ds}).

In the right panel of Fig.~\ref{fig_eta-s} we attempt a schematic estimate
of $\eta/s$ in the Quark-Gluon Plasma based on the 3 basic calculations
of the HQ diffusion coefficient discussed throughout this article. For
the LO-pQCD calculation, we adopt the estimate (\ref{etas-pqgp}) for a
weakly coupled gas, while for the resonance+pQCD model we use the
strong-coupling estimate (\ref{etas-cft}). The most realistic estimate
is presumably represented by the $T$-matrix+pQCD calculation, for which
we constructed a pertinent band in $\eta/s$ as follows: 
the lower limit of the band is based on the weak-coupling estimate,
while for the upper limit we adopt the strong-coupling estimate at the 
low temperature end ($T$=0.2~GeV), the LO-pQCD only result at the 
high-$T$ end ($T$=0.4~GeV) and a linear interpolation in between  
these 2 temperatures (there are additional uncertainties which are not 
displayed, e.g., due to the extraction of the lQCD-based interaction 
potentials).
\begin{figure}[!t]
\begin{center}
\includegraphics[width=0.45\linewidth]{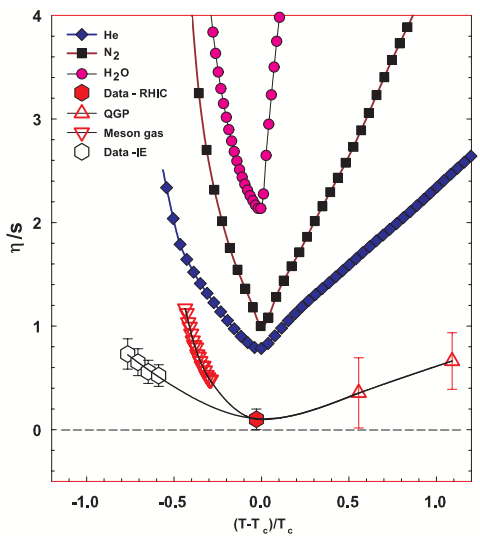}
\end{center}
\caption{Compilation of the ratio of shear viscosity to entropy density
  for various substances~\cite{Lacey:2006pn}: atomic He, molecular N$_2$
  and H$_2$O~\cite{Csernai:2006zz} (upper 3 symbols), pure-glue lattice
  QCD (upward triangles above $T_c$)\cite{Nakamura:2004sy}, pion gas
  (downward triangles below $T_c$)~\cite{Chen:2006iga} and empirical
  estimates from heavy-ion data (hexagons)~\cite{Lacey:2006pn}.}
\label{fig_etas-comp}
\end{figure}

A remarkable feature of the lattice-QCD potential based $T$-matrix
approach is that the interaction strength {\em decreases} with
increasing temperature - in other words, the most strongly coupled
regime appears to be close to the critical temperature. It turns out
that the occurrence of a maximal interaction strength at a phase
transition is a rather generic phenomenon which is present in a large
variety of substances: at their critical pressure, helium, nitrogen and
even water have a very pronounced minimum in $\eta/s$ at the critical
temperature, as is nicely demonstrated in Ref.~\cite{Csernai:2006zz} and
compiled in Fig.~\ref{fig_etas-comp} taken from
Ref.~\cite{Lacey:2006pn}. This plot also contains calculations for
$\eta/s$ in the hadronic phase, using free $\pi$-$\pi$ interactions in a
pion gas~\cite{Chen:2006iga} (similar results are obtained for a
$\pi$-$K$ gas with empirical (vacuum) scattering phase
shifts~\cite{Prakash:1993bt}). The decrease of $\eta/s$ in a hot meson
gas with increasing $T$ corroborates the presence of a minimum around
the critical temperature.  The finite-temperature QCD phase transition
at $\mu_q$=0 is presumably a cross-over, but the minimum structure of
$\eta/s$ close to $T_c$ is likely to persist, in analogy to the atomic
and molecular system above the critical pressure~\cite{Csernai:2006zz}.
If such a minimum is indeed intimately linked to the critical
temperature, the AdS/CFT correspondence may be of limited applicability
to establish rigorous connections between the sQGP (and RHIC
phenomenology) and CFTs, since, as mentioned above, the latter do not
possess an intrinsic scale.

The question of how a gluon plasma (GP) compares to a quark-gluon plasma
is not merely a theoretical one, but also of practical relevance.  In
heavy-ion collisions at collider energies (RHIC and LHC), the (very)
early phases of the reaction are presumably dominated by the (virtual)
gluon fields in the incoming nuclei. A lot of progress has been made in
recent years in the determination of these gluon distributions and their
early evolution in URHICs~\cite{McLerran:2005kk}. The modifications of
the HQ spectra due to these strong ``color'' fields should certainly be
addressed in future work. In the subsequent evolution, an early formed
GP is estimated to chemically equilibrate into a QGP rather rapidly, and
closer to $T_c$ quark coalescence models are suggestive for the
dominance of quark degrees of freedom - this is where the $T$-matrix
approach with lQCD-based potentials is predominantly operative (recall
that it hinges on the presence of quarks and antiquarks in the
medium). The underlying interactions could therefore provide a unified
and quantitative framework for HQ diffusion, quark coalescence and the
sQGP in the vicinity of $T_c$, based on input which, in principle, is
directly extracted from lQCD. Its further development should
aim at a better determination of potentials extracted from lQCD, include
constraints from lattice correlation functions and applications to
quarkonia, and implement contributions due to gluon radiation processes.

\section{Conclusion and Outlook}
\label{sec_concl}

The study of elementary particle matter has been an extremely active
research field over the last 20-30 years, and it may not have reached
its peak time yet. We are beginning to understand better what the key
features of media are whose forces are directly governed by gauge
theories. Here the strong nuclear force between quarks and gluons (as
described by Quantum Chromodynamics) occupies a special role due to its
large interaction strength and the self-couplings of its field quanta.
On the one hand, QCD gives rise to novel nonperturbative phenomena in
the vacuum, most notably the confinement of quarks into hadrons and the
chiral symmetry breaking (generating the major part of the visible mass
in the universe), whose underlying mechanisms are, however, not yet
understood. On the other hand, the Strong Force generates a very rich
phase structure of its different matter states (upon varying temperature
and baryon density), which are even less understood. Pertinent phase
transitions are, in fact, closely connected with deconfinement and 
chiral symmetry restoration. First principle numerical calculations of 
discretized (lattice) QCD at finite temperature have clearly established
a transition (or a rapid cross over) from hadronic matter into a
deconfined Quark-Gluon Plasma with restored chiral symmetry at a
temperature of $T_c\simeq 0.175$~GeV. This state of matter is believed
to have prevailed in the early Universe in the first few microseconds
after the Big Bang. A particularly exciting aspect of this research
field is that such kind of matter can be reproduced, at least for a
short moment, in present-day laboratory experiments, by accelerating and
colliding heavy atomic nuclei. However, to analyze these reactions, and
to extract possible evidence for QGP formation out of the debris of
hundreds to thousands of produced hadrons, is a formidable task that
requires a broad approach, combining information from lattice QCD,
effective models of QCD in a well-defined applicability range, and 
their implementation into heavy-ion phenomenology. Large progress has 
been made at the Relativistic Heavy-Ion Collider (RHIC), where clear 
evidence for the formation of thermalized QCD matter well above the 
critical energy density has been deduced. The apparently small 
viscosity of the medium is inconsistent with a weakly
interacting gas of quarks and gluons, but is possibly related to recent
results from lattice QCD which suggest the presence of resonant
correlations for temperatures up to $\sim$2~$T_c$. We have argued that
heavy quarks can serve as controlled probe of the transport properties
of the strongly coupled QGP (sQGP). The modifications of heavy-quark
spectra in Au-Au collisions at RHIC can be evaluated in a Brownian
motion framework which allows to establish quantitative connections
between the heavy-quark diffusion coefficients and observables, such as
the suppression of their spectra and especially their elliptic flow.
Theoretical analyses have confirmed that perturbative interactions are
too weak to account for the measured heavy-quark observables (i.e., 
their electron decay spectra), while an effective resonance model seems 
to furnish the required nonperturbative interaction strength. An
appealing framework to calculate heavy-quark interactions (and transport 
properties) in the medium is to extract interaction potentials 
from lattice QCD and iterate them in a nonperturbative $T$-matrix
equation. This approach is, in principle, free of adjustable parameters, 
but currently subject to significant uncertainties, primarily in the 
definition of the
potentials and their applicability to light quarks and at high
momentum. However, promising results have been obtained in that the
$T$-matrix builds up resonance-like structures close to the phase
transition, which could be instrumental in explaining the observed
elliptic flow. In addition, the resonance correlations naturally explain
the importance of quark coalescence processes for the hadronization of
the QGP (as indicated by the measured light hadron spectra). Clearly, a
lot more work is required to elaborate these connections more rigorously
and quantitatively, to implement additional components toward a more
complete description (e.g., energy loss via gluon radiation or the
effects of strong color fields), to scrutinize the results in comparison
to improved lattice QCD computations, and to confront
calculations with high precision RHIC (and LHC) data.  The latter are
expected to emerge in the coming years and will surely hold new 
surprises, further pushing the frontier of our knowledge of the
Quark-Gluon Plasma.
     
\vspace{1.0cm}

\noindent
{\bf \Large Acknowledgment}\\
We are indebted to our colleagues Vincenzo Greco, Che-Ming Ko and
Massimo Mannarelli for many fruitful discussions and collaboration,
and thank William Zajc for a careful reading of the manuscript.
This work is supported in part by a U.S. National Science Foundation
CAREER award under grant no. PHY-0449489.

\end{document}